\documentclass[times,final]{elsarticle}

\usepackage{jcomp}
\usepackage{framed,multirow}

\usepackage{amssymb}
\usepackage{latexsym}

\usepackage{url}
\usepackage{xcolor}
\definecolor{newcolor}{rgb}{.8,.349,.1}

\usepackage{graphicx}
\usepackage{epstopdf, epsfig}
\usepackage{hyperref}
\usepackage{hhtensor}
\usepackage{amsmath}

\newcommand{\reft}[1]{\textcolor{black}{#1}}
\newcommand{\refm}[1]{\textcolor{black}{#1}}
\newcommand{\jlvg}[1]{\textcolor{black}{#1}}
\newcommand{\jlvgp}[1]{\textcolor{black}{#1}}
\newcommand{\fsa}[1]{\left\langle{#1}\right\rangle}
\newcommand{\KNOSOS}{{\ttfamily KNOSOS}}
\newcommand{\DKES}{{\ttfamily DKES}}
\newcommand{\EUTERPE}{{\ttfamily EUTERPE}}

\journal{Accepted for publication in Journal of Computational Physics}

\begin{document}

\verso{J. L. Velasco \textit{et al.}}

\begin{frontmatter}

\title{KNOSOS: a fast orbit-averaging neoclassical code for stellarator geometry}

\author[1]{J. L. {Velasco}\corref{cor1}}
\cortext[cor1]{Corresponding author}
 \ead{joseluis.velasco@ciemat.es}
\author[1]{I. {Calvo}}
\author[2]{F. I. {Parra}}
\author[1]{J. M. {Garc\'ia-Rega\~na}}

\address[1]{Laboratorio Nacional de Fusi\'on, CIEMAT, 28040 Madrid, Spain}
\address[2]{Rudolf Peierls Centre for Theoretical Physics, University of Oxford, Oxford OX1 3PU, UK}

\received{}
\finalform{}
\accepted{}
\availableonline{}
\communicated{}

\begin{abstract}

{\ttfamily KNOSOS}~(KiNetic Orbit-averaging SOlver for Stellarators) is a freely available, open-source code (\href{https://github.com/joseluisvelasco/KNOSOS}{https://github.com/joseluisvelasco/KNOSOS}) that calculates neoclassical transport in low-collisionality plasmas of three-dimensional magnetic confinement devices \jlvg{by solving the radially local drift-kinetic and quasineutrality equations}. The main feature of {\ttfamily KNOSOS}~is that it relies on orbit-averaging to solve the drift-kinetic equation very fast.~{\ttfamily KNOSOS}~treats rigorously the effect of the component of the magnetic drift that is tangent to magnetic surfaces, and of the \jlvg{component of the electrostatic potential that varies on the flux surface, $\varphi_1$. Furthermore, the equation solved is linear in $\varphi_1$, which permits an efficient solution of the quasineutrality equation. As long as the radially local approach is valid, {\ttfamily KNOSOS} can be applied to the calculation of neoclassical transport in stellarators (helias, heliotrons, heliacs, etc.) and tokamaks with broken axisymmetry. In this paper, we show several calculations for the stellarators W7-X, LHD, NCSX and TJ-II that provide benchmark with standard local codes and demonstrate the advantages of this approach.}

\end{abstract}

\end{frontmatter}


\section{Introduction}\label{SEC_MOT}


Stellarators are non-axisymmetric devices in which the magnetic field is created basically by external magnets, without the need of any mechanism to drive current within the plasma. This provides them with an inherent capability for steady state operation and makes them less prone to plasma magnetohydrodynamic instabilities, but it also generally produces larger energy losses: at low collisionalities, the combination of \jlvg{magnetic} geometry and particle collisions leads to a variety of stellarator-specific neoclassical transport regimes, which usually give a large contribution to the radial energy and particle transport in the core of the device~\citep{dinklage2013ncval,dinklage2018np}. Of special relevance are the 1/$\nu$, the $\sqrt{\nu}$ and the superbanana-plateau regimes~\citep{hokulsrud1986neo,beidler2011icnts,calvo2017sqrtnu}, in which the energy transport coefficients show a positive temperature dependence, much more unfavourable than the negative $T^{-1/2}$ scaling of the banana regime of the axisymmetric tokamak. 

The fundamental reason for this behaviour is that in a generic stellarator, unlike in an axisymmetric tokamak, trapped particle orbits have non-zero secular radial drifts. The exception are omnigenous stellarators: in these magnetic configurations, the secular radial drifts vanish~\citep{cary1997omni,parra2015omni}, and the level of neoclassical transport is low, similar to that of the tokamak. Quasisymmetric stellarators~\citep{boozer1983qs} are a particular family of omnigenous stellarators, see e.g.~\citep{landreman2012omni}.

The two world's largest stellarators in operation, Wendelstein 7-X (W7-X)~\citep{klinger2017op11,wolf2017op11} and the Large Helical Device (LHD)~\citep{takeiri2017iaea}, have relied on optimization of neoclassical transport for their design and operation. The magnetic configuration of W7-X has been designed to be close to omnigeneity with poloidally-closed contours of the magnetic field strength; one of the goals of the project has been to prove the constructability and reliability of such designs~\citep{sunnpedersen2016nature}. In LHD, the plasma column can be shifted inwards so that the minimum values of the magnetic field along the field line have approximately the same value (see figure 2 of~\citep{beidler2011icnts}), a well-known geometric property of some omnigenous magnetic fields~\citep{mynick1982omni,landreman2012omni}; discharges performed using this magnetic configuration consistently show better energy confinement~\citep{yamada2005taue}. Finally, a particular kind of quasisymmetry, quasiaxisymmetry, was the design criterion of the National Compact Stellarator Experiment (NCSX)~\citep{zarnstorff2001ncsx}. Power reactor designs exist for these three stellarator concepts~\citep{sagara2010reactors}.

It is then clear that optimization of neoclassical transport is a crucial issue for a stellarator reactor. One of the most common goals of stellarator optimization efforts is the minimization of the so-called \textit{effective ripple}, a figure of merit that provides information of the level of transport in the 1/$\nu$ regime. While there is little doubt that minimization of this quantity should be a design criterion in any future stellarator, it has important limitations. On the one hand, empirical studies of the energy confinement time of several devices aimed at obtaining a unified International Stellarator Scaling law (ISS04) have not shown a very strong correlation between reduced effective ripple and improved energy confinement~\citep{yamada2005taue,fuchert2018taue}; on the other hand, self-consistent neoclassical transport simulations performed in the configuration space of W7-X, complemented with simplified anomalous modelling (accounting for non-negligible turbulent contributions to transport), have shown mild increases of the energy confinement time for configurations of significantly reduced effective ripple~\citep{geiger2014w7x}. This points towards one of the obvious limitations of the effective ripple: it is only an appropriate figure of merit for neoclassical transport if the plasma species are in the asymptotic 1/$\nu$ regime. However, bulk particles are distributed close to a Maxwellian that typically spans across several transport regimes. Even in cases in which the collisionality is low and the neoclassical predictions of the radial energy flux agree with the experiment, the parameter dependence of the experimental energy flux does not follow the scaling expected for any specific neoclassical transport regime, see e.g.~\citep{alonso2017eps}, \jlvg{because the flux is caused by a combination of transport regimes.}
 
The reason for choosing the effective ripple as a figure of merit is that the 1/$\nu$ regime is the low-collisionality regime of stellarators in which the effect of the magnetic geometry on transport can be encapsulated in a straightforward manner in a single quantity that is independent of density, \jlvg{temperature} and radial electric field. Furthermore, this quantity can be efficiently calculated by solving the bounce-averaged drift-kinetic equation, e.g. with the \texttt{NEO} code~\citep{nemov1999neo}. None of this has been possible so far for other low-collisionality regimes for arbitrary stellarator geometry.

Moreover, for other regimes such as the $\sqrt{\nu}$ and the superbanana-plateau regimes, the effect of the electric field (radial and tangential to the flux surface, the latter associated to the variation of the electrostatic potential on the flux surface, $\varphi_1$) has to be considered~\citep{calvo2017sqrtnu}, and this quantity is determined by imposing ambipolarity of the neoclassical particle fluxes and quasineutrality, which in turn depend on the plasma profiles, and specifically on the gradients. In order to address this issue, self-consistent neoclassical transport simulations have been performed in the last few years: the neoclassical fluxes are calculated with the~\DKES~code~\citep{hirshman1986dkes} and then the ambipolar and energy transport equations are solved (the latter with a prescribed energy source)~\citep{turkin2011predictive,geiger2014w7x}. Although we will see that~\DKES~makes use of the so-called monoenergetic approximation, which reduces the problem from five dimensions to three, using~\DKES~to self-consistently solve neoclassical energy transport is still computationally expensive at low collisionality. Moreover,~\DKES~is inaccurate at \jlvg{sufficiently} low collisionality: it uses an incompressible $E\times B$ drift~\citep{beidler2007icnts} and does not include the tangential magnetic drift or the radial $E\!\times\!B$ drift caused by the variation of the electrostatic potential within the flux surface (the latter makes the fluxes depend non-linearly on the plasma gradients~\citep{calvo2018jpp}). Some or all of these approximations are absent in more recent codes such as \texttt{SFINCS}~\citep{landreman2014sfincs}, \texttt{EUTERPE}~\citep{regana2013euterpe,regana2017phi1} or \texttt{FORTEC-3D}~\citep{satake2006fortec3d}, but at the expense of higher computational cost.

We have developed a new code, the KiNetic Orbit-averaging Solver for Optimizing Stellarators, \texttt{KNOSOS}, based on the analytical techniques developed in a series of papers~\citep{calvo2013er,calvo2014er,calvo2015flowdamping,calvo2017sqrtnu,calvo2018jpp}. It solves local drift-kinetic equations that will be summarized in the next section and that accurately describe neoclassical transport in the $1/\nu$, $\sqrt{\nu}$ and superbanana-plateau regimes. The equations include the effect of the magnetic drift tangential to flux surfaces and the radial $E\!\times\!B$ drift due to the variation of the electrostatic potential within the flux surface; the radial electric field $E_r$ and $\varphi_1$ are obtained by imposing ambipolarity and quasineutrality, respectively. Local drift kinetic equations are valid for large-aspect-ratio stellarators or configurations close to omnigeneity (see the discussion before equation~(\ref{EQ_LOCAL}) in \S\ref{SEC_EQUATIONS}). Unlike \jlvg{preliminary} versions of {\ttfamily KNOSOS}~\citep{velasco2018phi1,calvo2018jpp}, this version does not require an explicit split of the magnetic field magnitude into omnigeneous and non-omnigeneous pieces. The goal of this code is to be, at the same time, accurate and fast, so that it allows one to perform comprehensive parameter scans and to provide input to other codes or suites of codes. Generally speaking, the goal is to improve our confidence in neoclassical predictions, in light of recent theory developments, and to be able to fully exploit these predictive capabilities. To facilitate this objective, the code is freely-available and open-source.

The rest of this paper is organised as follows. \S\ref{SEC_EQUATIONS} presents the drift-kinetic and quasineutrality equations solved by {\ttfamily KNOSOS}. Then, \S\ref{SEC_ALGORITHMS} summarises how the equations are solved: the drift-kinetic equation is written in terms of a few integrals along the magnetic field lines in \S\ref{SEC_FINALDKE}, and these integrals are discussed in \S\ref{SEC_COEFFICIENTS}; the parameter space and discretization of the drift-kinetic equation is discussed in \S\ref{SEC_GRID} and \S\ref{SEC_SOLDKE}, and the consistent solution of the drift-kinetic and quasineutrality equations is presented in \S\ref{SEC_SOLQN}. \S\ref{SEC_RESULTS} shows several calculations for real magnetic confinement devices and comparisons with widely benchmarked neoclassical codes: the monoenergetic transport coefficients are compared with~\DKES~in \S\ref{SEC_DKES}; the effect of the tangential magnetic drift on the energy flux is discussed in \S\ref{SEC_TANGVM}; the variation of the electrostatic potential along the flux surface is compared with {\ttfamily EUTERPE}~in \S\ref{SEC_EUTERPE}. Finally, \S\ref{SEC_CONCLUSIONS} summarizes the conclusions. Additionally, there are three appendixes:~\ref{AP0} discusses the collision operator, and appendices~\ref{AP1}~and~\ref{AP2} describe algorithms employed to accelerate the calculation of the bounce integrals.


\section{Equations}\label{SEC_EQUATIONS}


In this section, we briefly present the equations solved by {\ttfamily KNOSOS}. Their derivation and further details can be found in previous work by~\citep{calvo2017sqrtnu,calvo2018jpp}. We first define the coordinate system that we will use. The flux surfaces are labelled by the radial coordinate
\begin{equation}
\psi=|\Psi_t|\,,
\end{equation}
where $2\pi\Psi_t$ is the toroidal magnetic flux. The magnetic field lines on the surface are labelled by an angular coordinate
\begin{equation}
\alpha = \theta -\iota\zeta\,,
\end{equation}
where $\theta$ and $\zeta$ are poloidal and toroidal Boozer angles, respectively, and $\iota$ is the rotational transform. Finally, $l$ is the arc-length along the magnetic field line. In these coordinates, the magnetic field $\mathbf{B}$ can be written as
\begin{align}
\mathbf{B}=\Psi_t'\nabla \psi \times\nabla\alpha\,,
\end{align}
where primes stand for derivatives with respect to $\psi$, and $\Psi_t'=\pm 1$ depending on whether the magnetic field is parallel or antiparallel to the direction of the Boozer toroidal angle (i.e. depending on the sign of $\mathbf{B}\cdot\nabla\zeta$).

As velocity coordinates, we choose the particle velocity
\begin{equation}
v = |\mathbf{v}|\,,
\end{equation}
the pitch-angle coordinate
\begin{equation}
\lambda=\frac{1}{B}\frac{v_\perp^2}{v^2}\,,
\end{equation}
and the sign of the parallel velocity
\begin{equation}
\sigma= \frac{v_\parallel}{|v_\parallel|}=\pm 1\,,
\end{equation}
where, as usual,
\begin{align}
v_\parallel &= \mathbf{v}\cdot \mathbf{b} = \mathbf{v}\cdot\frac{\mathbf{B}}{|\mathbf{B}|}= \mathbf{v}\cdot\frac{\mathbf{B}}{B}\,,\nonumber \\
v_\perp &= \sqrt{v^2-v_\parallel^2}\,.
\end{align}

For each species $b$ ($i$ will denote bulk ions and $e$ electrons), we need to calculate \jlvgp{the deviation of the distribution function from a Maxwellian for trapped particles, that we denote by $g_b(\psi,\alpha,l,v,\lambda,\sigma)$. The Maxwellian distribution function reads}
\begin{equation}
 F_{M,b}=n_b\left(\frac{m_b}{2\pi T_b}\right)^{3/2}\exp{\left(-\frac{m_bv^2}{2T_b}\right)}\,,
\end{equation}
\jlvgp{where $n_b$ is the density, $T_b$ the temperature and $m_b$ the mass}. \jlvg{Trapped particles are those for which $v_\parallel=0$ at some point along their trajectories. For them, $1/B_{max}\le\lambda\le1/B_{min}$, where $B_{max}$ and $B_{min}$ are the maximum and minimum values of the magnetic field strength on the flux surface, respectively.}

\begin{figure}
\centering
\includegraphics[angle=0,width=0.6\columnwidth]{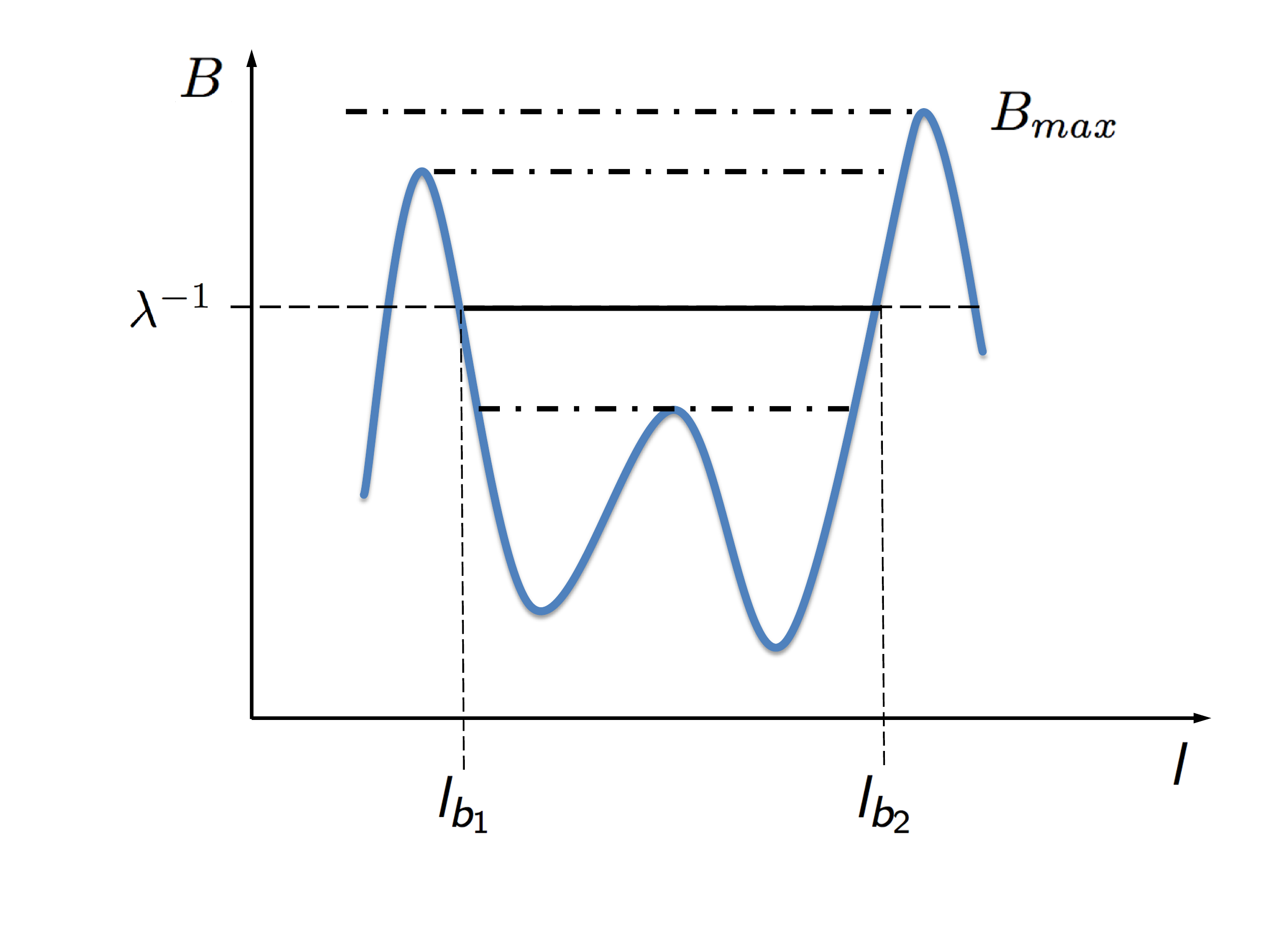}
\caption{Sketch of a particle trajectory at fixed $\alpha$. \reft{The horizontal thick line represents a standard trajectory, and dot-dashed horizontal lines depict trajectories (or parts of trajectories) with numerical divergencies (see text).}}
\label{FIG_ORBIT}
\end{figure}

The equation for $g_b(\psi,\alpha,v,\lambda)$ is
\begin{equation}
\int_{l_{b_1}}^{l_{b_2}} \frac{\mathrm{d}l}{|v_\parallel|} \mathbf{v}_{D,b}\cdot\nabla\alpha~\partial_\alpha g_b+\int_{l_{b_1}}^{l_{b_2}} \frac{\mathrm{d}l}{|v_\parallel|} \mathbf{v}_{D,b}\cdot\nabla \psi \Upsilon_b F_{M,b} =\int_{l_{b_1}}^{l_{b_2}} \frac{\mathrm{d}l}{|v_\parallel|} C_b^{\mathrm{lin}}[g_b]\,,
\label{EQ_DKEFINAL}
\end{equation}
complemented with the condition at the boundary between passing and trapped, $\lambda = 1/B_{max}$ ($B_{max}$ is the maximum value of the magnetic field strength on the flux-surface),
\begin{equation}
g_b (\lambda=1/B_{max})=0\,,
\label{EQ_CONT2}
\end{equation}
and the condition
\begin{equation}
\int_0^{2\pi}g_b~\mathrm{d}\alpha =0\,.
\label{EQ_CONT1}
\end{equation}
\jlvg{The coefficients of equation (\ref{EQ_DKEFINAL}) are integrals over the arc-length between the bounce points $l_{b_1}$ and $l_{b_2}$, i.e., between the points where the parallel velocity of the particle is zero (see a sketch in figure~\ref{FIG_ORBIT}). On the right-hand side of equation~(\ref{EQ_DKEFINAL})}, $C_b^{\mathrm{lin}}[g_b]$ is the linearized pitch-angle-scattering collision operator:
\begin{equation}
C_b^{\mathrm{lin}}[g_b]=
\frac{\nu_{\lambda,b} v_{||}}{v^2 B}\partial_\lambda\left(v_{||}\lambda\partial_\lambda g_b \right)\,.
\label{EQ_COLOP}
\end{equation}
For the ions, since $\sqrt{m_e/m_i}\ll 1$, this single-species collision operator is correct, but electron-ion collisions need to be retained in the electron drift-kinetic equation. For both species, we follow the common practice (see e.g.~\citep{beidler2011icnts}) of using equation (\ref{EQ_COLOP}) with an effective collision frequency accounting for inter-species collisions. This is discussed in more detail in~\ref{AP0}.
On the left-hand-side of equation~(\ref{EQ_DKEFINAL}), 
\begin{equation}
\Upsilon_b = \frac{\partial_\psi n_b}{n_b} + \frac{\partial_\psi T_b}{T_b}\left(\frac{m_bv^2}{2T_b}-\frac{3}{2}\right)+\frac{Z_be\partial_\psi\varphi_0}{T_b}
\end{equation}
is a combination of thermodynamical \jlvg{forces} \jlvgp{($Z_b$ is the charge number and the elementary charge is denoted by $e$)} and the drift velocity,
\begin{equation}
\mathbf{v}_{D,b} = \mathbf{v}_{M,b}+\mathbf{v}_E\,,
\end{equation}
is the sum of the (low $\beta$) magnetic drift and the $E\times B$ drift:
\begin{align}
\mathbf{v}_{M,b} &= \frac{m_bv^2}{Z_be}\left(1-\frac{\lambda B}{2}\right)\frac{\mathbf{B}\times\nabla B}{B^3}\,,\nonumber\\
\mathbf{v}_E &= -\frac{\nabla\varphi\times\mathbf{B}}{B^2}\,.
\end{align}
Here, $\varphi$ is the electrostatic potential, that can be split as
\begin{equation}
\varphi(\psi,\alpha,l)=\varphi_0(\psi)+\varphi_1(\psi,\alpha,l)\,,
\end{equation}
with
\begin{equation}
|\varphi_1|\ll|\varphi_0|\,,
\end{equation}
which means that $\varphi_0$ and $\varphi_1$ will be the dominant contribution to the radial and tangential components of the electric field, respectively (and in turn to the tangential and radial components of the $E\times B$ drift, respectively)\footnote{\reft{In Appendix B of~[28], two different expansions are discussed, depending on whether $\exp{(Z_ie\varphi_1/T_b)}$ is absorbed or not in the zeroth-order distribution function, and this leads to different expressions for the thermodynamical forces and the radial fluxes. For $Z_be\varphi_1/T_b\sim \varphi_1/\varphi_0\ll 1$, these differences are vanishingly small.}}. The potentials $\varphi_0$ and $\varphi_1$ can be determined by solving two additional equations.

The component of the electrostatic potential that varies on the flux surface, $\varphi_1$, is obtained from the quasineutrality equation, which for a pure plasma \jlvg{(i.e., composed of electrons and one ion species)} reads
\begin{equation}
\left(\frac{Z_i}{T_i}+\frac{1}{T_e}\right)\varphi_1 =\frac{2\pi}{en_e}\sum_b Z_b \int_0^\infty\mathrm{d} v \int_{B^{-1}_{{\rm max}}}^{B^{-1}}\mathrm{d}\lambda\frac{v^3 B}{|v_\parallel |}g_b\,.
\label{EQ_QNFINAL}
\end{equation}
The sum is done over kinetic species. Here, we have used that, in terms of our coordinates, velocity space integrals are of the form
\begin{equation}
\int\mathrm{d}^3v (...) = \pi \sum_\sigma\int_0^\infty\mathrm{d}v\,v^2\int_0^{B^{-1}}\mathrm{d}\lambda\frac{B}{\sqrt{1-\lambda B}} (...) \,,
\label{EQ_VELINT}
\end{equation}
that $g_b$ is even in $\sigma$ and that $g_b=0$ for $\lambda<B_{max}^{-1}$. We note that, since $\varphi_1$ and $g_b$ appear in equations (\ref{EQ_DKEFINAL}) and (\ref{EQ_QNFINAL}), both equations need to be solved consistently.

The radial electric field is given by the radial derivative of the piece of the electrostatic potential that is constant on the flux surface, 
\begin{equation}
E_r =-\partial_r\varphi_0\, = -\frac{\partial \psi}{\partial r}\partial_\psi\varphi_0\,,
\end{equation}
where $r=a\sqrt{\psi/\psi_{LCFS}}$, $\psi_{LCFS}$ being the flux label at the last closed flux surface and $a$ the minor radius of the device. The radial electric field is set by the ambipolarity of the neoclassical radial particle fluxes,
\begin{equation}
\sum_b Z_b\Gamma_b(\partial_\psi\varphi_0) =0\,.\label{EQ_AMB}
\end{equation}
In our variables,
\begin{equation}
\Gamma_b\equiv\fsa{\pmb{\Gamma}_b\cdot\nabla r} = 2\frac{\partial r}{\partial\psi}\left\langle\int_0^\infty\mathrm{d} v \int_{B^{-1}_{{\rm max}}}^{B^{-1}}\mathrm{d}\lambda\frac{v^2 B}{\sqrt{1-\lambda B}} g_b ~\mathbf{v}_{D,b}\cdot\nabla\psi \right\rangle\nonumber\,,\label{EQ_GAMMA}
\end{equation}
where $\fsa{...}$ denotes flux-surface average. Finally, the radial energy flux is given by
\begin{equation}
Q_b\equiv\fsa{\mathbf{Q}_b\cdot\nabla r} = 2\frac{\partial r}{\partial\psi}\left\langle \int_0^\infty\mathrm{d} v \int_{B^{-1}_{{\rm max}}}^{B^{-1}}\mathrm{d}\lambda\frac{v^3 B}{\sqrt{1-\lambda B}} g_b\frac{m_bv^2}{2}\mathbf{v}_{D,b}\cdot\nabla\psi\right\rangle \,.\label{EQ_GAMMAQ}
\end{equation}

\KNOSOS~solves equations~(\ref{EQ_DKEFINAL}) and (\ref{EQ_QNFINAL}), together with equation~(\ref{EQ_AMB}). These equations have been rigorously derived in~\citep{calvo2018jpp} under the hypotheses of low collisionality, large aspect ratio and closeness to omnigeneity (we note that large aspect-ratio is a common characteristic of real stellarators~\citep{beidler2011icnts} while, as noted in the introduction, closeness to omnigeneity is a property sought in present and future devices). At low collisionalities, the motion of particles along the magnetic field is much faster than collisions, and the distribution function does not depend on the arc length $l$. Closeness to omnigeneity makes neoclassical transport describable by a radially-local equation for the deviation of the distribution function of trapped particles from a Maxwellian. In particular, it guarantees that the bounce-averaged radial drift is small enough so that
\begin{equation}
\left|\int_{l_{b_1}}^{l_{b_2}} \frac{\mathrm{d}l}{|v_\parallel|} \mathbf{v}_{D,b}\cdot\nabla\psi~\partial_\psi g_b\right|\ll
\left|\int_{l_{b_1}}^{l_{b_2}} \frac{\mathrm{d}l}{|v_\parallel|} \mathbf{v}_{D,b}\cdot\nabla\alpha~\partial_\alpha g_b\right|\,
\label{EQ_LOCAL}
\end{equation}
even in situations of small $E\times B$ drift. Hence, for stellarators close to omnigenity, terms proportional to $\partial_\psi g$ do not appear in equation~(\ref{EQ_DKEFINAL}). Finally, the large-aspect ratio approximation allows us to \reft{neglect energy-scattering and} use the pitch-angle collision operator, equation~(\ref{EQ_COLOP}) \reft{(the field particle part of the collision operator has negligible effect on radial transport, which is determined by the part of the distribution function that is even in the parallel velocity~\citep{calvo2017sqrtnu})}.

\reft{As we will see in detail in section~\ref{SEC_ALGORITHMS}, the bounce points $l_{b_1}$ and $l_{b_2}$ in equation (\ref{EQ_DKEFINAL}) are determined along the field line, even though trapped particles experience tangential drifts described by the first term of the right-hand-side of said equation. This is not a contradiction, but is derived rigorously under the hypothesis of low collisionality, since the drifts are much slower than the motion of the particles along the magnetic field line. The hypotheses discussed in the previous paragraph also ensure that $\varphi_1$ is small enough not to affect the orbits of main species via electrostatic trapping (except for deeply trapped particles when the radial electric field is small, an effect that we briefly discuss in section~\ref{SEC_ALGORITHMS}.}

\jlvg{Let us finally discuss the neoclassical regimes that equations~(\ref{EQ_DKEFINAL}) and (\ref{EQ_QNFINAL}) can describe. The second term on the left-hand side of equation~(\ref{EQ_DKEFINAL}) includes the radial magnetic and $E\times B$ drifts caused by the inhomogeneity of the magnetic field strength and of the electrostatic potential on the flux surface, respectively. This means that equation~(\ref{EQ_DKEFINAL}) can model the $1/\nu$ regime and the transport caused by $\varphi_1$. The first term of the left-hand side includes the precession tangential to the flux surface caused by the radial variation of the electrostatic potential (i.e. the radial electric field $E_r$) and of the magnetic field strength. This implies that we can model the $\sqrt{\nu}$ and superbanana-plateau regimes. As discussed previously, radially global effects are not accounted for.}

\section{Solution of the equations}\label{SEC_ALGORITHMS}


In this section we provide an overview of how equations~(\ref{EQ_DKEFINAL}) and (\ref{EQ_QNFINAL}) are solved. We first give an explicit expression for equation~(\ref{EQ_DKEFINAL}) in \S\ref{SEC_FINALDKE} and we discuss how to calculate its bounce-averaged coefficients in \S\ref{SEC_COEFFICIENTS}. We then devote \S\ref{SEC_GRID} to build the grid in which we will evaluate the distribution function, and \S\ref{SEC_SOLDKE} to discuss the discretization of the equation. Finally, the solution of quasineutrality, equation (\ref{EQ_QNFINAL}), is addressed in \S\ref{SEC_SOLQN}.

\subsection{Final expression of the drift-kinetic equation}\label{SEC_FINALDKE}

Using the expressions of the pitch-angle scattering collision operator described in equation~(\ref{EQ_COLOP}) and of the magnetic and $E\times B$ drifts in \textit{right handed} Boozer coordinates, equation~(\ref{EQ_DKEFINAL}) \jlvg{can be written in terms of a few bounce integrals:}
\begin{align}
\left(I_{v_{M,\alpha}} (\alpha,\lambda)+\frac{1}{v_{d,b}}I_{v_E,\alpha}(\alpha,\lambda)\right)\partial_\alpha g_b &+ \left( I_{v_{M,\psi}}(\alpha,\lambda)+\frac{1}{v_{d,b}}I_{v_{E,\psi}}(\alpha,\lambda)\right) F_{M,b}\Upsilon_b = \frac{\nu_{\lambda,b}}{v_{d,b}} \partial_\lambda\left[ I_\nu(\alpha,\lambda) \partial_\lambda g_b\right]\,,
\label{EQ_NDKE}
\end{align}
\jlvg{with}
\begin{align}
v_{d,b}&\equiv\frac{m_bv^2}{Z_be}\,,\nonumber\\
I_{v_{E,\alpha}}&=\Psi_t'\partial_\psi\varphi_0\int_{l_{b_1}}^{l_{b_2}} \frac{\mathrm{d}l}{\sqrt{1-\lambda B}}\,,\nonumber\\
I_{v_{M,\alpha}}&=\int_{l_{b_1}}^{l_{b_2}}\frac{\mathrm{d}l }{\sqrt{1-\lambda B}}\left(1-\frac{\lambda B}{2}\right)\left[\Psi_t'\frac{\partial_\psi B}{B}+
 \frac{B_\zeta\partial_\theta B - B_\theta\partial_\zeta B}{B|B_\zeta+\iota B_\theta|}\zeta\partial_\psi \iota\right]\,,\nonumber\\
I_{v_{E,\psi}}&=\int_{l_{b_1}}^{l_{b_2}} \frac{\mathrm{d}l}{\sqrt{1-\lambda B}}\frac{B_\theta\partial_\zeta \varphi_1 - B_\zeta\partial_\theta \varphi_1}{|B_\zeta+\iota B_\theta|}\,,\nonumber\\
I_{v_{M,\psi}}&=\int_{l_{b_1}}^{l_{b_2}} \frac{\mathrm{d}l }{\sqrt{1-\lambda B}}\left(1-\frac{\lambda B}{2}\right)\frac{B_\theta\partial_\zeta B - B_\zeta\partial_\theta B}{B|B_\zeta+\iota B_\theta|}\,,\nonumber\\
I_\nu&=\int_{l_{b_1}}^{l_{b_2}} \mathrm{d}l\frac{\lambda\sqrt{1-\lambda B}}{B}\,,
\label{EQ_BINT}
\end{align}
where $B_\psi$, $B_\theta$ and $B_\zeta$ are the covariant components of $\mathbf{B}$, and $B_\psi=0$ in the low-$\beta$ approximation. We note that only $v_{d,b}$, $\nu_{\lambda,b}$, $F_{M,b}$ and $\Upsilon_b$ depend on the species: the bounce-integrals are only determined by the magnetic configuration and the electrostatic potential. \jlvg{The magnetic shear appears explicitly in $I_{v_{M,\alpha}}$.}

\jlvg{Equation~(\ref{EQ_NDKE}) is a differential equation in} two variables only, $\alpha$ and $\lambda$, which is the origin of the fast performance of~\KNOSOS~that will be demonstrated in~\S\ref{SEC_RESULTS}. The radial coordinate $\psi$ is a parameter, since we are solving radially local equations; $v$ is a parameter as well, since $\varphi_1\ll \varphi_0$; and finally $l$ has disappeared since the coefficients are bounce-averages of certain quantities. The calculation of these coefficients is described in \S\ref{SEC_COEFFICIENTS}.


\subsection{Calculation of the coefficients of the drift-kinetic equation}\label{SEC_COEFFICIENTS}


The integrals in $l$ are done using an extended midpoint rule \citep[see e.g.][subroutine {\ttfamily midpnt}]{numericalrecipes}. This open formula is appropriate for integrals that are improper in the sense that they have an integrable singularity at the integration limits. This is our case, since by definition $\lambda B(l_{b_1}) = \lambda B(l_{b_2}) =1$. The number of points that we use is not pre-defined: starting from being one, it is tripled until the integral converges.

Let us now note that integrals such as those of equations~(\ref{EQ_BINT}) may be difficult to converge if the numerator does not go to zero in the integration limits, or it does, but slower than the denominator. This may happen, first, if $\lambda$ is such that $l_{b_1}$ \reft{and/or} $l_{b_2}$ are close to a point $l_T$ where $B(l)$ has a local maximum $B(l_T)$ for fixed $\alpha$ \reft{(e.g., the dot-dashed lines at the top of figure~\ref{FIG_ORBIT}}; second, if the interval ($l_{b_1},l_{b_2}$) contains \reft{such point $l_T$ and $\lambda$ is close to $1/B(l_T)$ (e.g. the bottom dot-dashed line in figure~\ref{FIG_ORBIT})}. In such cases, \reft{the bounce integral} may become very large; if the inverse of $\lambda$ is equal to the corresponding maximum of $B$, the integral diverges logarithmically. We can physically identify these situations in the example of figure~\ref{FIG_ORBIT}: divergences happen at \textit{bifurcations}, where orbits go from being trapped in a particular region in $l$ to be trapped, for smaller $\lambda$, in a wider region (the boundary between passing and trapped particles is a particular case of this). 

One can ease the convergence, and thus make the calculation faster, by removing the divergence and solving it analytically as explained in Appendix C of~\citep{calvo2017sqrtnu}. This is described more in detail in our~\ref{AP1}. Additionally, in~\ref{AP2} we will discuss how the fact that field lines are straight in magnetic coordinates is used to accelerate the evaluation of the magnetic field strength at each point ($\alpha, l$) without loss of accuracy.


\subsection{Spatial and velocity grid}\label{SEC_GRID}


In \S\ref{SEC_COEFFICIENTS} we have seen how the integrals of equations~(\ref{EQ_BINT}) are calculated. These integrals will be evaluated at the points ($\alpha,\lambda$) in which we want to determine the distribution function $g_b$. The selection of these points constitute the subject of this subsection.

Let us start with the spatial grid. We have seen that $\psi$ is a parameter, and $l$ does not appear in the bounce-averaged drift-kinetic equation, which leaves us with the field line label $\alpha$. There are, however, two complications: first, at a given $\alpha$ and $\lambda$, several wells may exist (in other words, several pairs of $l_{b_1}$ and $l_{b_2}$), which means that we need to use an integer label $w$ for them (as we will discuss more in detail in the following subsection). Second, even if $g_b$ does not depend on $l$, its integrals over velocities (needed e.g. to compute $\varphi_1$, see equation~(\ref{EQ_QNFINAL})) do, so we must define a two-dimensional angular grid. As a general rule, when doing so, we try to minimize the number of points at which $g_b$ needs to be solved, in order to save computing resources. With this in mind, we make use of periodicity and align the grid points with the field lines. The grid points are also aligned with $\zeta=0$.

\begin{figure}\vskip-1.5cm
\centerline{\includegraphics[angle=0,width=\columnwidth]{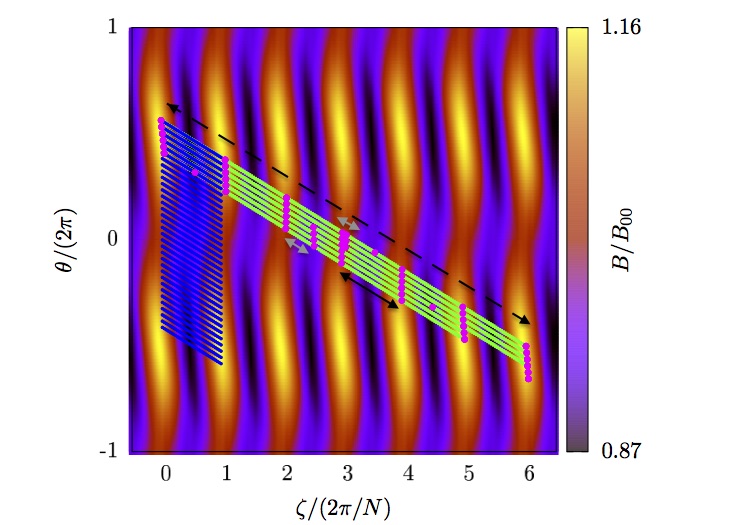}}
\centerline{\includegraphics[angle=0,width=\columnwidth]{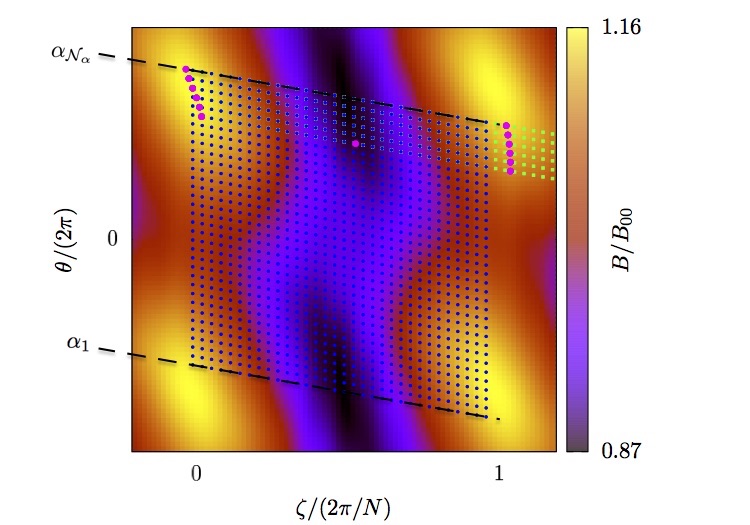}}
\caption{Construction of the angular grid (see text) for a flux surface of W7-X (top); zoom (bottom).}
\label{FIG_GRID}
\end{figure}

We use figure~\ref{FIG_GRID} (top), which shows one example of stellarator flux surface (the W7-X case discussed in \S\ref{SEC_DKES}) to describe how the angular grid is built. We follow several field lines until they have completed a full poloidal turn. \reft{For a flux-surface characterized by $\iota$ and the number of toroidal periods $N$, this means that we follow these field lines until they have traversed approximately $N/\iota$ toroidal periods, 6 in our example. The distance between two consecutive field lines $\Delta\alpha$ is taken to be an integer fraction of} $2\pi\iota/N$. This is how the green points are located, with uniform spacing in the toroidal angle. Along the field lines, several maxima of the magnetic field are found, plotted with magenta circles. It is observed that we are dealing with a relatively optimized configuration, in the sense that most trapped particles are so in a major well that coincides with one field period (black continuous arrow). In other words, their bounce points $l_{b_1}$ and $l_{b_2}$ are two consecutive magenta points, separated toroidally by a characteristic angular distance $\sim 2\pi/N$ (smaller for large values of $\lambda$, close to the bottom of the magnetic well). In the example, several ripple wells are found (grey arrows). For small enough values of $\lambda$, trajectories trapped in more than one field-period exist. \reft{In this example, there exist ripple-trapped particles and particles trapped in 1, 2, 3, 4, 5 and 6 periods whose trajectories are all computed; the latter (black dashed arrow) may move between $\zeta=0$ and $\zeta=6\frac{2\pi}{N}$. T}rajectories with smaller $\lambda$ (that is, trapped in more than 6 toroidal periods) are ignored in this case; this procedure effectively sets the boundary between passing and trapped particles. \reft{Following field lines until they have completed more than one poloidal turn (i.e., more than  $N/\iota$ toroidal periods) would allow us to describe trajectories with smaller $\lambda$, but this is not necessary in the light of the good agreement with~\DKES~shown in \$\ref{SEC_DKES}.}

Periodicity allows us to project all these grid points onto the first period. The result is a bidimensional grid in $\alpha$ and $l$, with ${\cal{N}}_\alpha$ and ${\cal{N}}_l$ points in each direction. ${\cal{N}}_\alpha$ is the integer quantity such that ${\cal{N}}_\alpha<\frac{2\pi}{\Delta\alpha}\le{\cal{N}}_\alpha+1$. The ${\cal{N}}_l$ points along the field line are distributed uniformly in the toroidal angle along a toroidal period, and ${\cal{N}}_l$ is the largest power of 2 that is smaller than or equal to ${\cal{N}}_\alpha$. This will be useful for a fast computation of the Fourier transform, needed when solving quasineutrality. Toroidal periodicity is also enforced at the corners of the grid: for instance, in figure~\ref{FIG_GRID} bottom, point $\alpha=\alpha_{{\cal{N}}_\alpha-4}$, $\zeta=0$, is not contained in the wells marked in magenta. Using periodicity, the value of the distribution function at this point will be taken to be equal to the value at \reft{a point of the grid close to $\alpha=\alpha_1$ and $\zeta=2\pi/N$}. The number of points where this has to be done can be minimized by putting one of the corners of the grid close to the global maximum of $B$ on the flux surface. For each of the nodes of this grid (and for each of the possible values of $\lambda$) the points along the trajectory and the bounce points of particles trapped in one or several field-periods are now clearly identified, and the integrals of equation~(\ref{EQ_BINT}) can be evaluated. 

Let us turn our attention to the velocity grid, where we are using $\lambda$ and $v$ as coordinates. Since we have seen in \S\ref{SEC_EQUATIONS} that only trapped particles need to be calculated, an obvious choice for the former is a uniform grid\footnote{When the particles are in the $1/\nu$ regime, special attention should be paid to bifurcations, where $g_b$ has discontinuous first $\lambda$-derivatives~\citep{nemov1999neo,calvo2014er}, and a non-uniform grid, adapted to the structure of maxima and minima at fixed $\alpha$, is a more efficient choice~\citep{kernbichler2016neo2}. The same applies to very low collisionalities, when the contribution to the flux is concentrated on very thin $\lambda$ layers. For the wide parameter range that will be studied with~\KNOSOS, the uniform grid is considered appropriate.}, with ${\cal{N}}_\lambda+1$ values between $\lambda_1\equiv 1/B_{max}$ and $\lambda_{{\cal{N}}_\lambda+1}\equiv 1/B_{min}$. The distribution function will not be evaluated at $\lambda_{{\cal{N}}_\lambda+1}$, which will be \textit{ghost} points employed for imposing the boundary conditions at the bottom of the well. \reft{Note that, since particles trapped in more than (in the above example) 6 periods are considered passing, there exist values of $\lambda$ close to $\lambda_{{\cal{N}}_\lambda+1}$ where the distribution function is not evaluated for some values of $\alpha$ either.} When integrating in $\lambda$, we will use the extended trapezoidal rule~\citep[][]{numericalrecipes}.

Finally, $v$ is a parameter in our calculations: equation~(\ref{EQ_NDKE}) will be solved for several values $v_i$ of the velocity and the solution will be numerically integrated in $v$. Since the integrand of equations~(\ref{EQ_VELINT}) contains an exponential coming from the Maxwellian distribution, we will use Gauss-Laguerre of order 64~\citep[][]{numericalrecipes}:
\begin{equation}
\int_0^\infty\,\mathrm{d}(v^2/v_{th,b}^2) f(v^2/v_{th,b}^2) \exp{(-v^2/v^2_{th,b})} \approx \sum_{i=1}^{n} \omega_i f(v_i^2/v_{th,b}^2)\,,
\label{EQ_CONV}
\end{equation}
being $v_{th,b}$ the thermal velocity of species $b$, and $\omega_i$ a set of tabulated real numbers. This procedure requires solving the monoenergetic drift-kinetic equation for $n=64$ values of $v/v_{th,b}$, typically from $\sim 10^{-2}$ to $\sim 10^2$. However, the contribution of the largest $v_i$ to the integral can be usually neglected, and this allows for an important reduction of computing time. Let us finally note that this is a standard and well-tested choice in neoclassics and gyrokinetics~\citep[e.g.][]{velasco2011bootstrap,barnes2019stella}, although other velocity-space discretization methods have been proposed in recent years~\citep{landreman2013intv} that could be easily implemented in~\KNOSOS.


\subsection{Discretization of the drift-kinetic equation}\label{SEC_SOLDKE}


In \S\ref{SEC_GRID} we have built a grid in variables $\alpha$ and $\lambda$. Three integers can be used to label any point $(\alpha_i,\lambda_j,w)$ of this grid: $i$ runs from 1 to ${\cal{N}}_\alpha$, $j$ from 1 to ${\cal{N}}_\lambda$ and $w=I,II...$ is an integer that labels wells for a given $\alpha$ and $\lambda$. At a given point, we define $g_{i,j,w}\equiv g_b(\alpha_i,\lambda_j,w)$, $I_{\nu,i,j,w}\equiv I_\nu(\alpha_i,\lambda_j,w)$ and so on (in order to ease the notation, $g_{i,j,w}$ does not contain a species index). The final step in the discretization of the drift-kinetic equation is how we approximate the derivatives of $g_b$ of equation~(\ref{EQ_NDKE}) at each point of this grid.

\begin{figure}
\centering
\includegraphics[angle=0,width=0.8\columnwidth]{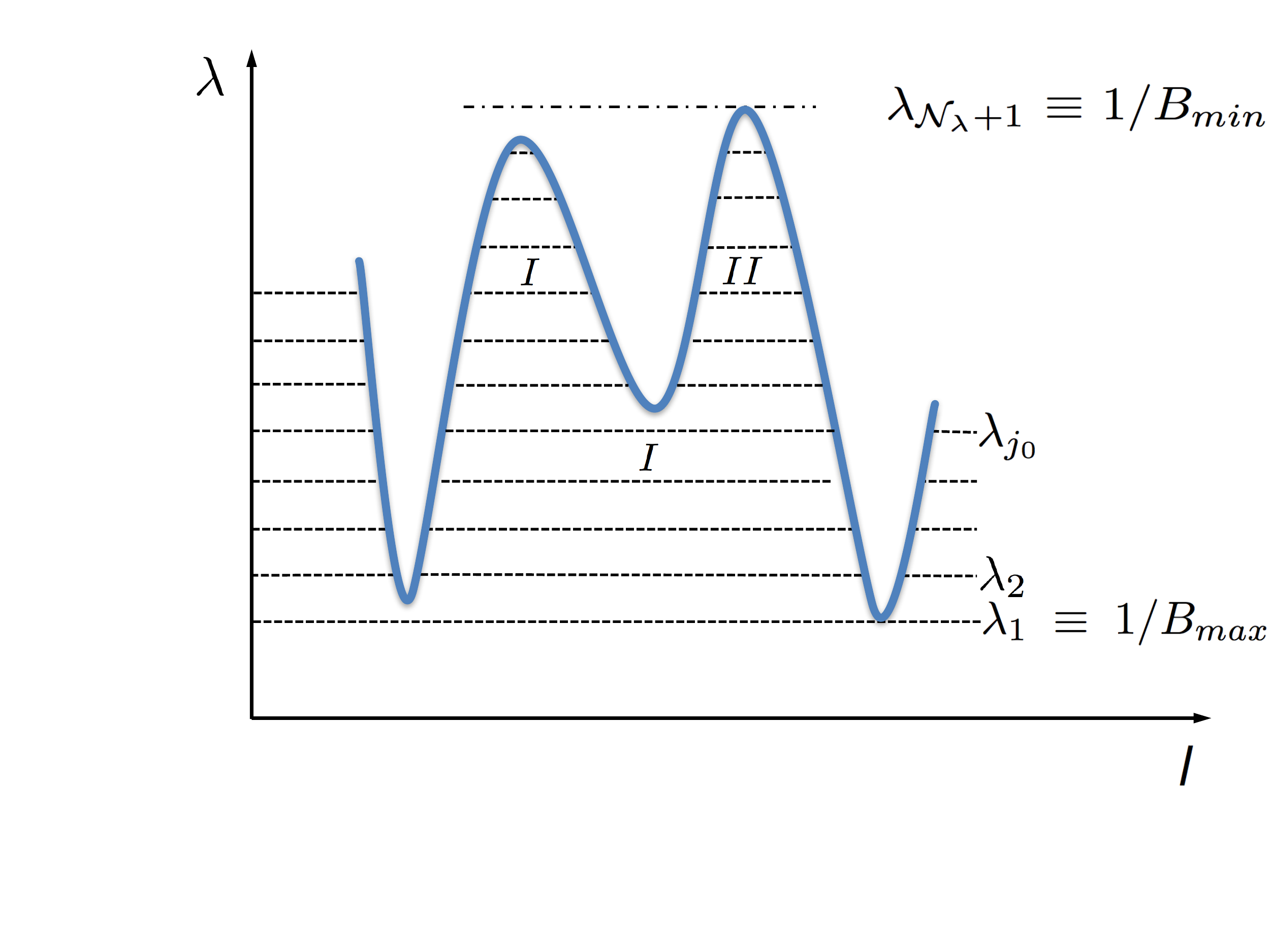}\
\caption{Sketch of grid in $\lambda$ space at fixed $\alpha$. The collision operator is discretized as in equation~(\ref{EQ_D2LAMBDA}) except at the top ($\lambda_1$) or bottom ($\lambda_{{\cal{N}}_\lambda+1}$) of the well and at bifurcations (e.g. $\lambda_{j_0}$); there, equations~(\ref{EQ_BOTTOM}),~(\ref{EQ_TOP}) and~(\ref{EQ_DLAMBDABIF}), respectively are used instead.}
\label{FIG_LAMBDA}
\end{figure}

Let us start with the collision operator, which divided by $ \frac{\nu_{\lambda},b}{v_{d,b}}$ reads
\begin{equation}
\partial_\lambda\left[ I_\nu \partial_\lambda g_b\right]\,,
\label{EQ_COL}
\end{equation}
and can be expanded into two terms
\begin{equation}
\left[ I_\nu \partial^2_\lambda + (\partial_\lambda I_\nu) \partial_\lambda \right] g_b\,. 
\label{EQ_CO_EXP}
\end{equation}
We represent the $\lambda$ grid at fixed $\alpha$ in figure~\ref{FIG_LAMBDA}. Here, $\lambda_1$ is the boundary between passing and trapped particles. In this example, only one complete well is plotted at $\lambda_2$, labelled $I$. If one moves to larger $\lambda$, a bifurcation appears in the vicinity of $\lambda_{j_0}$, with two wells labelled $I$ and $II$. At a larger value of $\lambda$, there are the bottoms of the wells, where the wells have their minimum magnetic field (different in $I$ than in $II$) and beyond which no orbits are allowed.

 At a generic point, we make use of equation~(\ref{EQ_CO_EXP}) and then employ central finite differences with second-order accuracy
\begin{align}
\left[ I_\nu \partial^2_\lambda + (\partial_\lambda I_\nu) \partial_\lambda \right] g_b|_{i,j,w}&= I_\nu,_{i,j,w} \frac{g_{i,j+1,w}+g_{i,j-1,w}-2g_{i,j,w}}{(\Delta\lambda)^2}+\partial_\lambda I_\nu |_{i,j,w} \frac{g_{i,j+1,w}-g_{i,j-1,w}}{2\Delta\lambda}\,,
\label{EQ_D2LAMBDA}
\end{align}
with $\Delta\lambda=\lambda_{j+1}-\lambda_j$. Differentiation is done at fixed $\alpha$ and well-label $w${. At a bifurcation, such as the one near $\lambda_{j_0}$ in figure~\ref{FIG_LAMBDA}, we use finite differences with second-order accuracy directly over equation~(\ref{EQ_COL}) and summing over wells, }
\begin{align}
\partial_\lambda\left[ I_\nu \partial_\lambda g_b\right]|_{i,j_0,I}&= \frac{[I_\nu \partial_\lambda g_b]|_{i,j_0+1,I} + [I_\nu \partial_\lambda g_b]|_{i,j_0+1,II} -  [I_\nu \partial_\lambda g_b]|_{i,j_0-1,I}}{2\Delta\lambda}\,\nonumber\\
&= I_\nu,_{i,j_0+1,I}\frac{g_{i,j_0+2,I}-g_{i,j_0,I}}{4(\Delta\lambda)^2}+ I_\nu,_{i,j_0+1,II}\frac{g_{i,j_0+2,II}-g_{i,j_0,I}}{4(\Delta\lambda)^2}-I_\nu,_{i,j_0-1,I}\frac{g_{i,j_0,I}-g_{i,j_0-2,I}}{4(\Delta\lambda)^2}\,.\label{EQ_DLAMBDABIF}
\end{align}
This discretization is designed to obtain the expected relation between different values of $\partial_\lambda g$ at the bifurcation for the $1/\nu$ regime~\citep{nemov1999neo,calvo2014er}. Finally, we have two kinds of boundary conditions: one at the boundary between passing and trapped particles, corresponding to equation~(\ref{EQ_CONT2}),
\begin{equation}
g_{i,1,w}=0\,,
\label{EQ_TOP}
\end{equation}
and one at the bottom, corresponding to regularity~\citep{calvo2013er},
\begin{equation}
\partial_\lambda\left[ I_\nu \partial_\lambda g_b\right]|_{i,{\cal{N}}_\lambda,w} = -I_{\nu,i,{\cal{N}_\lambda}-1,w}\frac{g_{i,{\cal{N}}_\lambda,w}-g_{i,{\cal{N}}_\lambda-2,w}}{4(\Delta\lambda)^2}\,.
\label{EQ_BOTTOM}
\end{equation}
Here we have employed a ghost point $\lambda_{{\cal{N}}_\lambda+1}$ at exactly the bottom of the well, where $I_{\nu,i,{\cal{N}}_\lambda+1,w}=0$. One precision must be made: while in omnigenous magnetic fields the values of the maxima and minima of $B$ are the same when moving in $\alpha$, and equation~(\ref{EQ_BOTTOM}) can be used as such for all $\alpha$, this ceases to be true in a generic stellarator. For instance, the distance from $\lambda_{{\cal{N}_\lambda}}$ to the local bottom will be exactly $\Delta\lambda$ for one field line and smaller elsewhere (it may even happen that the contour condition must not be applied to $\partial_\lambda\left[ I_\nu \partial_\lambda g_b\right]|_{i,{\cal{N}}_\lambda,w}$, but to $\partial_\lambda\left[ I_\nu \partial_\lambda g_b\right]|_{i,j,w}$ with a smaller $j$). This requires introducing straightforward corrections to equations~(\ref{EQ_D2LAMBDA}), (\ref{EQ_DLAMBDABIF}), (\ref{EQ_TOP}) and (\ref{EQ_BOTTOM}).

\begin{figure}
\centering
\includegraphics[angle=0,width=0.8\columnwidth]{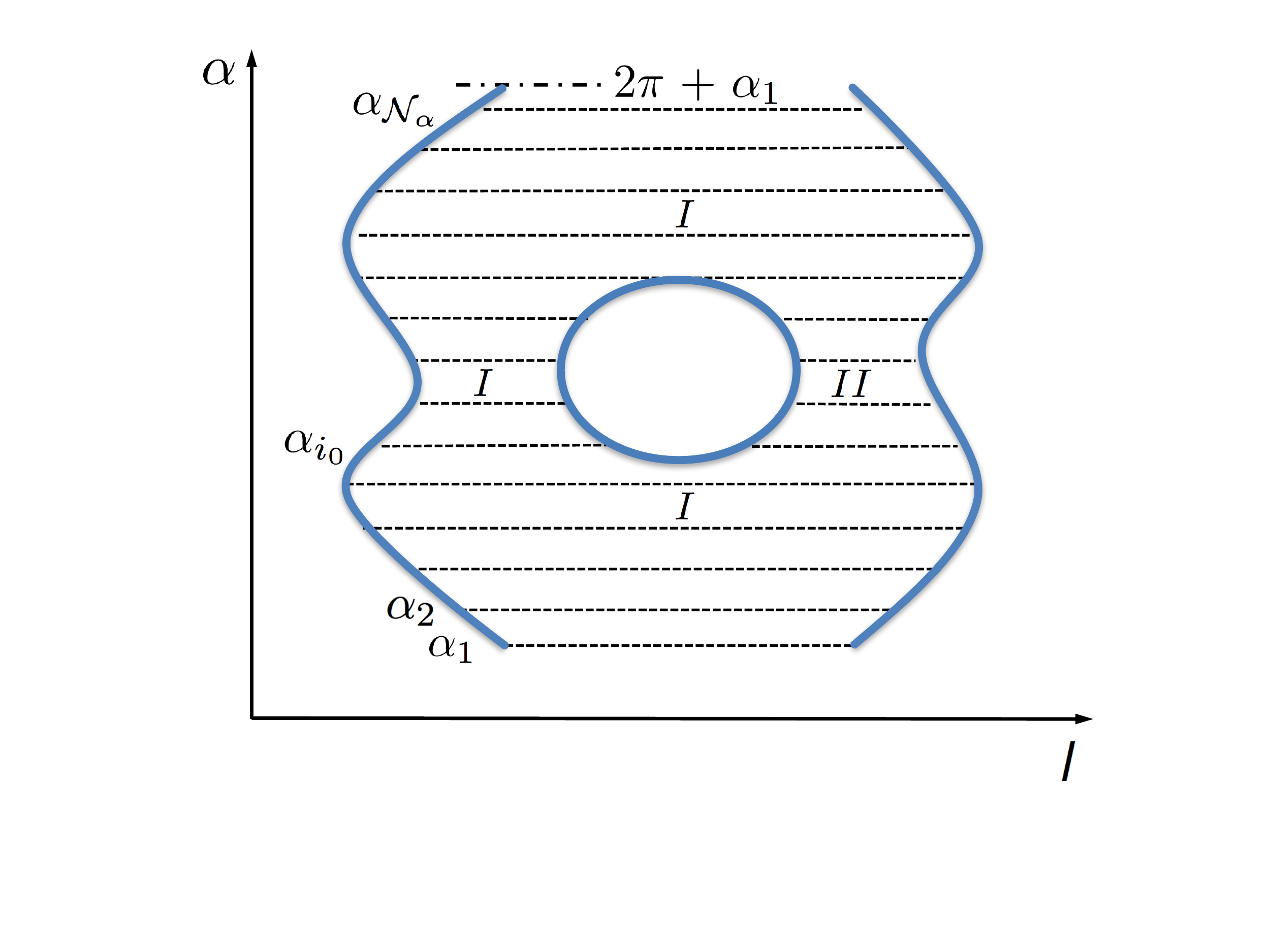}\vskip-1.5cm
\includegraphics[angle=0,width=0.8\columnwidth]{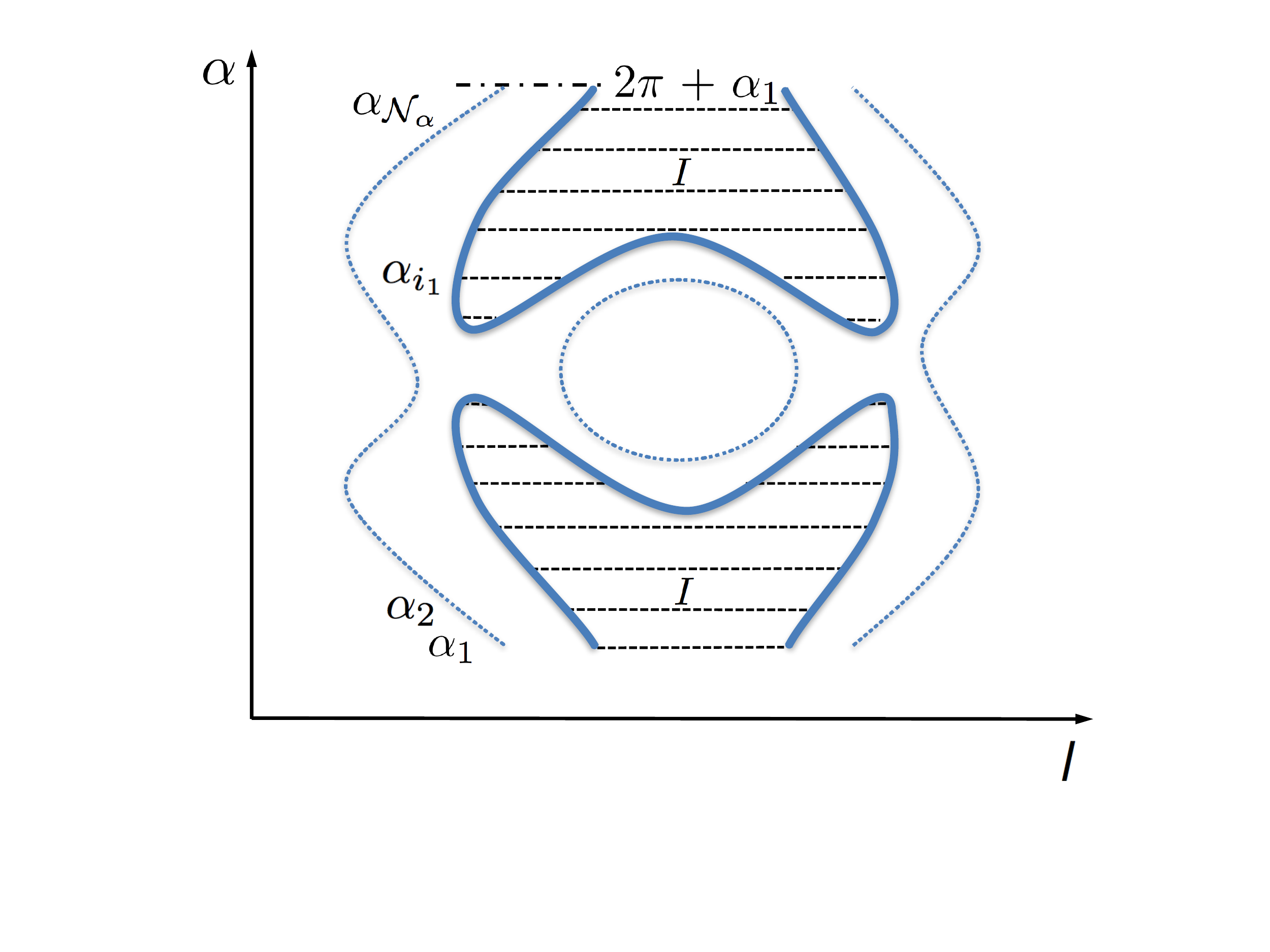}\vskip-0.5cm
\caption{Top: sketch of grid in $\alpha$ space at fixed $\lambda$. The tangential derivatives are discretized as in equations~(\ref{EQ_FDALPHA}) and (\ref{EQ_BDALPHA}) except close to the limits of the grid ($\alpha_1$ and $\alpha_{{\cal{N}}_\alpha}$) and to bifurcations (e.g. $\alpha_{i_0}$); there, equations~(\ref{EQ_FDALPHAA}),~(\ref{EQ_FDALPHAB}),~(\ref{EQ_BDALPHAA}),~(\ref{EQ_BDALPHAB}) and~(\ref{EQ_BIFALPHA}) are used instead. Bottom: sketch of grid in $\alpha$ space at larger $\lambda$ (the grid at smaller $\lambda$ is plotted for reference in dashed thin blue line). {$\alpha_{i_1}$} is a point where the backward derivative is discretized as discussed in equation~(\ref{EQ_NOALPHA}).}
\label{FIG_ALPHA}
\end{figure}

Let us now turn our attention to the terms with the first derivative in $\alpha$ in equation~(\ref{EQ_NDKE}), which we multiply by $v_{d,b}$:
\begin{equation}
\left(v_{d,b}I_{v_{M,\alpha}} +I_{v_E,\alpha}\right)\partial_\alpha g_b\,. 
\label{EQ_DALPHA}
\end{equation}
We represent the $\alpha$ grid at fixed $\lambda$ in figure~\ref{FIG_ALPHA} (top). In this example, there is only one well at $\alpha_1$, labelled $I$. If one moves from smaller to larger $\alpha$, a bifurcation appears in the vicinity of $\alpha_{i_0}$, with two wells labelled $I$ and $II$. At a larger value of $\alpha$, the wells merge into a single region labelled again $I$. The last point of the grid, $\alpha_{{\cal{N}}_\alpha}$, is close to $\alpha_1+2\pi$.

Non-centered finite differences with second-order accuracy are used. For a given flux surface, for each solution of the drift-kinetic equation, the sign of the coefficient in front of $\partial_\alpha g_b$ (i.e. the direction of the flow in the $\alpha$ direction) indicates whether forward
\begin{equation}
\partial_\alpha g_b|_{i,j,w}=\frac{-g_{i+2,j,w}+4g_{i+1,j,w}-3g_{i,j,w}}{2\Delta\alpha}\,,
\label{EQ_FDALPHA}
\end{equation}
 or backward differences 
 \begin{equation}
\partial_\alpha g_b|_{i,j,w}=\frac{g_{i-2,j,w}-4g_{i-1,j,w}+3g_{i,j,w}}{2\Delta\alpha}\,,
\label{EQ_BDALPHA}
\end{equation}
should be used, with $\Delta\alpha=\alpha_{i+1}-\alpha_i$. {To construct the derivatives with respect to $\alpha$ without much computational cost, we discretize separately the terms $I_{v_E,\alpha}\partial_\alpha g$ and $v_{d,b}I_{v_M,\alpha}\partial_\alpha g$ using a total of four matrices for a given flux surface. One corresponds to forward differences being used everywhere, and another one corresponds to backward differences everywhere. When solving equation ~(\ref{EQ_NDKE}), one of these two matrices will describe the $I_{v_E,\alpha}\partial_\alpha g$ term, depending on the sign of $E_r$. The other two matrices correspond to two $\lambda$ (and $w$)-dependent discretizations, in which forward (backward) differences are used according to the sign of $I_{v_{M,\alpha}}$. One of these two matrices will describe the $v_{d,b}I_{v_M,\alpha}\partial_\alpha g$ term, depending on the sign of $v_{d,b}$. Any matrix appropriate for describing equation~(\ref{EQ_DALPHA}) will thus be a linear combination of two of the four pre-calculated matrices, and a neoclassical simulation including ions and electrons and/or different values of the radial electric field will generally make use of the four of them.}

Periodicity in $\alpha$ is easily imposed by replacing equation (\ref{EQ_FDALPHA}) at $i\ge{\cal{N}}_\alpha-1$ with
\begin{align}
\partial_\alpha g_b|_{{\cal{N}}_\alpha-1,j,w}&=\frac{(g_{{\cal{N}}_\alpha,j,w}-g_{{\cal{N}}_\alpha-1,j,w})(2\pi+\alpha_1-\alpha_{{\cal{N}}_\alpha-1})}{(2\pi+\alpha_1-\alpha_{{\cal{N}}_\alpha})\Delta\alpha} -\frac{(g_{1,j,w}-g_{{\cal{N}}_\alpha-1,j,w})\Delta\alpha}{(2\pi+\alpha_1-\alpha_{{\cal{N}}_\alpha-1})(2\pi+\alpha_1-\alpha_{{\cal{N}}_\alpha})}\,,\label{EQ_FDALPHAA}\\
\partial_\alpha g_b|_{{\cal{N}}_\alpha,j,w}&=\frac{(g_{1,j,w}-g_{{\cal{N}}_\alpha,j,w})(2\pi+\alpha_{2}-\alpha_{{\cal{N}}_\alpha})}{(2\pi+\alpha_{1}-\alpha_{{\cal{N}}_\alpha})\Delta\alpha} -\frac{(g_{2,j,w}-g_{{\cal{N}}_\alpha,j,w})(2\pi+\alpha_{1}-\alpha_{{\cal{N}}_\alpha})}{(2\pi+\alpha_{i}-\alpha_{{\cal{N}}_\alpha})\Delta\alpha}\,,
\label{EQ_FDALPHAB}
\end{align}
respectively, and equation (\ref{EQ_BDALPHA}) at $i\le 2$ with
\begin{align}
\partial_\alpha g_b|_{2,j,w}&=-\frac{(g_{1,j,w}-g_{2,j,w})(\alpha_{{\cal{N}}_\alpha}-\alpha_2-2\pi)}{(\alpha_{{\cal{N}}_\alpha}-\alpha_1-2\pi)\Delta\alpha} +\frac{(g_{{\cal{N}}_\alpha,j,w}-g_{2,j,w})\Delta\alpha}{(\alpha_{{\cal{N}}_\alpha}-\alpha_2-2\pi)(\alpha_{{\cal{N}}_\alpha}-\alpha_1-2\pi)}\,,\label{EQ_BDALPHAA}\\
\partial_\alpha g_b|_{1,j,w}&=-\frac{(g_{{\cal{N}}_\alpha,j,w}-g_{1,j,w})(\alpha_{{\cal{N}}_\alpha-1}-\alpha_1-2\pi)}{(\alpha_{{\cal{N}}_\alpha}-\alpha_1-2\pi)\Delta\alpha} +\frac{(g_{{\cal{N}}_\alpha-1,j,w}-g_{1,j,w})(\alpha_{{\cal{N}}_\alpha}-\alpha_1-2\pi)}{(\alpha_{{\cal{N}}_\alpha-1}-\alpha_1-2\pi)\Delta\alpha}\,,
\label{EQ_BDALPHAB}
\end{align}
respectively. We note that, since $\iota$ is generally irrational, $2\pi+\alpha_1-\alpha_{{\cal{N}}_\alpha}$ will e.g. be slightly smaller than $\Delta\alpha$

We also note that bifurcations do not pose a problem for $\alpha$-derivatives, due to $g_b$ being continuous in $\alpha$. For example, in the vicinity of $\alpha_{i_0}$ in figure~\ref{FIG_ALPHA} (top) the forward derivative is discretized
\begin{align}
\partial_\alpha g_b|_{i_0-2,j,I}&=\frac{-g_{i_0,j,I}+4g_{i_0-1,j,I}-3g_{i_0-2,j,I}}{2\Delta\alpha}=\frac{-g_{i_0,j,II}+4g_{i_0-1,j,I}-3g_{i_0-2,j,I}}{2\Delta\alpha}\,,\nonumber\\
\partial_\alpha g_b|_{i_0-1,j,I}&=\frac{-g_{i_0+1,j,I}+4g_{i_0,j,I}-3g_{i_0-1,j,I}}{2\Delta\alpha}=\frac{-g_{i_0+1,j,II}+4g_{i_0,j,II}-3g_{i_0-1,j,I}}{2\Delta\alpha}\,,\nonumber\\
\partial_\alpha g_b|_{i_0,j,I}&=\frac{-g_{i_0+2,j,I}+4g_{i_0+1,j,I}-3g_{i_0,j,I}}{2\Delta\alpha}\,,\nonumber\\
\partial_\alpha g_b|_{i_0,j,II}&=\frac{-g_{i_0+2,j,II}+4g_{i_0+1,j,II}-3g_{i_0,j,II}}{2\Delta\alpha}\,,\nonumber\\
\partial_\alpha g_b|_{i_0+1,j,I}&=\frac{-g_{i_0+3,j,I}+4g_{i_0+2,j,I}-3g_{i_0+1,j,I}}{2\Delta\alpha}\,,\nonumber\\
\partial_\alpha g_b|_{i_0+1,j,II}&=\frac{-g_{i_0+3,j,II}+4g_{i_0+2,j,II}-3g_{i_0+1,j,II}}{2\Delta\alpha}\,,\nonumber\\
\partial_\alpha g_b|_{i_0+2,j,I}&=\frac{-g_{i_0+4,j,I}+4g_{i_0+3,j,I}-3g_{i_0+2,j,I}}{2\Delta\alpha}\,,\nonumber\\
\partial_\alpha g_b|_{i_0+2,j,II}&=\frac{-g_{i_0+4,j,II}+4g_{i_0+3,j,II}-3g_{i_0+2,j,I}}{2\Delta\alpha}\,,\nonumber\\
\partial_\alpha g_b|_{i_0+3,j,I}&=\frac{-g_{i_0+5,j,I}+4g_{i_0+4,j,I}-3g_{i_0+3,j,I}}{2\Delta\alpha}\,,\nonumber\\
\partial_\alpha g_b|_{i_0+3,j,II}&=\frac{-g_{i_0+5,j,II}+4g_{i_0+4,j,I}-3g_{i_0+3,j,I}}{2\Delta\alpha}\,
\label{EQ_BIFALPHA}
\end{align}
\reft{We note that there exist two alternative discretizations in the first two expressions of equation~(\ref{EQ_BIFALPHA}). Continuity of $g_b$ ensures that they give the same result for small $\Delta\alpha$.} Equivalent expressions can be obtained for the backward derivative.

One final caveat has to be made. In an omnigenous magnetic field, the contours of minimum $B$ on a flux surface must encircle the plasma (toroidally, poloidally, or helically). This is not true for a generic stellarator, in which local minima of $B$ exist on the flux surface. Close to these minima, moving in $\alpha$ at constant large $\lambda$ is not always possible, as these trajectories may not exist. This situation is illustrated in figure~\ref{FIG_ALPHA} (bottom), at $\alpha_{i_1}$. At, $\alpha_{i_1}$, instead of equation (\ref{EQ_BDALPHA}), we use
 \begin{equation}
\partial_\alpha g_b|_{i_1,j,w}=\frac{g_{i_1-2,j_0,w}-4g_{i_1-1,j,w}+3g_{i_1,j,w}}{2\Delta\alpha}\,.\label{EQ_NOALPHA}
\end{equation}
and we have implemented two models: in one, $\lambda_{j_0}$ is the value of $\lambda$ closest to $\lambda_j$ in which trajectories exist for all $\alpha$; in the second model, $\lambda_{j_0}$ is the closest value of $\lambda$ in which trajectories exist at $\alpha_{i_0-2}$. The relative differences between the two models are smaller than the error bars of~\DKES~in figure~\ref{FIG_D11PROF}. We note that (with different manifestations for other choices of velocity coordinates) an incorrect treatment of this kind of particles is common to all existing radially local codes. \reft{In general stellarators and in stellarators close to omnigeneity, both the tangential magnetic drift and the trapping due to $\varphi_1$ must be retained to reproduce the correct trajectories for these particles. This fact is usually ignored in local codes, although there are notable exceptions that include either the trapping by $\varphi_1$~\citep{mollen2018phi1} or several models of tangential magnetic drift~\citep{paul2017iter,matsuoka2015tangential}. Finally, in the limit of large aspect ratio stellarators with $\mathbf{E} \times \mathbf{B}$ drift much larger than the magnetic drift, it is possible to construct local equations that treat these deeply trapped particles correctly. The equations in \DKES\ \citep{hirshman1986dkes} are one such model, and an equivalent formulation for large aspect ratio stellarators is being developed for \KNOSOS\ \citep{dherbemont2020fow}.}

%
%
%
%

For each of the species $b$, we end up with an equation that is linear in $g_b$ and can be written as a linear problem in \jlvg{matrix form. The matrix that represents}
\begin{equation}
\left(I_{v_{M,\alpha}}+\frac{1}{v_{d,b}}I_{v_{E,\alpha}}\right)\partial_\alpha +\frac{\nu_{\lambda,b}}{v_{d,b}} \partial_\lambda \nu\partial_\lambda 
\end{equation}
is square with approximately ${\cal{N}}_\lambda\times {\cal{N}}_\alpha$ elements per row, and sparse, with $\sim 6$ non-zero elements per row: between 3 and 5 for the $\alpha$ derivatives, and typically 2 additional points for the collision operator. Although their relative weight varies with $\nu_{\lambda,b}$, $v_{d,b}$ and $\partial_\psi\varphi_0$, the non-zero elements are always at the same position for a given flux surface, which can be used to save computing time, by using the four pre-computed matrices described above.

We solve the linear problem with a direct solver from the {\ttfamily PETSc} library~\citep{petsc-efficient,petsc-web-page,petsc-user-ref} based on LU factorization. The reason is that the matrix is not large enough to require iterative methods, and reusing the LU factorization greatly accelerates the solution of the quasineutrality equation, as discussed in \S\ref{SEC_SOLQN}.


\subsection{Solution of the quasineutrality equation}\label{SEC_SOLQN}


We will solve the quasineutrality equation by means of a response matrix approach (similar methods are used in gyrokinetics for the calculation of the electrostatic potential fluctuations~\citep{kotschenreuther1995response}). Let us first rewrite equations (\ref{EQ_NDKE}) and~(\ref{EQ_QNFINAL}) making explicit the dependence on $\varphi_1$:

\begin{align}
\left(I_{v_{M,\alpha}} +\frac{I_{v_{E,\alpha}}}{v_{d,b}}\right)\partial_\alpha g_b-\frac{\nu_{\lambda,b}}{v_{d,b}} \partial_\lambda I_\nu \partial_\lambda g_b&={\refm{-}}\left(I_{v_{M,\psi}}-\int_{l_{b_1}}^{l_{b_2}} \frac{\mathrm{d}l}{\sqrt{1-\lambda B}}\frac{B_\theta\partial_\zeta \varphi_1 - B_\zeta\partial_\theta \varphi_1}{|B_\zeta+\iota B_\theta|} \right) F_{M,b}\Upsilon_b\,,\label{EQ_DKEPHI1}\\
\left(\frac{Z_i}{T_i}+\frac{1}{T_e}\right)\varphi_1 &=\frac{2\pi}{en_e}\sum_b Z_b \int_0^\infty\mathrm{d} v \int_{B^{-1}_{{\rm max}}}^{B^{-1}}\mathrm{d}\lambda\frac{v^3 B}{|v_\parallel |}g_b\,.\label{EQ_QNPHI1}
\end{align}
It can be observed that equation~(\ref{EQ_DKEPHI1}) is linear in $\varphi_1$, and therefore the response of the distribution function $g_b$ (and of its velocity integral) of species $b$ to certain $\varphi_1$ can be calculated as a superposition of the responses to a complete set of harmonics that parametrize $\varphi_1(\theta,\zeta)$. We can perform this parametrization efficiently thanks to the Fast Fourier Transform, using ${\cal{N}}=2(2{\cal{N}}_n+1)({\cal{N}}_m+1)$ coefficients:
\begin{equation}
\varphi_1(\theta,\zeta) =\sum_{-{\cal{N}}_n<n<{\cal{N}}_n} \sum_{0<m<{\cal{N}}_m} \left(\varphi^{(c)}_{mn}\cos(m\theta+Nn\zeta)+\varphi^{(s)}_{mn}\sin(m\theta+Nn\zeta)\right)\,
\label{EQ_FFTPHI1}
\end{equation}
(the grid defined in \S\ref{SEC_GRID} is not uniform in $\theta$, so an interpolation is done before the Fourier transform). We can now denote $u_k(\theta,\zeta)$ each of the ${\cal{N}}$ basis elements (e.g. $\cos(\theta+2N\zeta)$) and the combined system of drift-kinetic and quasineutrality equation can be symbolically written as
\begin{align}
\pmb{\varphi_1} = \pmb{\varphi_1^{0}} + \mathbf{A}\pmb{\varphi_1}\,,
\end{align}
where $\pmb{\varphi_1}$ is a vector whose ${\cal{N}}$ components are the coefficients of the expansion of $\varphi_1$ in equation~(\ref{EQ_FFTPHI1}) and $\mathbf{A}$ is a generally dense ${\cal{N}}\times{\cal{N}}$ matrix. In this linear, system, the right-hand side $\pmb{\varphi_1^{0}}$ can be obtained by solving equation~(\ref{EQ_DKEPHI1}) for all the kinetic species {(and for several values of $v$)} with $\varphi_1=0$, inserting the solution into equation~(\ref{EQ_QNPHI1}) and then Fourier-transforming the result following equation~(\ref{EQ_FFTPHI1}). Next, we fill the matrix $\mathbf{A}$: the \textit{kth} row is obtained by solving equation~(\ref{EQ_DKEPHI1}) with $\varphi_1= u_k$, inserting the solution into equation~(\ref{EQ_QNPHI1}), Fourier-transforming and then substracting $\pmb{\varphi_1^{0}}$ from the result. Once $\pmb{\varphi_1^{0}}$ and $\mathbf{A}$ have been filled, the new linear system can easily be solved, e.g. using a new LU decomposition, to obtain $\pmb{\varphi_1}$, i.e., the set of coefficients $\varphi^{(c)}_{mn}$ and $\varphi^{(s)}_{mn}$ that parametrize the solution to quasineutrality. Finally, since the response of $g_b$ to every basis element has already been computed, a simple linear combination yields the distribution function that is solution of the drift-kinetic and quasineutrality equations, without requiring an additional solve of the former.

In summary, the drift-kinetic equation is solved a total of ${\cal{N}}+1$ times (for each species), but LU factorization is done once (for each value of $v$). The linearity of the system of equations due to the smallness of $\varphi_1$, together with the method that we have chosen for solving the drift-kinetic equation, yields a large reduction of the computing time needed to solve the system of equations: the code is roughly ${\cal{N}}$ faster (with ${\cal{N}}$ ranging from 100 to 1000), with respect to an equivalent code that allowed the particle orbits be modified by $\varphi_1$.


\section{Results}\label{SEC_RESULTS}


In this section, we show calculations for a variety of three-dimensional magnetic configurations in order to compare~\KNOSOS~with widely-benchmarked codes and to illustrate its performance. In \S\ref{SEC_DKES}, we will solve a simplified drift-kinetic equation, without the magnetic drift and electric field components tangent to the flux surface, and we will compare our results with bidimensional databases of~\DKES ~monoenergetic transport coefficients. The effect of the tangential magnetic drift in the energy flux, calculated for realistic kinetic profiles, will be discussed in~\S\ref{SEC_TANGVM}. Finally, solutions of the quasineutrality equation will be compared with~\EUTERPE~calculations in~\S\ref{SEC_EUTERPE}.
 

\subsection{\DKES-like monoenergetic transport coefficients}\label{SEC_DKES}


\begin{figure}
\centering
\includegraphics[angle=0,width=0.45\columnwidth]{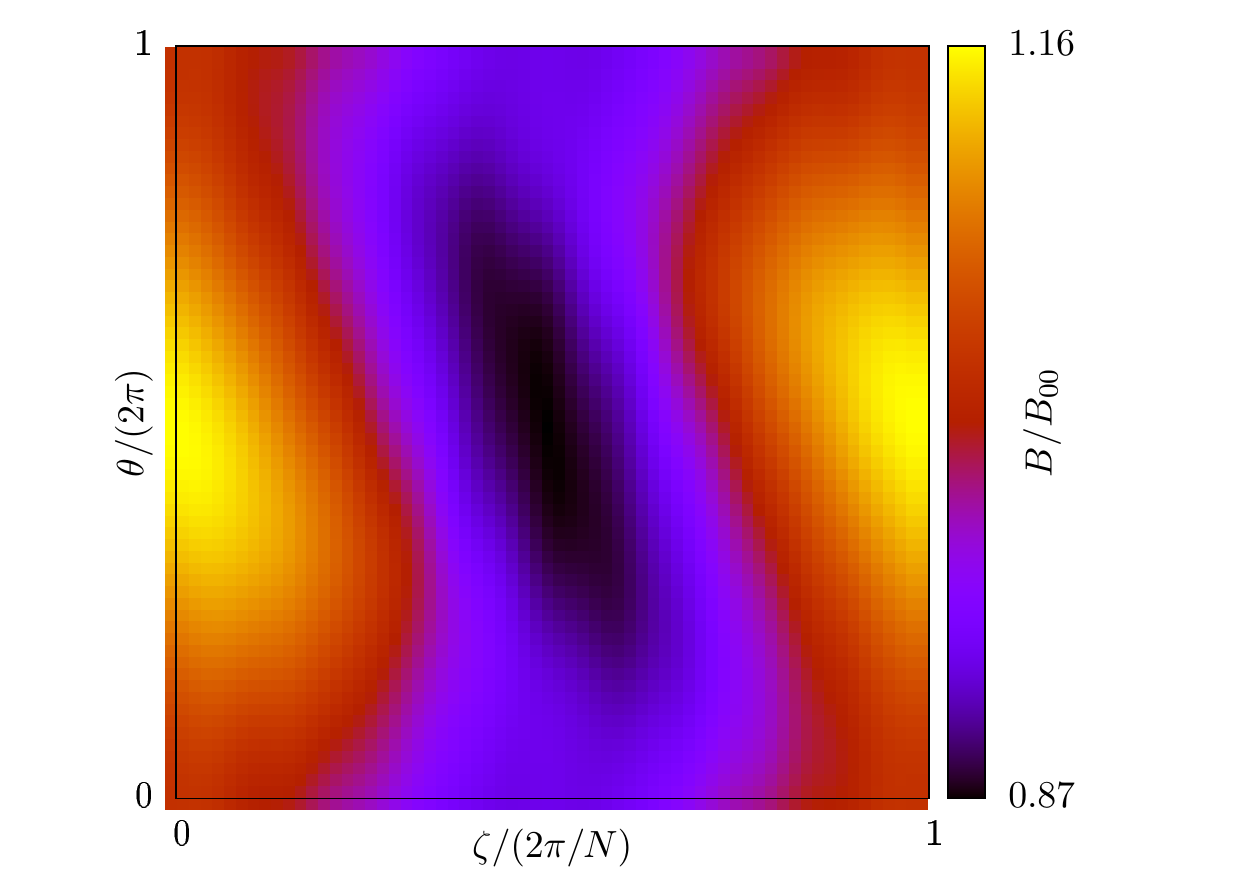}
\includegraphics[angle=0,width=0.45\columnwidth]{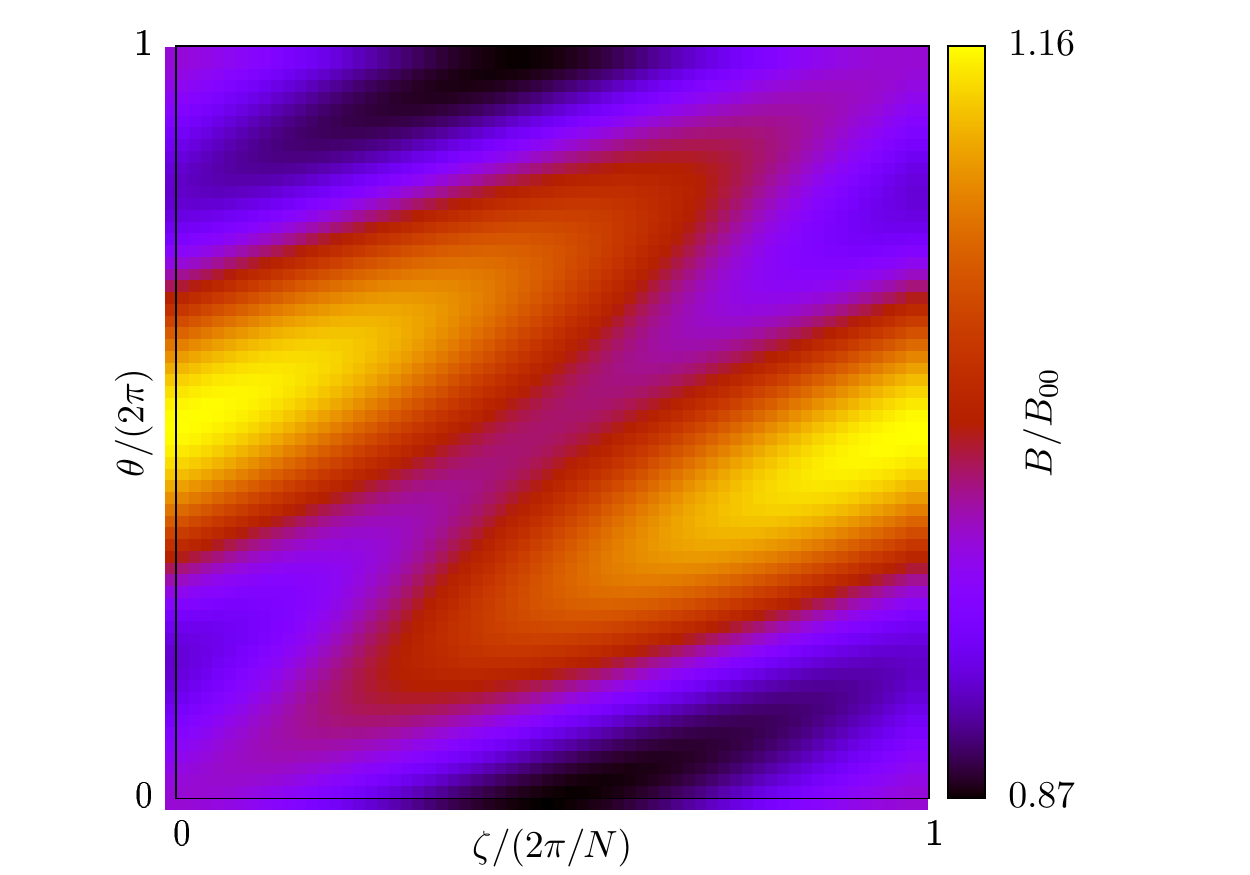}
\includegraphics[angle=0,width=0.45\columnwidth]{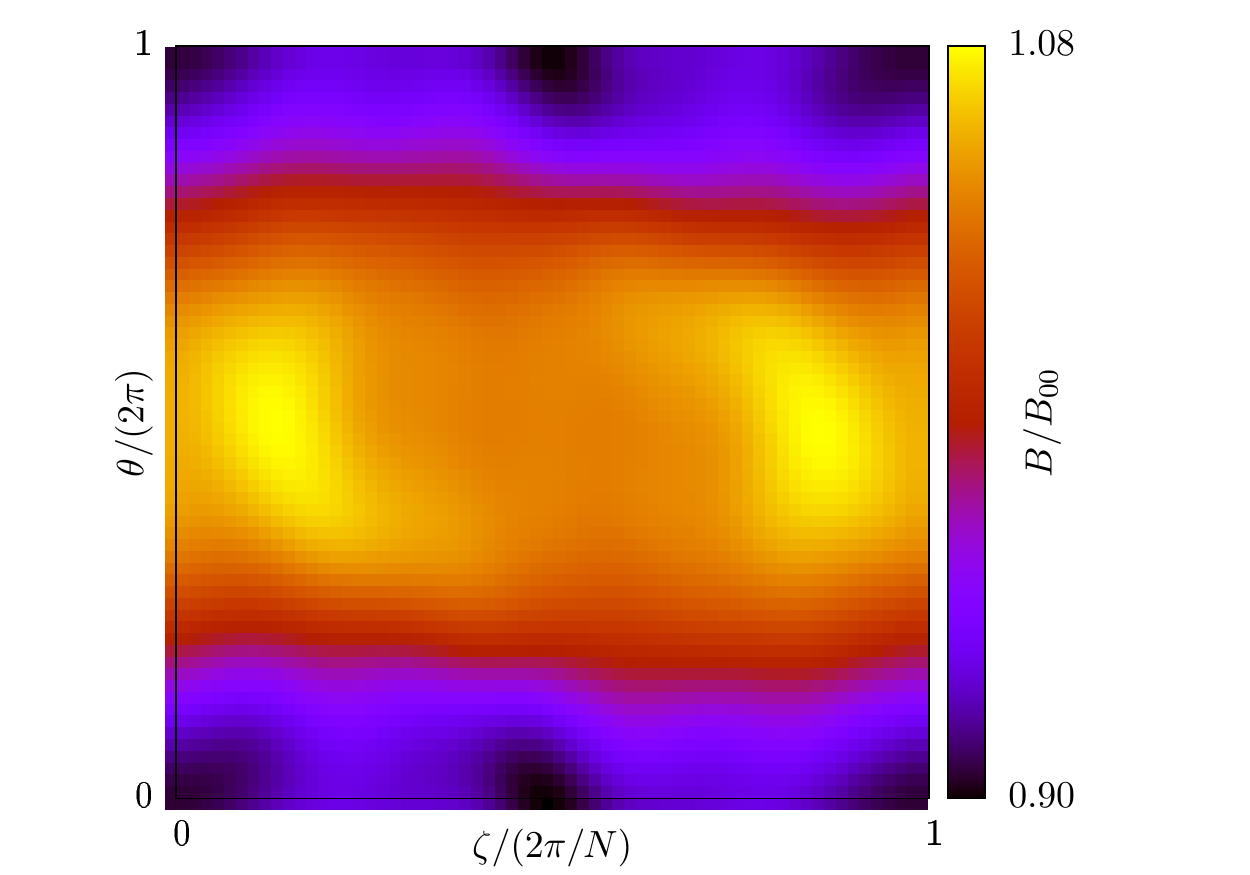}
\includegraphics[angle=0,width=0.45\columnwidth]{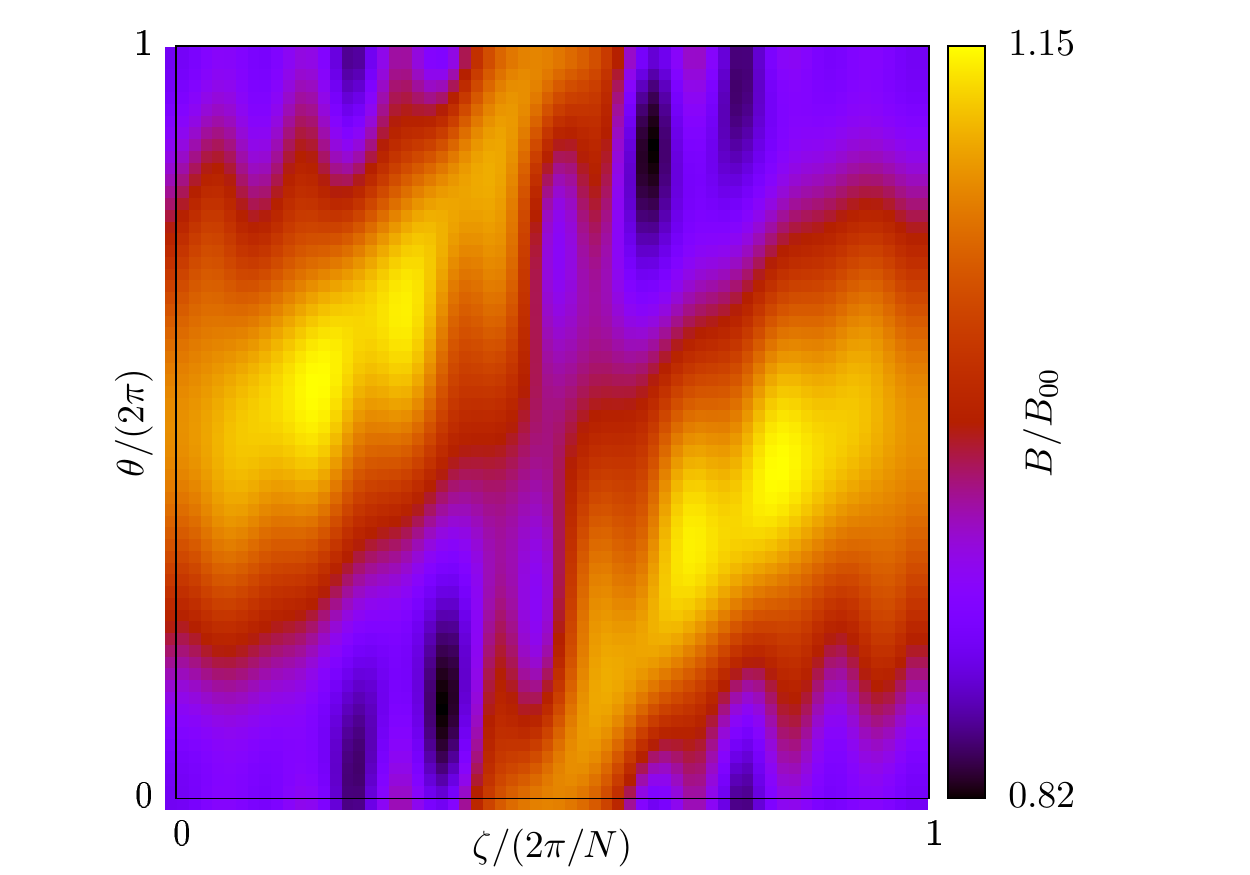}
\caption{Magnetic field strength for surface {$\psi/\psi_{LCFS}=0.5$} of the W7-X high-mirror configuration (top left), the LHD $R_{ax}=3.75\,$m configuration (top right), an NCSX equilibrium (bottom left) and the TJ-II standard configuration (bottom right).}
\label{FIG_MAP}
\end{figure}

In this subsection, we will show that~\KNOSOS~can be used for creating a \DKES-like database of monoenergetic transport coefficients at low collisionalities. We will compare our calculations with~\DKES, both in results and computing time. Let us first discuss the rationale behind the monoenergetic approach, which is not specific to~\DKES, and the particular simplifications involved in~\DKES. More details can be found in the overview paper~\citep{beidler2011icnts}.

Predictive transport simulations solve the energy transport equation for every species:
\begin{equation}
\frac{3}{2}\frac{\partial n_bT_b}{\partial t}+ \frac{1}{r}\frac{\partial}{\partial r}(r Q_b)= \fsa{P_b}\,,
\end{equation}
where $P_b$ is the net energy input to species $b$ and the energy flux $Q_b$ contains a turbulent contribution, at least close to the edge, that is currently provided by simplified models~\citep{turkin2011predictive}. Calculating the time evolution of the energy, as in~\citep{sunnpedersen2015op11}, or finding the steady-sate solution as in~\citep{geiger2014w7x}, requires evaluating the neoclassical contribution to $Q_b$ a large number of times. The monoenergetic approach, together with some simplifications to the drift-kinetic equation, provides a way out of solving the drift-kinetic equation many times.

Strictly speaking, monoenergetic transport coefficients can always be calculated if the velocity $v$ is a parameter in the drift-kinetic equation that is being solved, as in the case of equation~(\ref{EQ_NDKE}): one can rewrite
\begin{equation}
Q_b = \int_0^\infty\mathrm{d} v D_{11,b}\frac{m_bv^2}{2} F_{M,b} \Upsilon_b\frac{\partial\psi}{\partial r}
\end{equation}
as a {\textit{convolution}} of monoenergetic transport coefficients
\begin{equation}
D_{11,b} = 2\left(\frac{\partial r}{\partial\psi}\right)^2\left\langle\int_{B^{-1}_{{\rm max}}}^{B^{-1}}\mathrm{d}\lambda\frac{v^3 B}{|v_\parallel |}\frac{g_b}{F_{M,b}\Upsilon_b}\mathbf{v}_{D,b}\cdot\nabla\psi\right\rangle \,,
\label{EQ_D11}
\end{equation}
where $g_b$ is the solution of equation~(\ref{EQ_NDKE}). Up to this point, the reduction in computation time associated to the monoenergetic approach stems from the fact that $v$ is a parameter in equation (\ref{EQ_NDKE}), which is then easier to solve than a drift-kinetic equation with energy diffusion in the collision operator.

Additionally, some fundamental simplifications are done by~\DKES\jlvg{: instead} of $Q_b$, it calculates
\begin{equation}
\hat Q_b=\fsa{\mathbf{\hat Q_b}\cdot\nabla r} = \int_0^\infty\mathrm{d} v \hat D_{11,b}\frac{m_bv^2}{2} F_{M,b} \Upsilon_b\frac{\partial\psi}{\partial r}\,,\label{EQ_HATQ}
\end{equation}
with 
\begin{equation}
\hat D_{11,b} = 2\left(\frac{\partial r}{\partial\psi}\right)^2\left\langle\int_{B^{-1}_{{\rm max}}}^{B^{-1}}\mathrm{d}\lambda\frac{v^3 B}{|v_\parallel |}\frac{\hat g_b}{F_{M,b}\Upsilon_b}\mathbf{v}_{M,b}\cdot\nabla\psi\right\rangle \,.
\end{equation}
Here, $\hat g_b$ is the solution of a modified version of equation~(\ref{EQ_NDKE}), simplified as
\begin{equation}
\hat I_{v_E,\alpha}(\alpha,\lambda)\partial_\alpha \hat g_b + I_{v_{M,\psi}}(\alpha,\lambda) {v_{d,b}} F_{M,b}\Upsilon_b
= \nu_{\lambda,b} \partial_\lambda\left[ I_\nu(\alpha,\lambda) \partial_\lambda \hat g_b\right]\,.
\label{EQ_DKES}
\end{equation}
With respect to equations (\ref{EQ_NDKE}) and (\ref{EQ_D11}), we have set 
\begin{align}
\mathbf{v}_E\cdot\nabla\psi&=0\,,\nonumber\\
I_{v_{M,\alpha}}&=0\,,\nonumber\\
I_{v_{E,\psi}}&=0\,,\label{EQ_BINTDKES}
\end{align}
and replaced $I_{v_{E,\alpha}}$ with
\begin{align}
\hat I_{v_{E,\alpha}}&=\Psi_t'\partial_\psi\varphi_0\int_{l_{b_1}}^{l_{b_2}}\frac{B^2}{\fsa{B^2}} \frac{\mathrm{d}l}{\sqrt{1-\lambda B}}\,.
\end{align}
In other words, the effect of the tangential electric field and the tangential magnetic drift is ignored, and an incompressible $\mathbf{E}\times\mathbf{B}$ tangential drift is used (this \jlvg{last} simplification is specific of~\DKES~and is not used by other codes in~\citep{beidler2011icnts}). While it is well known~\citep{calvo2017sqrtnu} that these effects need to be kept in the drift-kinetic equation for an accurate computation of the radial fluxes, there is a range of situations in which $\hat Q_b\approx Q_b$ (this will be discussed in detail in \S\ref{SEC_TANGVM}) and this inaccuracy allows for a very large reduction of the computing time. The reason is that, for a given flux surface, when normalized by the plateau value
\begin{align}
\hat D_{11}^*&\equiv \frac{\hat D_{11,b}}{D_{11,b}^p}\,,\nonumber\\
D_{11,b}^p&= \frac{\pi v_{d,b}^2 R_0}{4v\iota}\,,
\end{align}
the transport coefficients $\hat D_{11}^*$ only depend on two $v$-dependent dimensionless parameters, the collisionality 
\begin{equation}
\nu_{*}=\frac{R_0\nu_\lambda}{\iota v}\,,
\end{equation}
and the normalized radial electric field
\begin{equation}
v_{E*}=\frac{E_r}{v B_{0,0}}\,.
\end{equation}
Here, $R_0$ is the major radius, and the main Fourier mode of $B$ (see~\ref{AP2}) is $B_{0,0}\sim 1\,$T in all the simulations presented in this paper. Since there is no species dependence, in the rest of the subsection we follow the common practice of dropping the species index when discussing monoenergetic calculations. A predictive transport simulation thus requires to precompute a so-called database of (\DKES-like) monoenergetic coefficients $\hat D_{11}^*( \nu_*,v_{E*}$). Once this is done, the calculation of $\hat Q_b$ for given $n_b$, $T_b$ and $E_r$ using equation~(\ref{EQ_HATQ}) requires a few bidimensional interpolations and an integral in $v$. The problem then lies in the computation of the database $\hat D_{11}^*( \nu_*,v_{E*})$ for every new magnetic configuration, which typically takes hours, due to the poor convergence of~\DKES~(and most neoclassical codes~\citep{beidler2011icnts}) at low collisionalities. We will show that the bounce-average technique greatly reduces the computing time by using in~\KNOSOS~equation (\ref{EQ_DKES}) and comparing the results with~\DKES. Calculations without the simplifications made by~\DKES~are left for \S\ref{SEC_TANGVM}.

In order to illustrate the performance of~\KNOSOS~\jlvg{in a variety of three-dimensional configurations}, we choose four very different types of stellarators. Figure~\ref{FIG_MAP} shows the map of the magnetic field strength on the flux surface $\psi/\psi_{LCFS}=0.3$ of the high-mirror configuration of the helias W7-X (top left), the $R_{ax}=3.75\,$m configuration of the heliotron LHD (top right), an equilibrium of NCSX close to quasiaxisymmetry (bottom left) and the standard configuration of the heliac TJ-II~(bottom right)\citep{ascasibar2019iaea}. 

\begin{figure}
\centering
\includegraphics[angle=0,width=0.45\columnwidth]{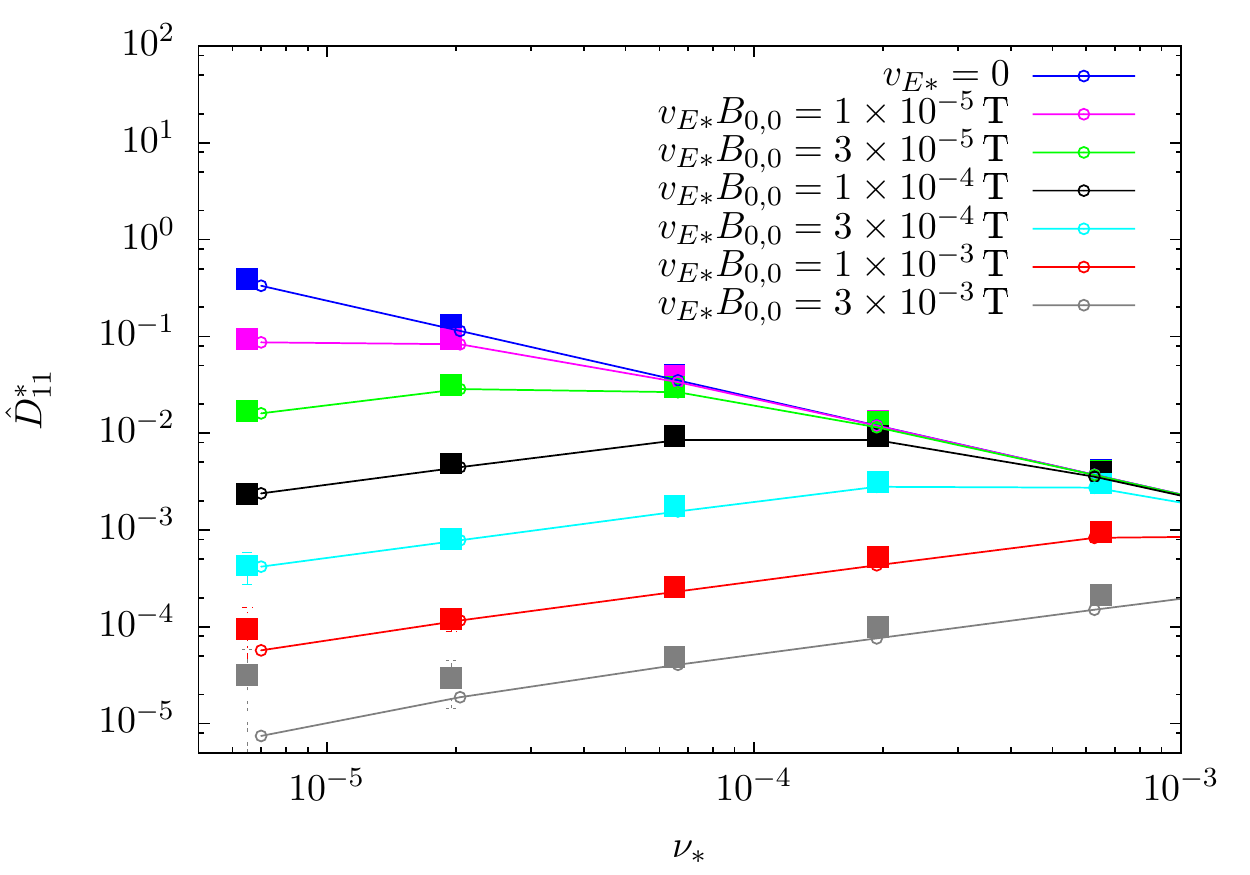}
\includegraphics[angle=0,width=0.45\columnwidth]{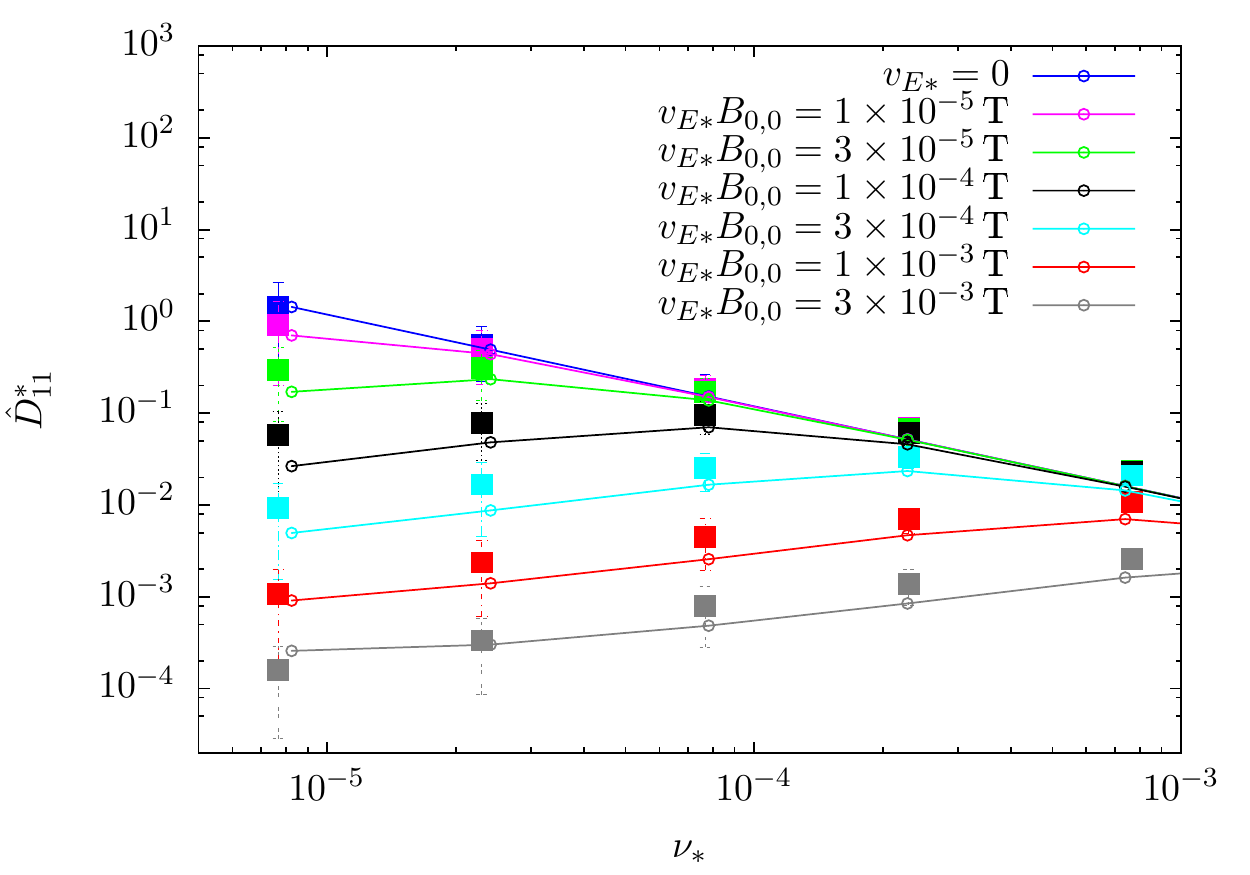}
\includegraphics[angle=0,width=0.45\columnwidth]{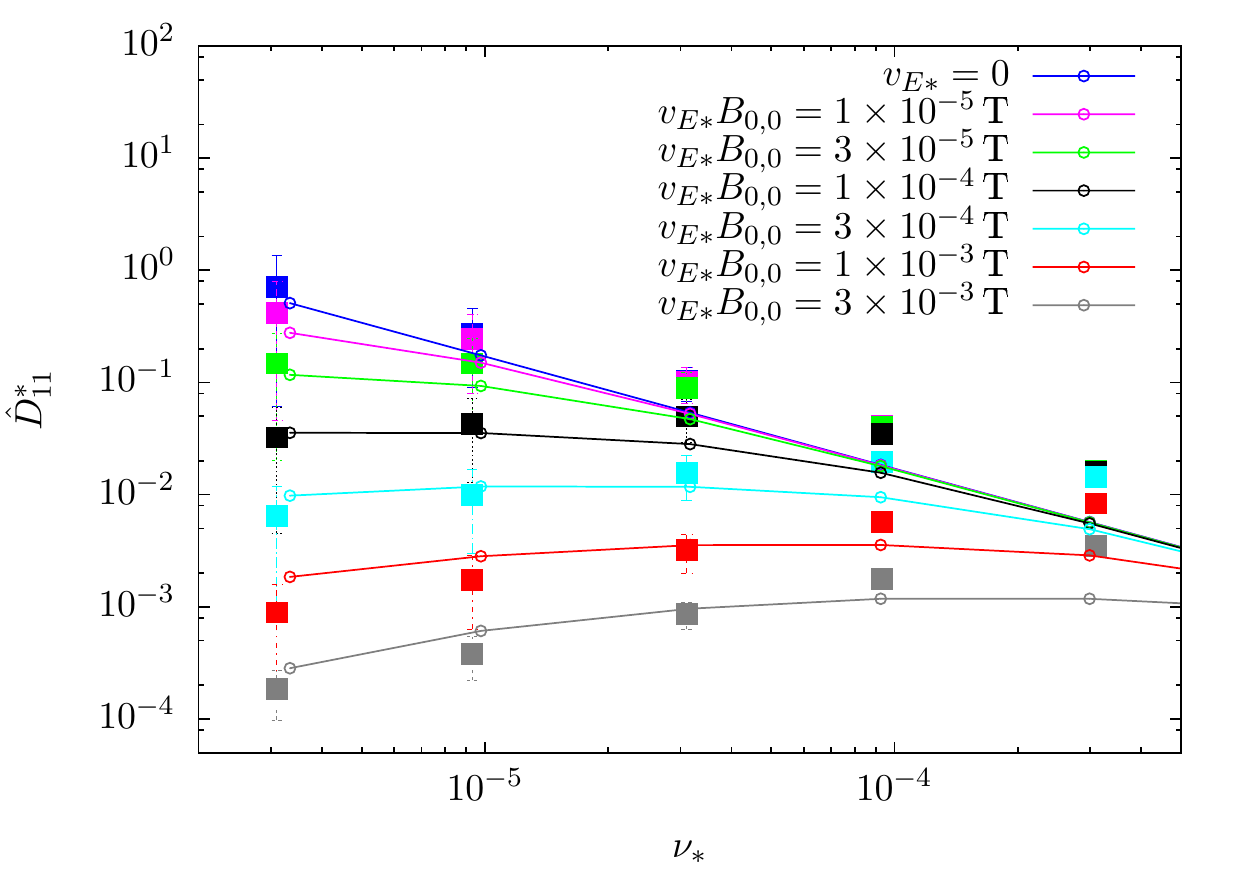}
\includegraphics[angle=0,width=0.45\columnwidth]{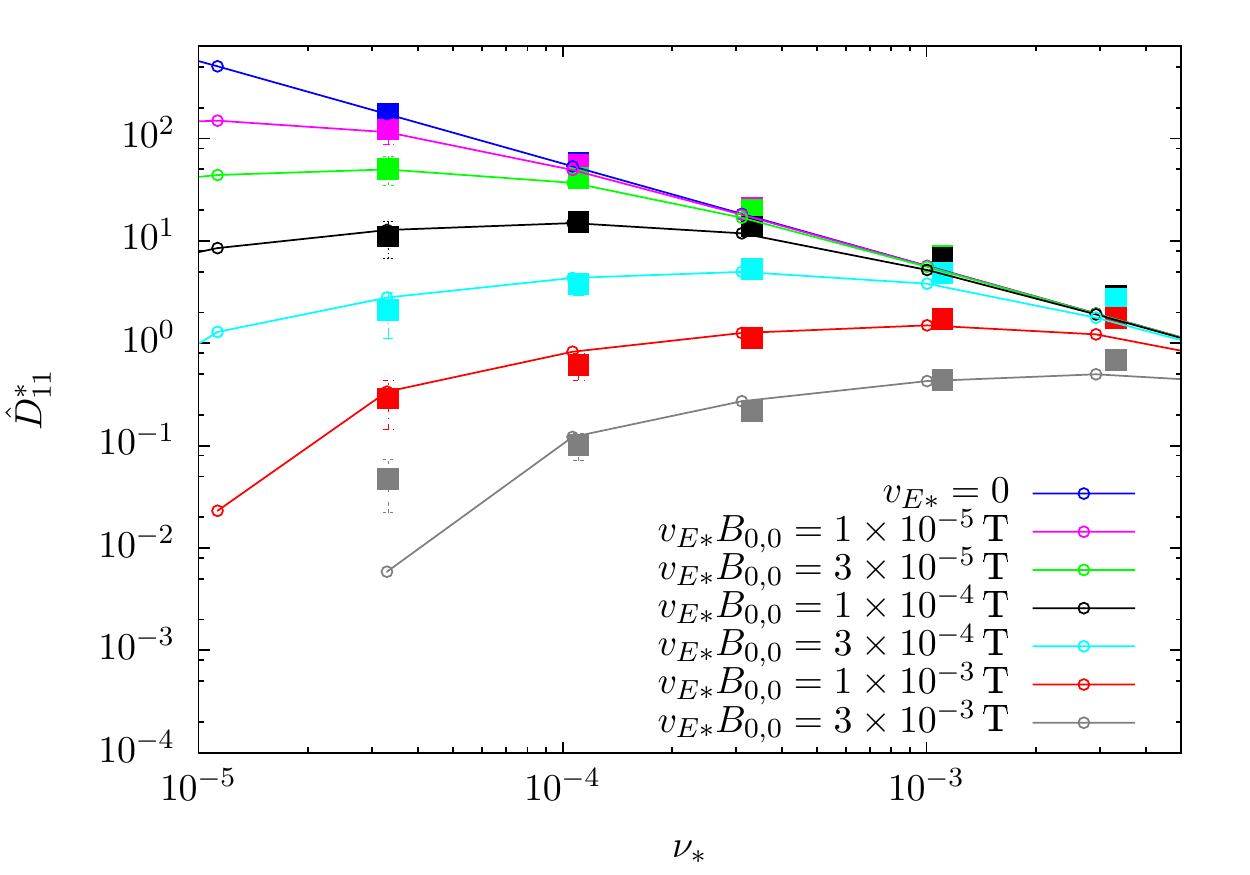}
\caption{Monoenergetic transport coefficients calculated with~\DKES~(full squares) and~\KNOSOS~(small open circles with lines) as a function of the collisionality at $\psi/\psi_{LCFS}=0.5$ surface of W7-X (top left), LHD (top right), NCSX (bottom left) and TJ-II (bottom right). The colour code is: $v_{E*}B_{0,0}=0$ (blue), $1\times 10^{-5}\,$T (magenta), $3\times 10^{-5}\,$T (green), $1\times 10^{-4}\,$T (black), $3\times 10^{-4}\,$T (cyan), $1\times 10^{-3}\,$T (red), and $3\times 10^{-3}\,$T (grey).}
\label{FIG_D11}
\end{figure}

Figure~\ref{FIG_D11} shows the first comparisons between~\KNOSOS~and~\DKES, in which the normalized monoenergetic transport coefficient $\hat D_{11}^*$ is calculated for several values of the collisionality and the normalized radial electric field. Figure~\ref{FIG_D11} (top left) contains data for the W7-X high-mirror configuration, which we discuss in more detail. The expected $1/\nu$ dependence is observed at the highest collisionalities and, due only to the absence of tangential magnetic drift, for small values of $v_{E*}$. There is $\sqrt{\nu}$ characteristic behaviour elsewhere, with smaller levels of transport for larger $|E_r|$. The comparison between~\KNOSOS~and~\DKES~is satisfactory, with agreement within the error bars of the~\DKES~calculation \reft{(for a discussion on how the error bars of~\DKES~are determined, see page 14 of~\citep{beidler2011icnts})}, and only at the highest collisionalities, and for the largest values of $E_r$, there are very small differences. The calculation for all the points of this case was made with ${\cal{N}}_\alpha=32$  and ${\cal{N}}_\lambda=64$, and it took $2.0$ seconds in a single standard CPU. Of this time, around $0.7$ seconds were used for setting the grid and performing the bounce-averages, and then it took less than $0.04$ seconds to calculate each point. This number may be reduced even further using smaller ${\cal{N}}_\lambda$ for the cases of largest collisionality and smallest radial electric field. In the $\sqrt{\nu}$ regime, transport is given by a small layer close to the boundary between passing and trapped particles. The size in $\lambda$ of this layer is proportional to $\sqrt{\nu_\lambda/E_r}$~\citep{calvo2017sqrtnu}, and this determines the required number of grid points ${\cal{N}}_\lambda$ in the low collisionality cases with radial electric field.
\begin{figure}
\centering
\includegraphics[angle=0,width=0.45\columnwidth]{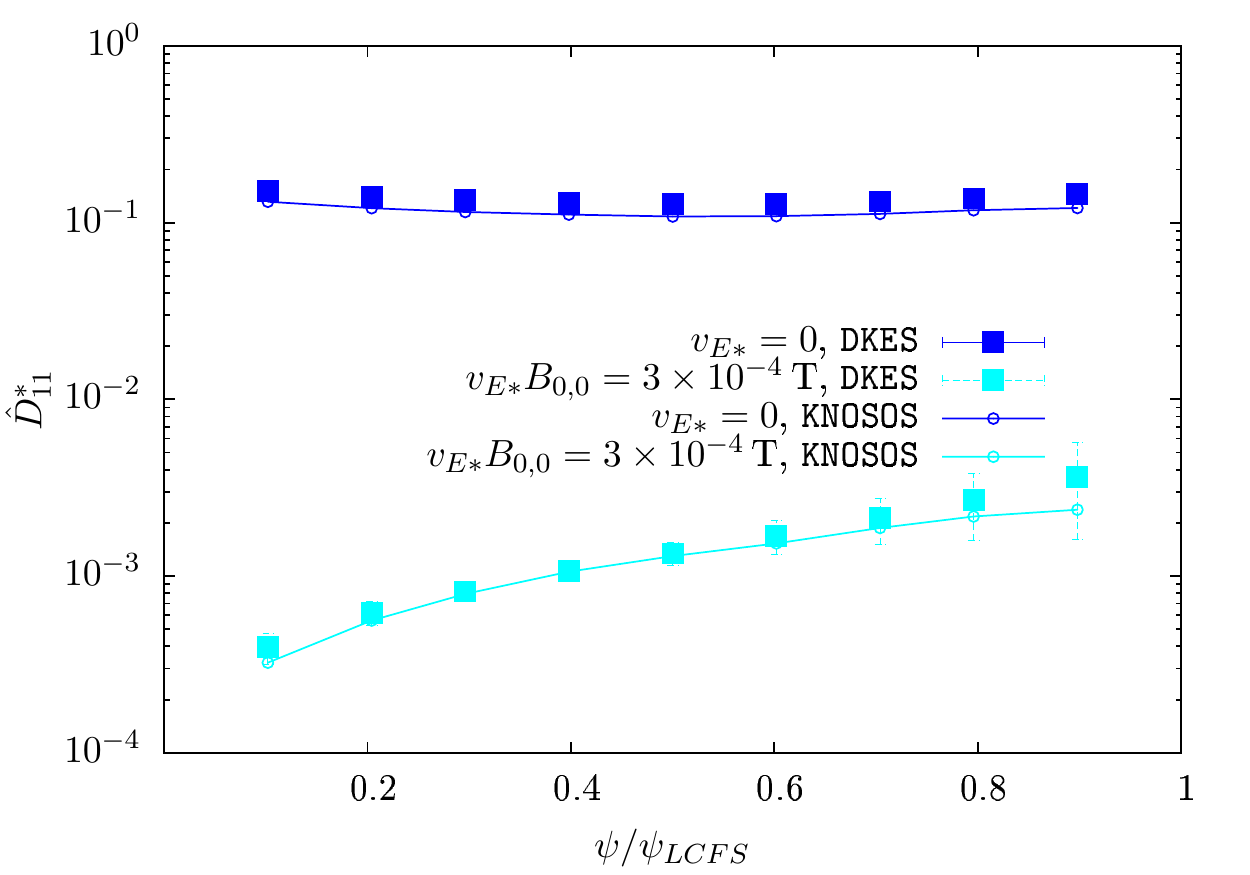}
\includegraphics[angle=0,width=0.45\columnwidth]{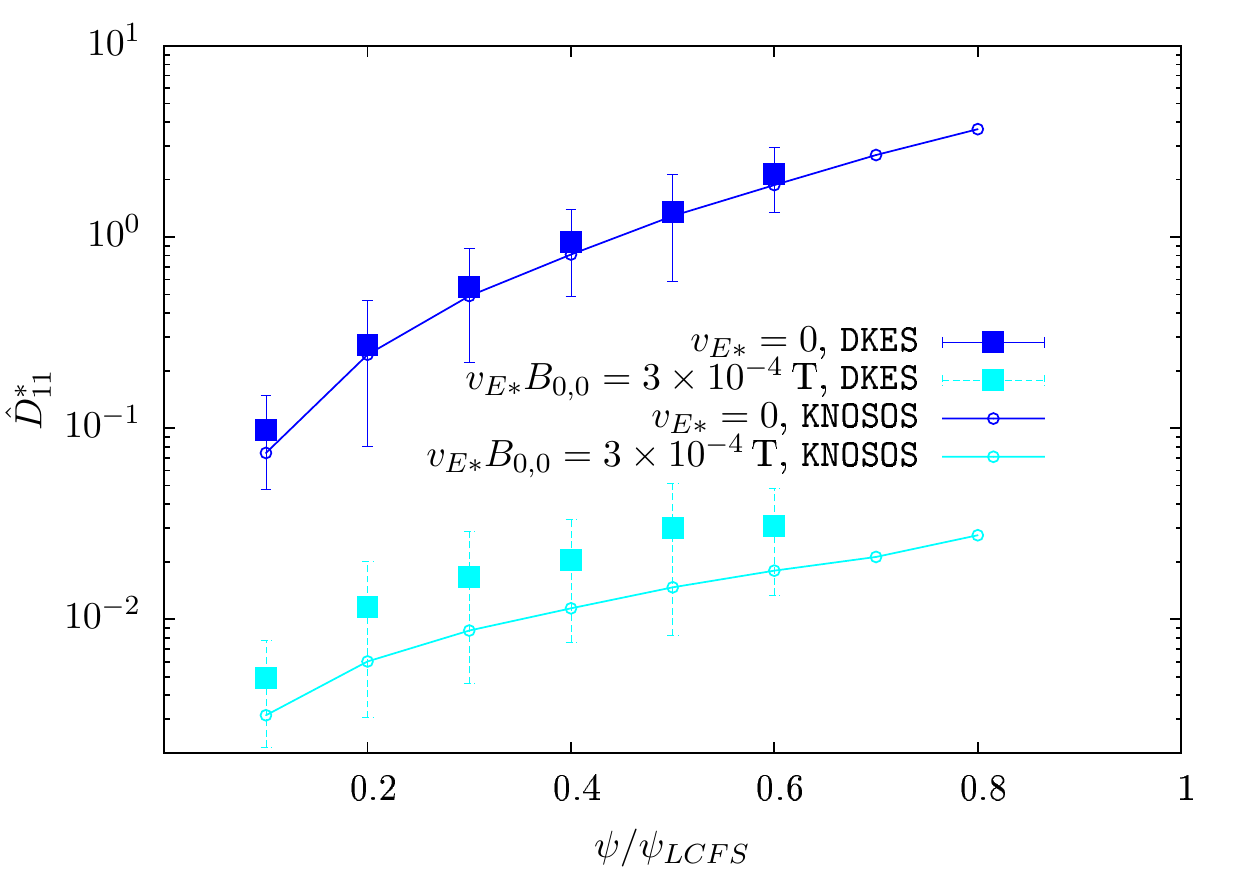}
\includegraphics[angle=0,width=0.45\columnwidth]{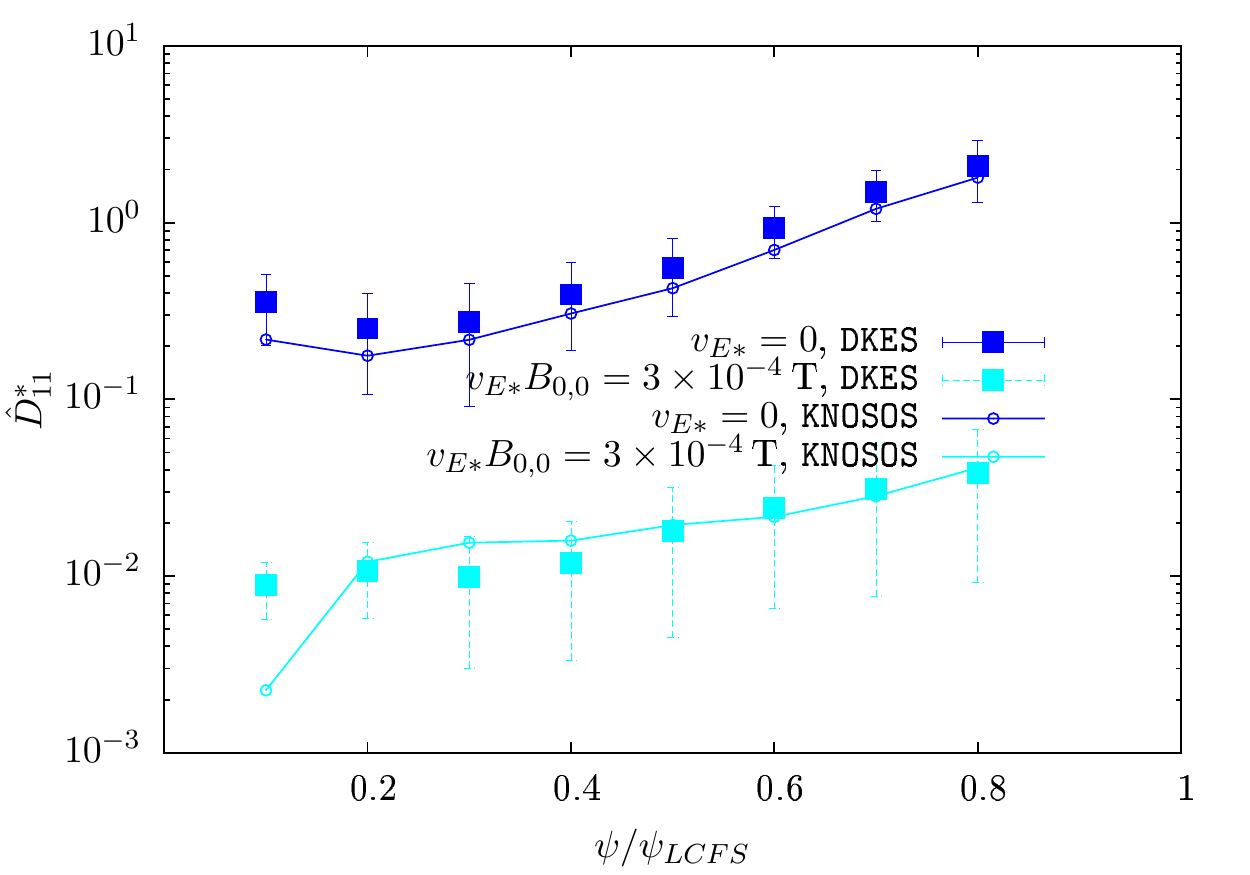}
\includegraphics[angle=0,width=0.45\columnwidth]{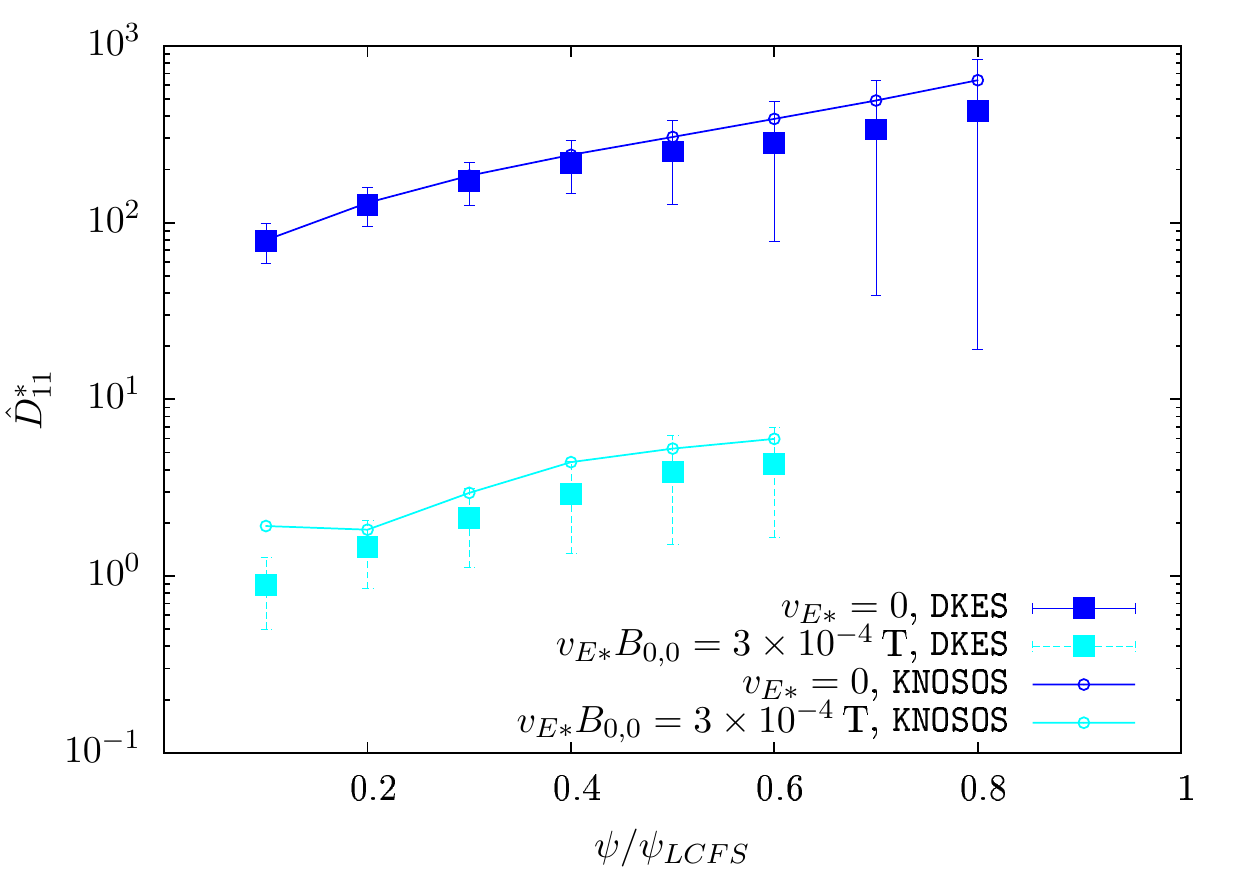}
\caption{Radial profile of normalized monoenergetic transport coefficient calculated with~\DKES~(full squares) and~\KNOSOS~(small open circles with lines) for W7-X (top left), LHD (top right), NCSX (bottom left) and TJ-II (bottom right). Cyan corresponds to the $\sqrt{\nu}$ regime ($v_{E*}B_{0,0}=3\times 10^{-4}\,$T) and blue to the $1/\nu$ regime ($v_{E*}=0$).}
\label{FIG_D11PROF}
\end{figure}

Similar results can be seen for LHD in figure~\ref{FIG_D11} (top right). ${\cal{N}}_\alpha=32$ and ${\cal{N}}_\lambda=64$ grid points were used, and the total computation time was 2.1 seconds. For NCSX, figure~\ref{FIG_D11} (bottom left), the agreement is good except for the higher collisionalities, where the $1/\nu$ regime should connect with a banana regime (see figure 15 of~\citep{beidler2011icnts}). This regime\reft{, which cannot be not described by a bounce-averaged drift-kinetic equation}, could be easily \reft{added to}~\KNOSOS~following~\citep{landreman2012omni}. ${\cal{N}}_\alpha=32$ and ${\cal{N}}_\lambda=64$ grid points were used, and the total computation time was 1.0 seconds. Finally, figure~\ref{FIG_D11} (bottom right) contains the results for TJ-II, the hardest case due to its complicated magnetic geometry, see figure~\ref{FIG_MAP} (bottom right). ${\cal{N}}_\alpha=32$ and ${\cal{N}}_\lambda=128$ were used, and the simulation took 157 seconds. \reft{The points corresponding to $v_{E*}B_{0,0} \geq 10^{-3}\,$T and $\nu_*< 10^{-4}$ do not agree with~\DKES: this would have required a finer grid, and it is an indication of how cases deeper in the $\sqrt{\nu}$ regime are more difficult to compute.} The rest of the simulations agree with~\DKES~and reach even lower collisionalities than those typically required for describing a TJ-II plasma, whose ion temperature never exceeds a few hundred eV.

Figure~\ref{FIG_D11PROF} contains, for each of the four configurations, radial profiles of the transport coefficient $\hat D^*_{11}$ for two cases, $v_{E*}=0$ and $v_{E*}B_{0,0}=3\times 10^{-4}\,$T, for a given collisionality. They are meant to represent the level of transport in the $1/\nu$ regime ($\hat D^*_{11}$ is by definition proportional to $\epsilon_{eff}^{3/2}$, being $\epsilon_{eff}$ the effective ripple) and the $\sqrt{\nu}$ regime, respectively. We choose $\nu_*=2\times 10^{-5}$ for W7-X (top left) and LHD (top right), $\nu_*=10^{-5}$ for NCSX (bottom left) and $\nu_*=3\times 10^{-5}$ for TJ-II (bottom right). It can be observed that the good agreement holds for all cases at all radial positions. The comparison of the different parts of figure~\ref{FIG_D11PROF} provides additional information that may be relevant when devising a stellarator optimization strategy: in general, configurations with lower $1/\nu$ transport show lower $\sqrt{\nu}$ transport as well. This is not surprising considering that both quantities are connected to the bounce-averaged radial component of the magnetic drift, which appears in the source of the drift-kinetic equation~(\ref{EQ_DKEFINAL}) in both regimes, and which is in turn proportional to the variation of the second adiabatic invariant on the flux surface, $\partial_\alpha J$. As long as the optimization procedure actually reduces the size of $\partial_\alpha J$, both the $1/\nu$ and $\sqrt{\nu}$ (and superbanana-plateau) regimes will generally be optimized. Nevertheless, using directly the effective ripple as figure of merit of neoclassical transport does not automatically guarantee a reduction of $\partial_\alpha J$, and the $\sqrt{\nu}$ transport may remain unoptimized. Figure~\ref{FIG_D11PROF} (top) may represent an example of this situation: while this W7-X configuration is designed to have low level of $1/\nu$ transport at an intermediate radial position (where the plasma volume is relatively large and neoclassical transport is expected to be at least comparable to anomalous transport), the $\sqrt{\nu}$ transport is smallest exactly at the magnetic axis. A fast computation of the $\sqrt{\nu}$ and superbanana-plateau opens the possibility of a more efficient optimization with respect to neoclassical transport. \reft{In the next subsection, we will see that the regimes of collisionality lower than the $1/\nu$ play a role in the transport of relevant plasmas. For this reason, their fast computation opens the possibility of a more efficient stellarator optimization with respect to neoclassical transport.}


\subsection{Effect of the tangential magnetic drift on the radial transport of energy}\label{SEC_TANGVM}


In \S\ref{SEC_DKES}, we have shown solutions of equation~(\ref{EQ_DKES}), a simplified drift-kinetic equation that is not accurate when the tangential components of the magnetic drift and of the electric field play a role. In this section, we will demonstrate the importance of solving equation~(\ref{EQ_NDKE}) instead of equation~(\ref{EQ_DKES}), i.e., of computing $Q_b$ and not $\hat Q_b$, when calculating the radial energy flux in real plasmas. It must be noted that the solution of equation~(\ref{EQ_NDKE}) with~\KNOSOS~is not computationally more expensive than that of equation (\ref{EQ_DKES}): in the superbanana-plateau regime, that may arise in the presence of the tangential magnetic drift for certain values of $E_r$, transport is dominated by a resonant layer whose size decreases with $(\nu_\lambda/E_r)^{1/3}$, i.e., slower than the boundary layer that determines the $\sqrt{\nu}$ transport~\citep{calvo2017sqrtnu}. Calculating $Q_b$ instead of $\hat Q_b$ does not require a larger value of ${\cal{N}}_\lambda$ in general.

In this section, we focus on characterizing the effect of the tangential magnetic drift for the particular case of $\varphi_1=0$. We advance one of the salient results: this effect will be non-negligible even at not very low collisionalities. The reason is that the calculation of the energy flux for a given plasma, characterized by the kinetic profiles, requires the solution of the drift-kinetic equation for several values of the velocity, see equation~(\ref{EQ_CONV}), with the normalized particle energy $(v/v_{th,b})^2$ spanning several orders of magnitude. This means that, even if the thermal particles are in $1/\nu$ regime, there are particles with higher $v$ that are in lower collisionality regimes. 

\begin{figure}
\centering
\includegraphics[angle=0,width=0.6\columnwidth]{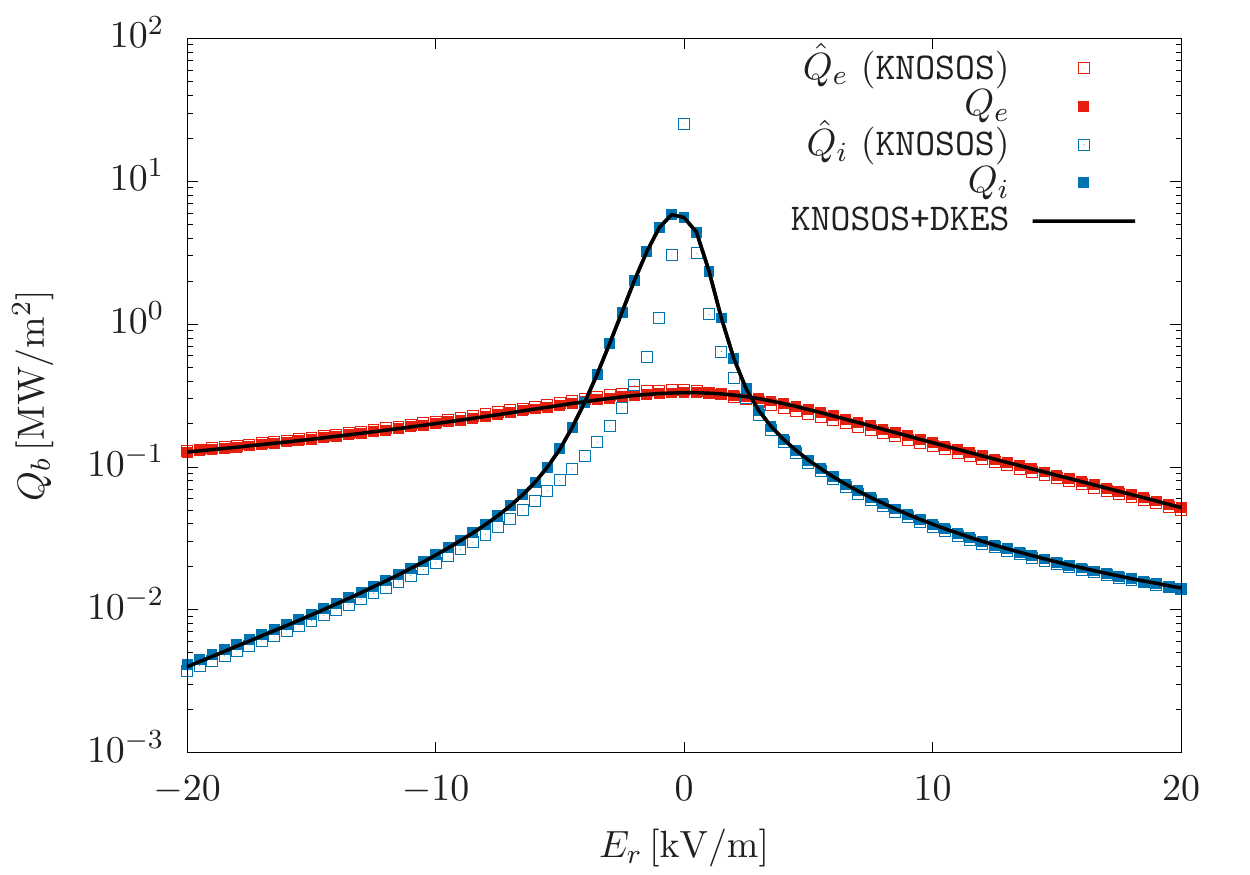}
\includegraphics[angle=0,width=0.6\columnwidth]{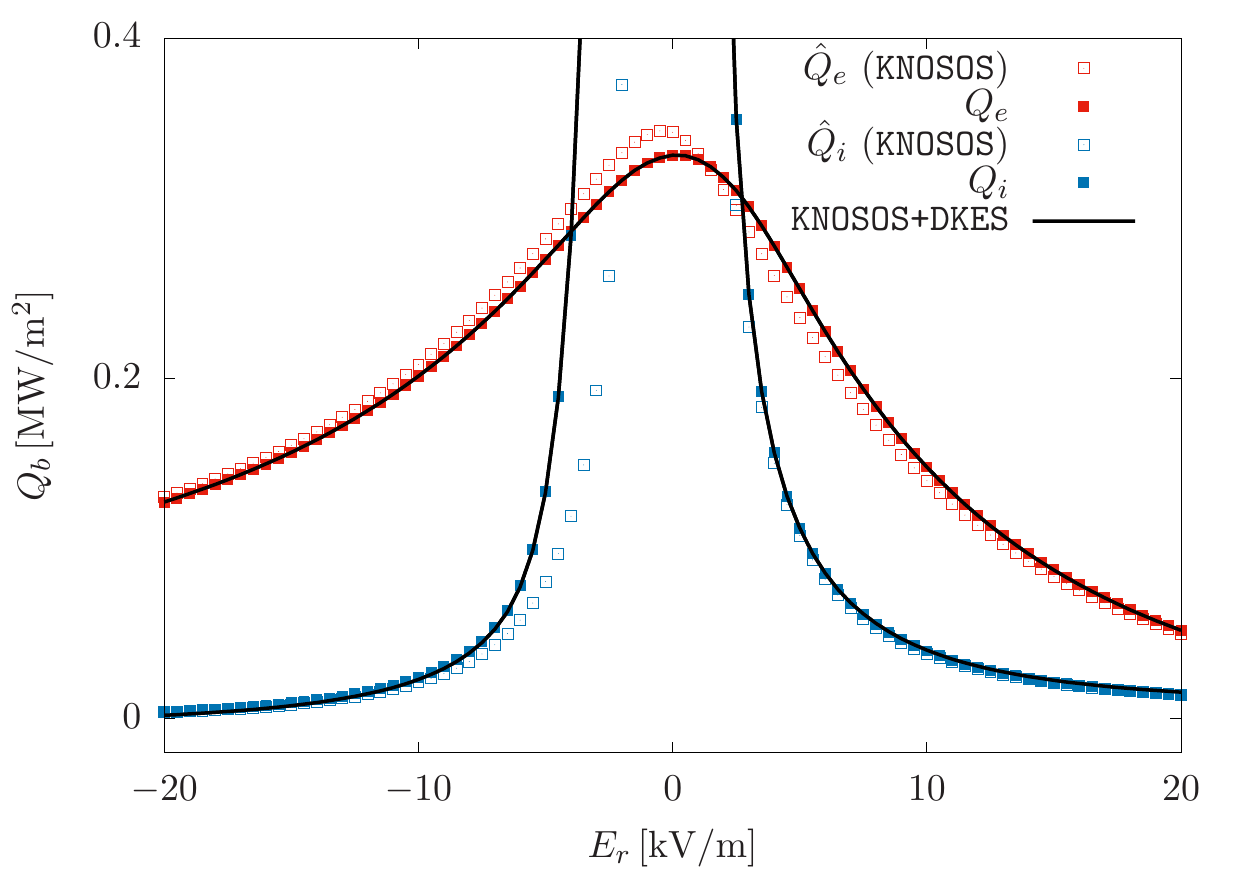}
\caption{Radial energy flux as a function of the radial electric field for a W7-X high-performance plasma: logarithmic (top) and linear (bottom) scale.}
\label{FIG_QER1}
\end{figure}

Figure~\ref{FIG_QER1} contains simulations for the high-mirror configuration of W7-X at $\psi/\psi_{LCFS}=0.25$, which corresponds to $r/a=0.5$. We choose a pure hydrogen plasma, with $n_e=8.0\times 10^{19}\,$m$^{-3}$, $\partial_r n_e/n_e=-2.0\,$m$^{-1}$, $T_e = T_i = 4.0\,$keV, $\partial_r T_e/T_e =\partial_r T_i/T_i = -3.0\,$m$^{-1}$. These are values comparable to those measured in high-performance OP1.2 plasmas of W7-X~\citep{klinger2019op12} in the region of crossover between positive and negative radial electric field, corresponding to electron and ion root solutions of the ambipolarity equation~\citep{pablant2019ionroot}. In these plasmas, neoclassical transport calculated neglecting the tangential magnetic drift typically accounts for around half the total experimental transport. Figure~\ref{FIG_QER1} (top) contains a plot, in logarithmic scale, of the ion and electron radial energy flux as a function of the radial electric field. Empty and full blue boxes correspond to $\hat Q_i$ and $Q_i$ respectively, both calculated with~\KNOSOS. We immediately see that $\hat Q_i$ overestimates the radial energy flux at small values of the radial electric field, specially at $E_r = 0$ (strictly the only point of the figure where $\hat Q_i$ is proportional to $\varepsilon_{eff}^{3/2}$). The tangential drifts make the ion flux decrease, differently in the case of $\hat Q_i$ and $Q_i$, as we will discuss below. Finally, empty and full red boxes correspond to $\hat Q_e$ and $Q_e$ calculated with~\KNOSOS. In this plot, is difficult to notice any difference between the different electron calculations. Figure~\ref{FIG_QER1} (top) contains additional black lines that are the result of combining calculations with~\DKES~and~\KNOSOS. We will leave the discussion of these results for the end of the section.

Figure~\ref{FIG_QER1} (bottom) contains a blowup in linear scale of the most relevant range of the data in figure~\ref{FIG_QER1} (top). Here, the effect of the tangential magnetic drift on the energy flux can be observed more clearly: the size of the peak at small $|E_r|$ is reduced and displaced to positive (negative) values in the case of electrons (ions). The effect is larger for the ions due to their larger normalized Larmor radius $\rho_{i*}$, which makes them leave the $1/\nu$ regimes at relatively higher collisionalities. We have mentioned that these plasmas are close to the crossover between ion and electron root, and this figure can help us discuss some features of transport in both situations. In electron root, the radial electric field is expected to be positive and large, and the electrons are expected to give the largest contribution to energy transport. According to figure~\ref{FIG_QER1} (bottom), $Q_e$ provides a minor, although systematic, correction to $\hat Q_e$, below 10\% for this plasma profiles and configuration. The situation is different in ion root, typically characterized by a negative radial electric field that is small in size, and dominant ion transport. Here, including the tangential magnetic drift can lead to large corrections, above 50\% in some cases.

\begin{figure}
\centering
\includegraphics[angle=0,width=0.6\columnwidth]{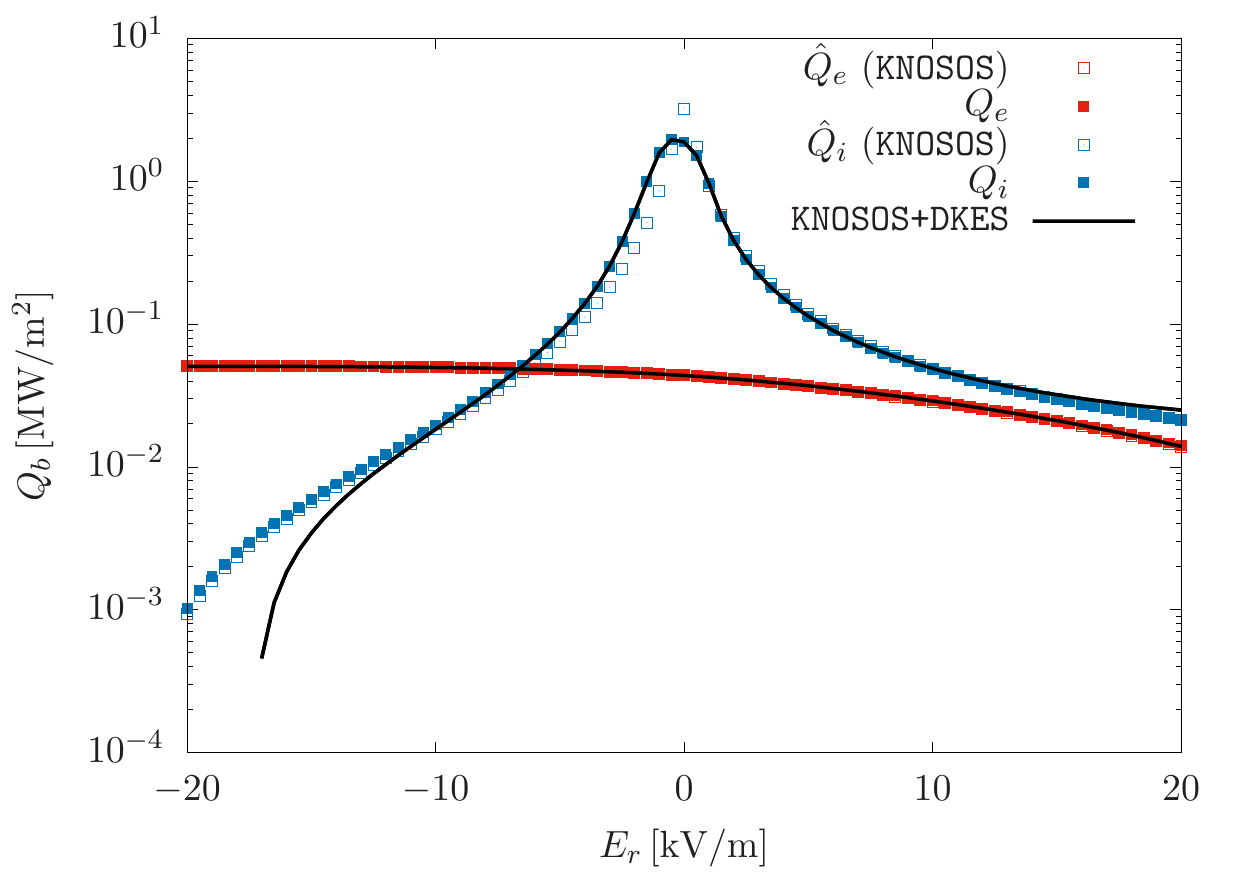}
\includegraphics[angle=0,width=0.6\columnwidth]{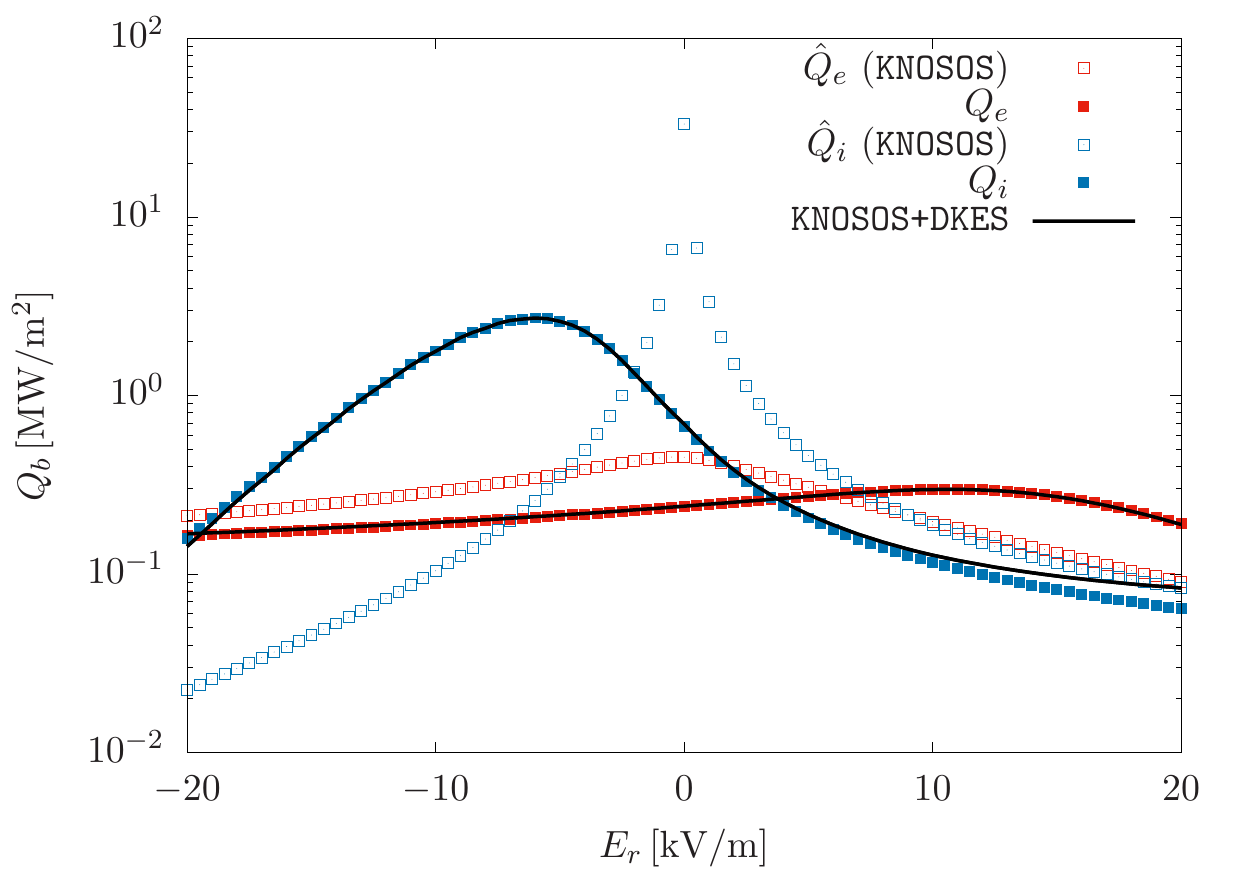}
\caption{Radial energy flux as a function of the radial electric field for a W7-X high density plasma (top) and an LHD plasma (bottom).}
\label{FIG_QER2}
\end{figure}

For the sake of completeness, figure~\ref{FIG_QER2} contains two more cases. In figure~\ref{FIG_QER2} (top) we repeat the calculation for a W7-X plasma of much higher collisionality, choosing $n_e=1.6\times 10^{20}\,$m$^{-3}$, $\partial_r n_e/n_e=-2.0\,$m$^{-1}$, $T_e = T_i = 2.5\,$keV, $\partial_r T_e/T_e =\partial_r T_i/T_i = -3.0\,$m$^{-1}$. We first note that the electrons are deep in the $1/\nu$ regime, since $\nu_{e*} = 3.4\times 10^{-2}$ and $\rho_{e*} = 1.4\times 10^{-5}$. Nevertheless, $Q_e(E_r)$ does not show the linear dependence expected when the $1/\nu$ dominates. This is an indication of what we advanced at the beginning of this section: even in plasmas nominally in the $1/\nu$ regime, the contribution of the $\sqrt{\nu}$ regime is not negligible, and should not be neglected in the optimization procedure. For the ions, even at these higher collisionalities and low temperatures, $\nu_{i*} = 1.6\times 10^{-2}$ is not much larger than $\rho_{i*} = 6.0\times 10^{-4}$ divided by the inverse aspect ratio. This means that, for ions slightly more energetic than the thermal ions, the tangential magnetic drift is relevant at small values of $|E_r|$~\citep{calvo2018jpp}. Figure~\ref{FIG_QER2} (top) shows indeed systematic differences between $\hat Q_i$ and $Q_i$. 

Finally, figure~\ref{FIG_QER2} (bottom) contains a calculation with the same kinetic profiles of figure~\ref{FIG_QER1} (top) for the inward-shifted configuration of LHD. It can be observed that the effects discussed in figure~\ref{FIG_QER2} (top) are even more pronounced, to the extent of changing qualitatively the $Q_b(E_r)$ dependence (and making it more similar to that reported in~\citep{matsuoka2015tangential}): while practically any increase of $|E_r|$ causes a reduction of $Q_i$ in W7-X, this is not the case for LHD. For finite ion-root values of $E_r$, $Q_i(E_r)$ has a peak whose height is determined by superbanana-plateau transport. 

In light of these results, two comments related to stellarator optimization can be made. First, the fact that the monoenergetic transport coefficients respond to small tangential $\mathbf{E}\times\mathbf{B}$ drifts differently in the inward-shifted LHD, with respect to other configurations, was already discussed in~\citep{beidler2011icnts}, and it can be observed more clearly when calculating the energy flux including the tangential magnetic drift. We also note that part of the neoclassical optimization of W7-X comes from its large aspect-ratio, which tends to make the tangential magnetic drift smaller, when compared with the $E\times B$ drift. It is then clear than a systematic study of the different low-collisionality regimes, and their different configuration dependence, should be addressed when devising an stellarator optimization strategy. Second, a comprehensive optimization strategy will involve, at least, solving energy transport consistently with ambipolarity and quasineutrality. Along this subsection, we have compared $Q$ and $\hat Q$ at fixed $E_r$, but a more systematic study applied to real discharges of W7-X, including the experimental validation of $E_r$ predictions, is ongoing~\citep{carralero2019irw}.

Let us finally discuss the black lines of figures~\ref{FIG_QER1} and~\ref{FIG_QER2}, which correspond to combining simulations of~\DKES~and~\KNOSOS. As we have argued at the beginning of this section, calculating the radial energy flux requires solving the drift-kinetic equation for velocities $(v/v_{th,b})^2$ spanning from $\sim 10^{-2}$ to $\sim 10^{2}$, typically. Similarly to what we discussed for $v\gg v_{th,b}$, this means that particles with $v\ll v_{th,b}$ could be in the plateau regime, and they would not be described by equation~(\ref{EQ_NDKE}). In order to quantify this effect, and to show that it is negligible for the high-performance plasmas of W7-X, we perform calculations of $Q_i(E_r)$ and $Q_e(E_r)$ combining~\KNOSOS~with~\DKES. This can be done by rewriting equation~(\ref{EQ_GAMMAQ}) as
\begin{equation}
Q_b = D_{11,b}^p\int_0^\infty\mathrm{d} v \left[ H(v_0-v) \hat D^*_{11}(v) + H(v-v_0) D^*_{11}(v) \right] \frac{m_bv^2}{2} F_{M,b} \Upsilon_b\,,
\label{EQ_DKESpKNOSOS}
\end{equation}
where $H$ is the Heaviside function, $v_0$ is a cut-off velocity, $\hat D^*_{11}(v)$ comes from~\DKES~in this case and
\begin{equation}
D^*_{11}(v) = \frac{D_{11,b}}{D_{11,b}^p}
\end{equation}
from~\KNOSOS. The latter is calculated according to equation~(\ref{EQ_D11}) solving the drift-kinetic equation that is correct at low collisionalities with $\varphi_1$ set to zero. In other words, monoenergetic transport coefficients $\hat D^*_{11}$ coming from~\DKES~are used above certain collisionality when performing the velocity integral and monoenergetic transport coefficients $D^*_{11}$ coming from~\KNOSOS~are used below that collisionality. The cut-off velocity $v_0$ must correspond to particles in the $1/\nu$ regime, which is correctly described by the two codes\reft{, in order to guarantee} that both codes are employed in the parameter region where they are accurate (and fast). \reft{Here, $v_0$ is a value of $v$ for which $D^*_{11}$~shows a clear $1/\nu$ dependence for non-zero $E_r$ and lies above the plateau value provided by~\DKES.}

In figures~\ref{FIG_QER1} and \ref{FIG_QER2} (bottom), the black lines corresponding to using equation~(\ref{EQ_DKESpKNOSOS}) barely separate from the solution of equation~(\ref{EQ_NDKE}). This means that the contribution of the plateau regime to the energy flux is negligible. Only for ions in the presence of very negative values of the radial electric field, in the high-density W7-X calculation, starts the black line to separate from the blue signs. This is to be expected: the contribution of low collisionalities to transport is reduced for very large values of the tangential magnetic drift (something that happens for smaller values of $|E_r|$ if $E_r$ is negative), and therefore the contribution of the plateau becomes non-negligible.


\subsection{Tangential electric field}\label{SEC_EUTERPE}


\begin{figure}
\begin{center}
\includegraphics[width=0.3\columnwidth,angle=0]{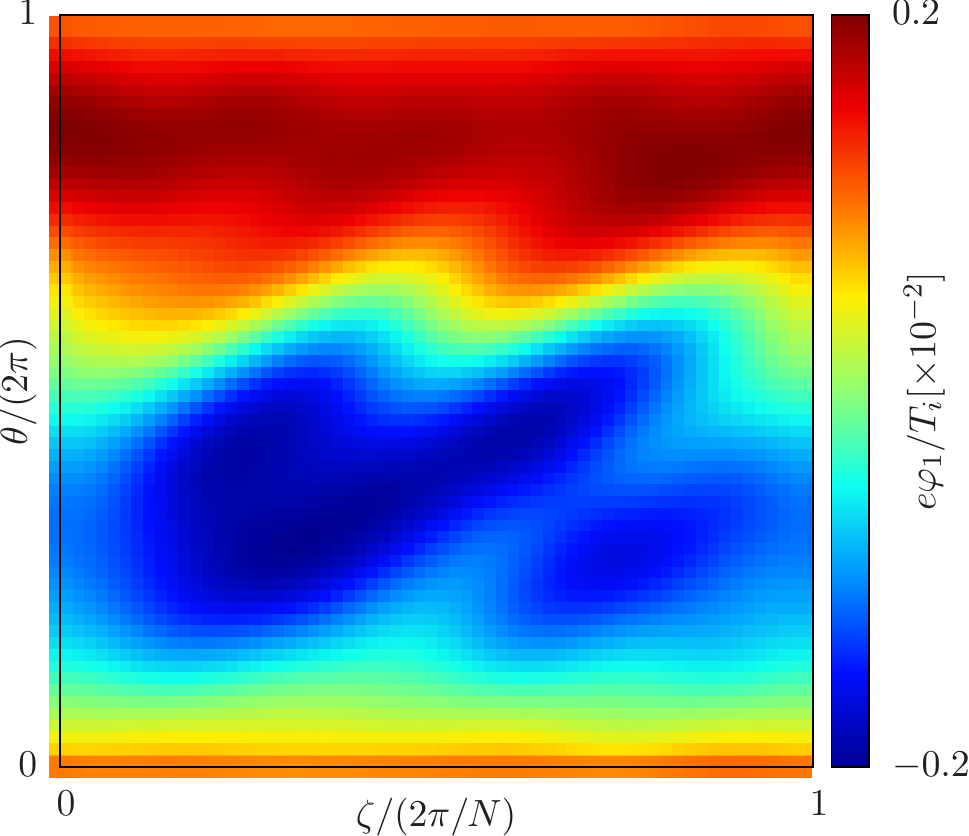}
\includegraphics[width=0.3\columnwidth,angle=0]{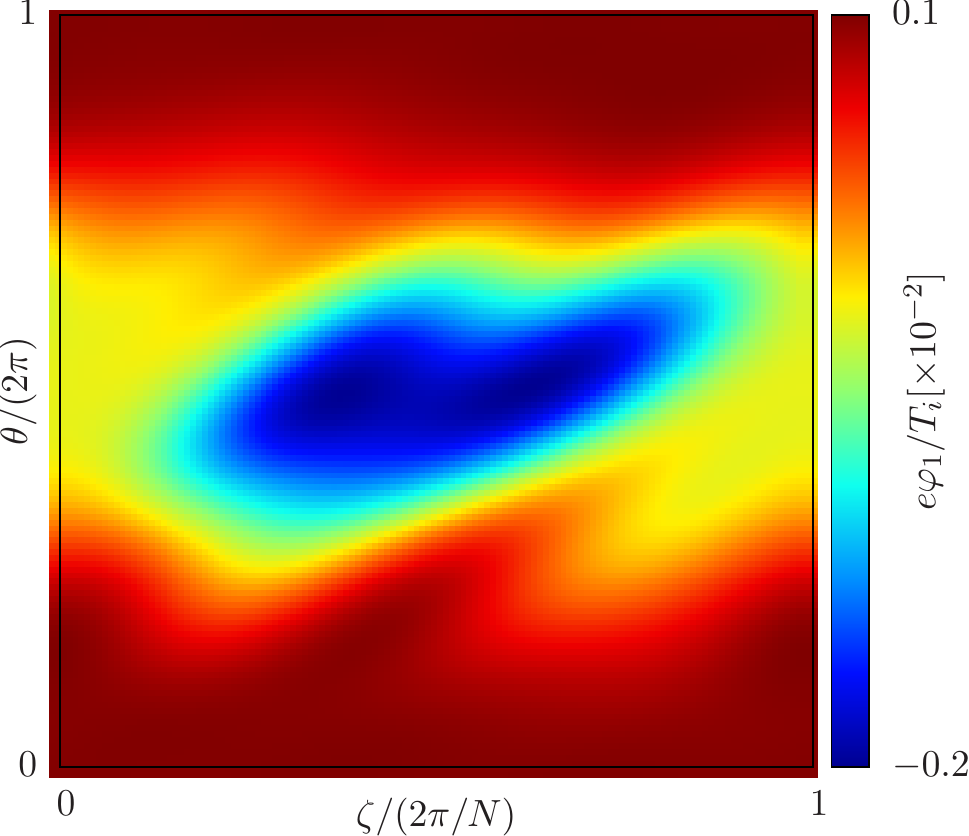}
\includegraphics[width=0.3\columnwidth,angle=0]{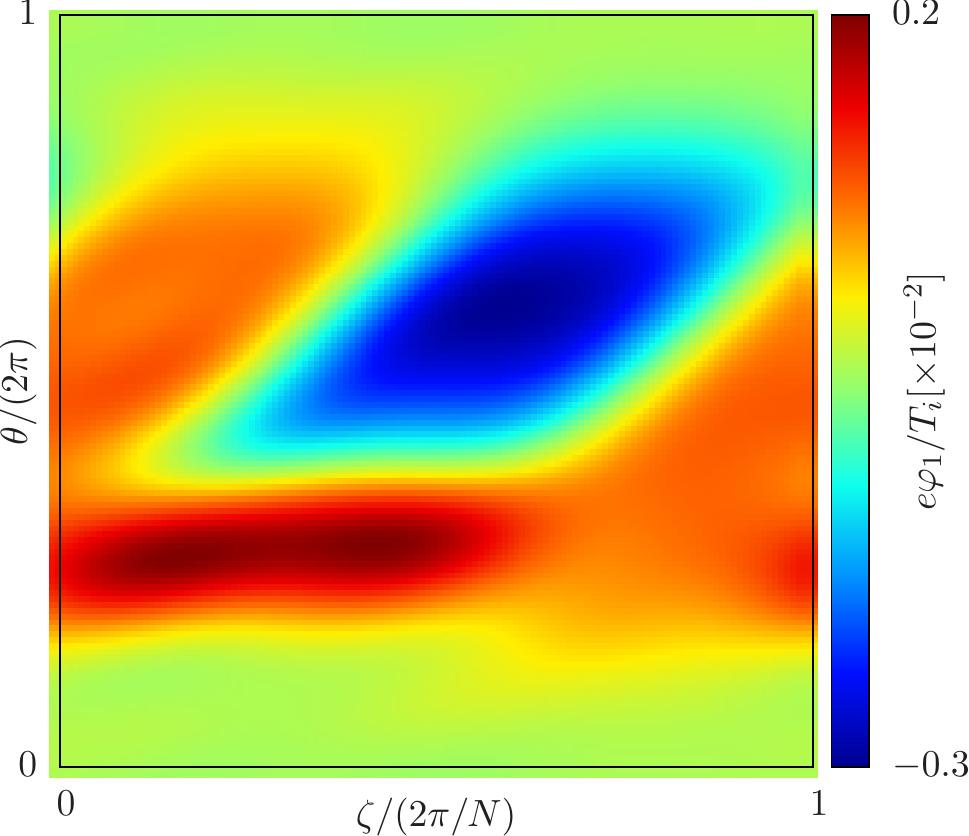}

\

\includegraphics[width=0.3\columnwidth,angle=0]{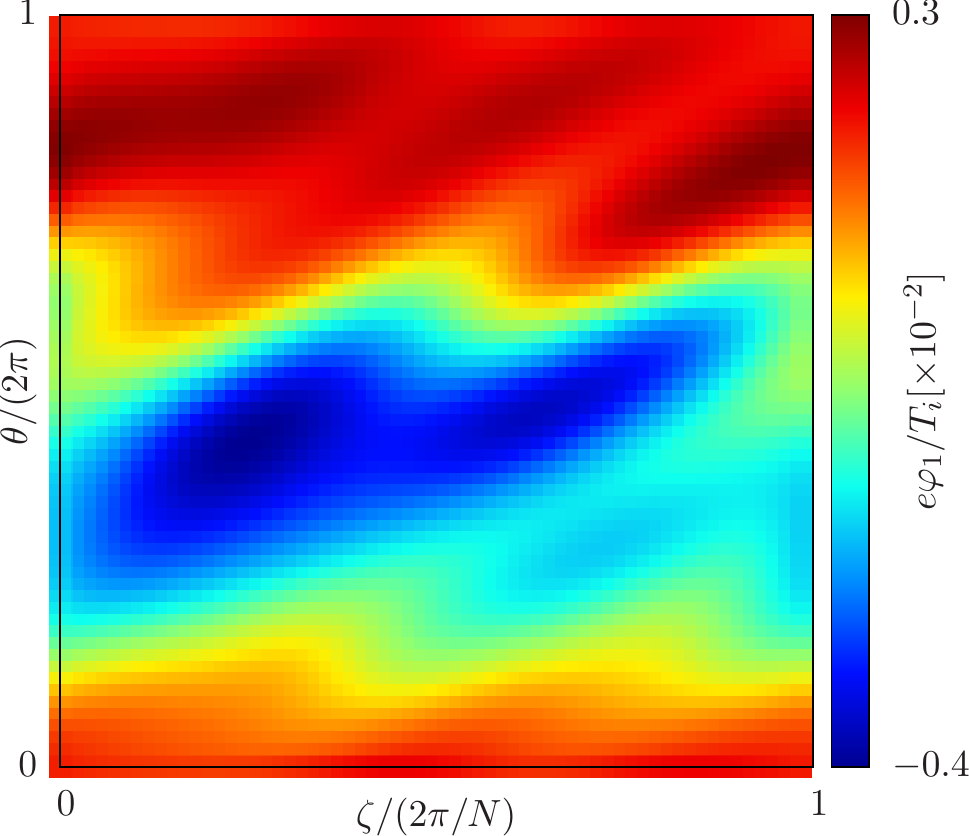}
\includegraphics[width=0.3\columnwidth,angle=0]{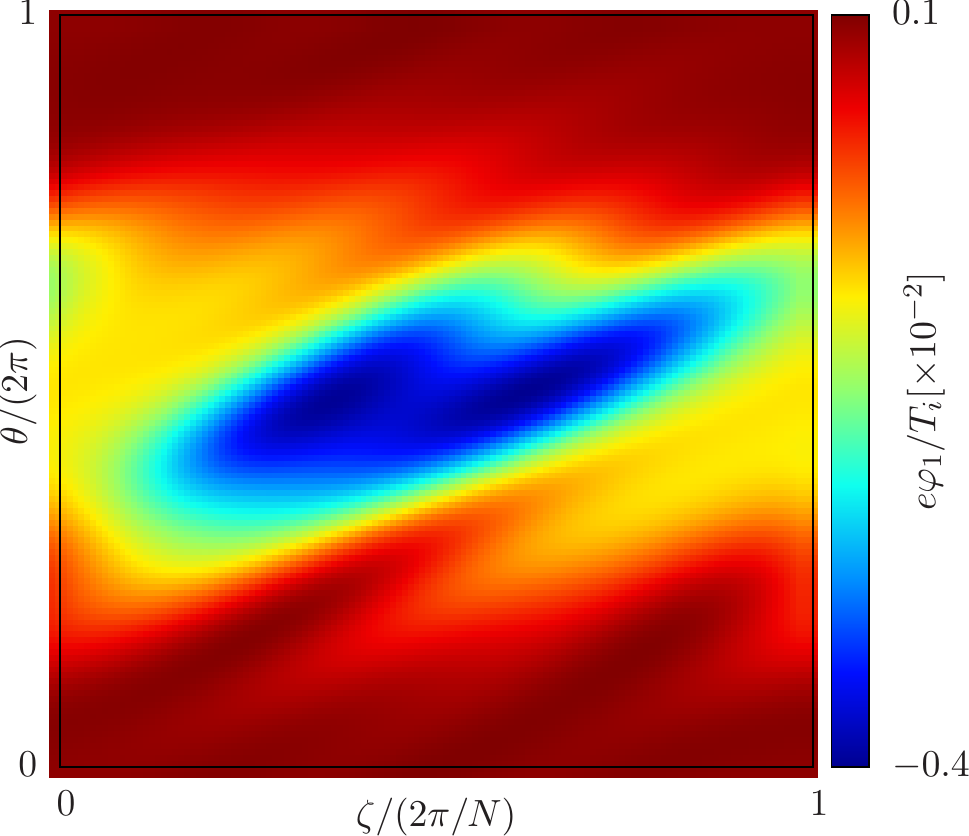}
\includegraphics[width=0.3\columnwidth,angle=0]{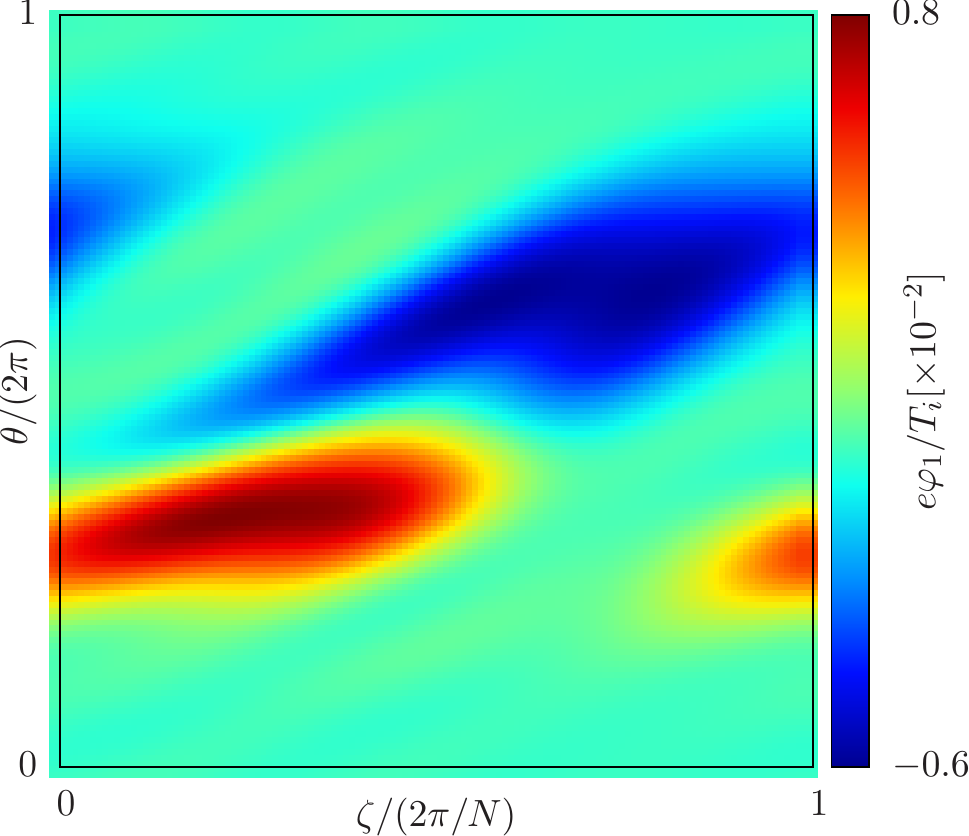}

\

\includegraphics[width=0.3\columnwidth,angle=0]{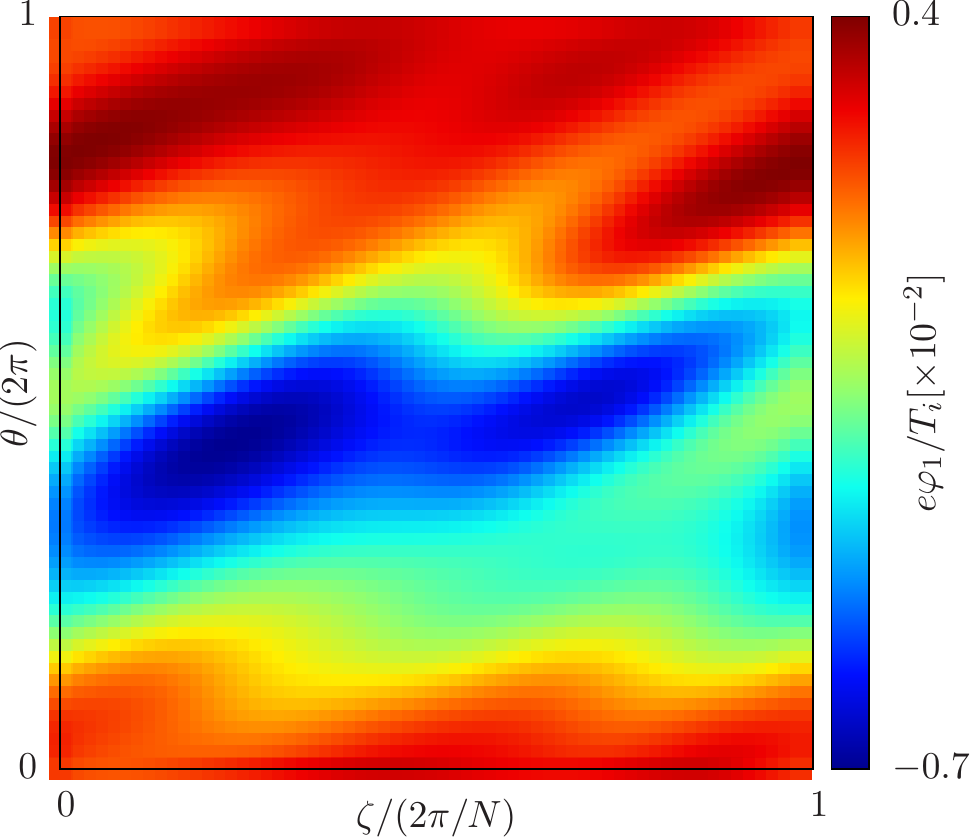}
\includegraphics[width=0.3\columnwidth,angle=0]{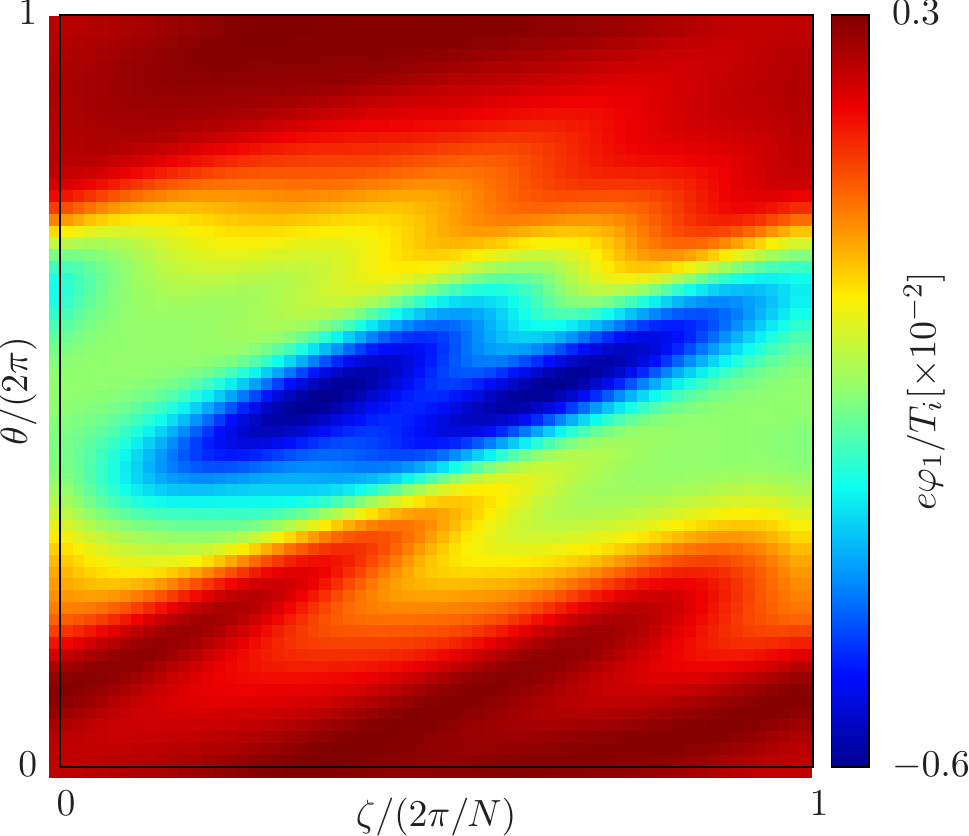}
\includegraphics[width=0.3\columnwidth,angle=0]{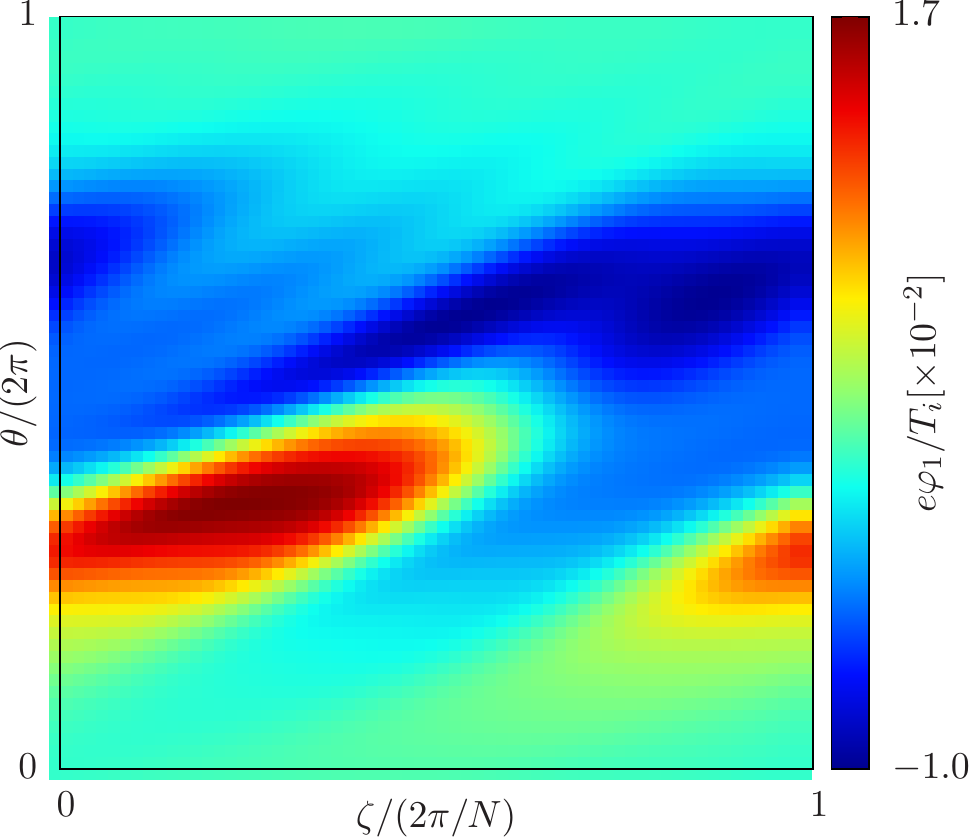}

\

\includegraphics[width=0.3\columnwidth,angle=0]{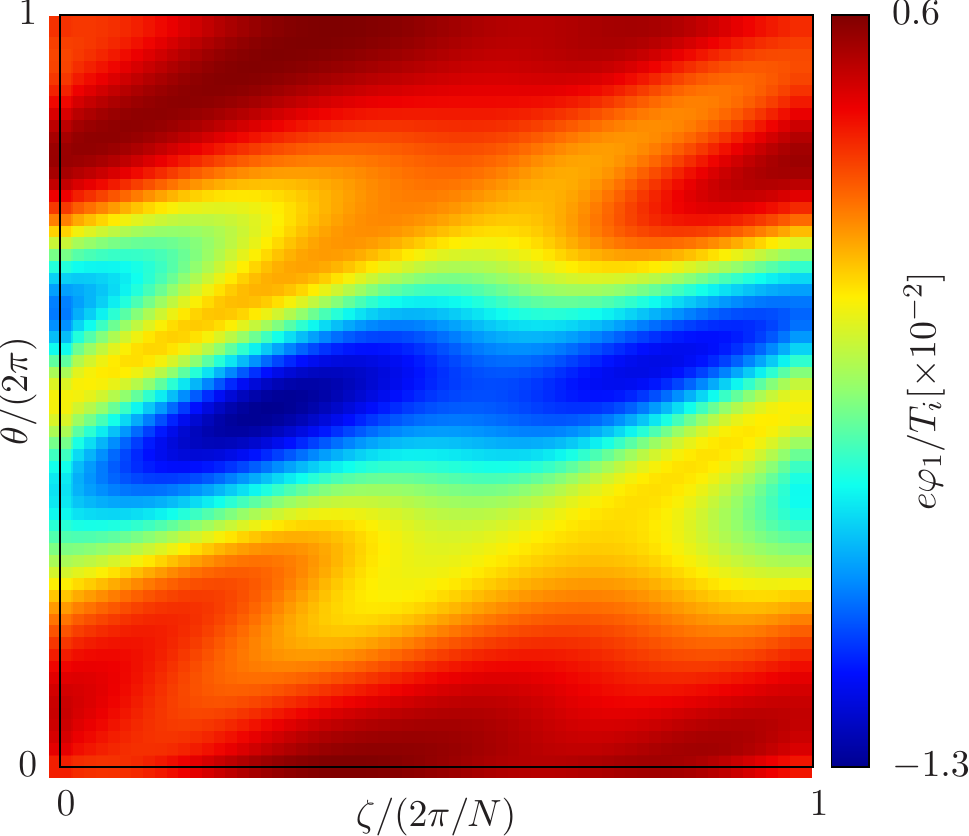}
\includegraphics[width=0.3\columnwidth,angle=0]{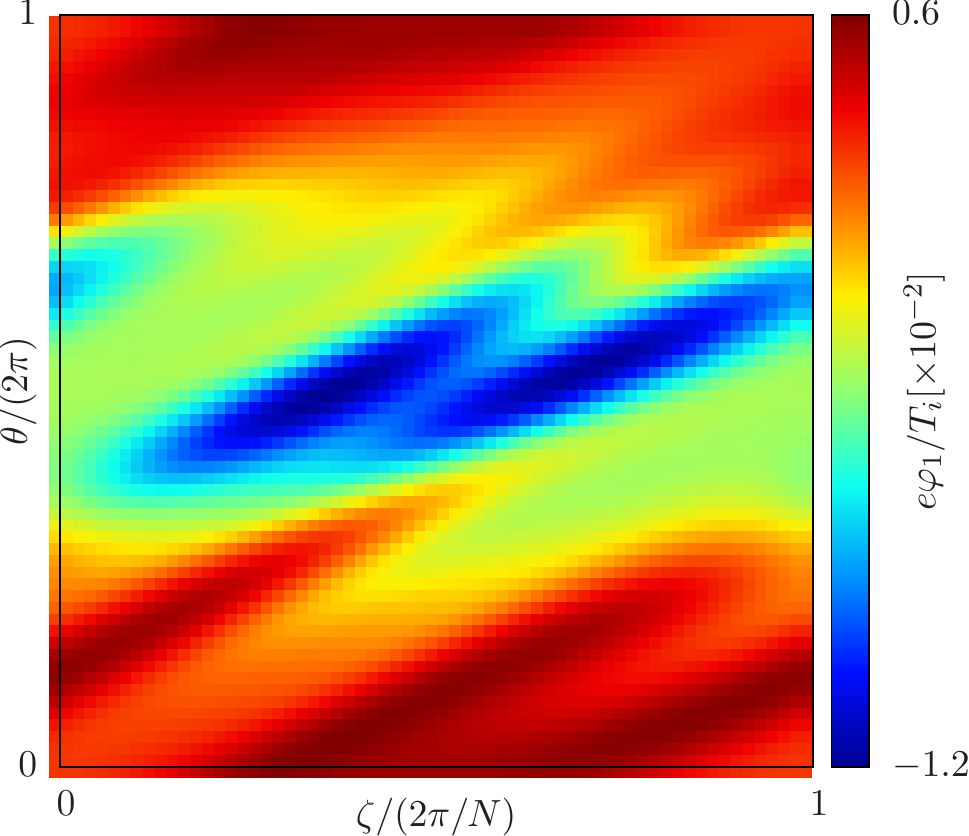}
\includegraphics[width=0.3\columnwidth,angle=0]{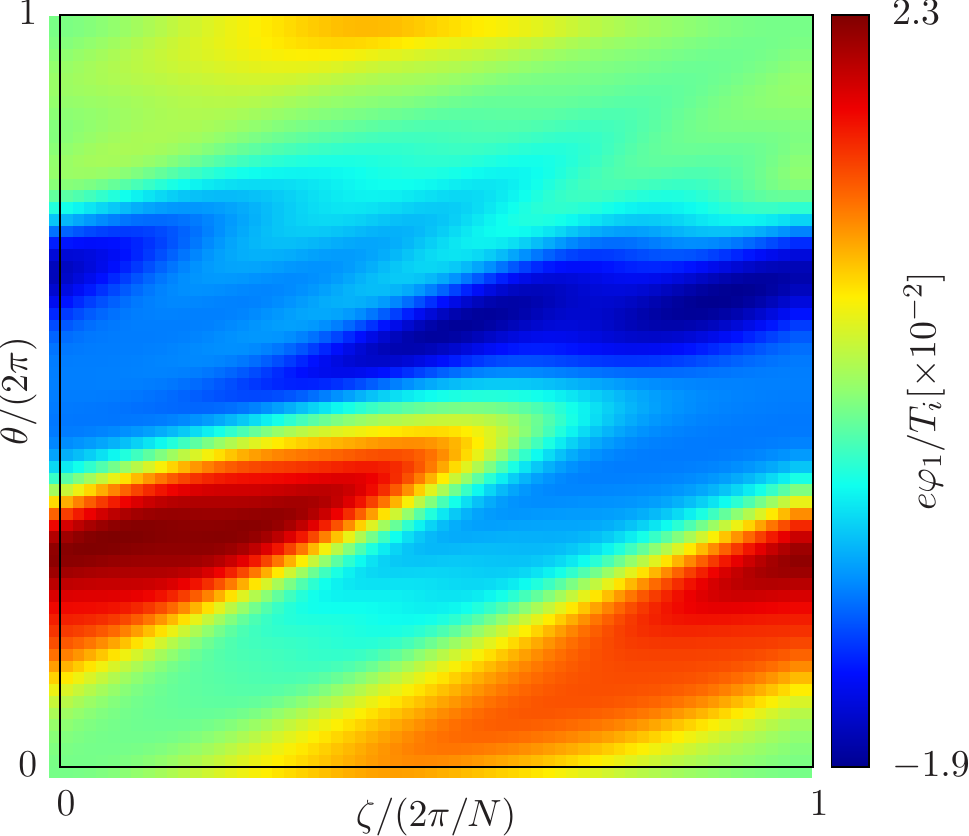}
\end{center}
\caption{Electrostatic potential variation on the flux surface calculated fort the LHD plasma with \texttt{EUTERPE} (left) and \texttt{KNOSOS} neglecting (center) and including (right) the tangential magnetic drift. The four rows correspond to radial positions $r/a\,=\,$0.2, 0.4, 0.6 and 0.8.}
\label{FIG_PHI1LHD}
\end{figure}

\begin{figure}
\begin{center}
\includegraphics[width=0.3\columnwidth,angle=0]{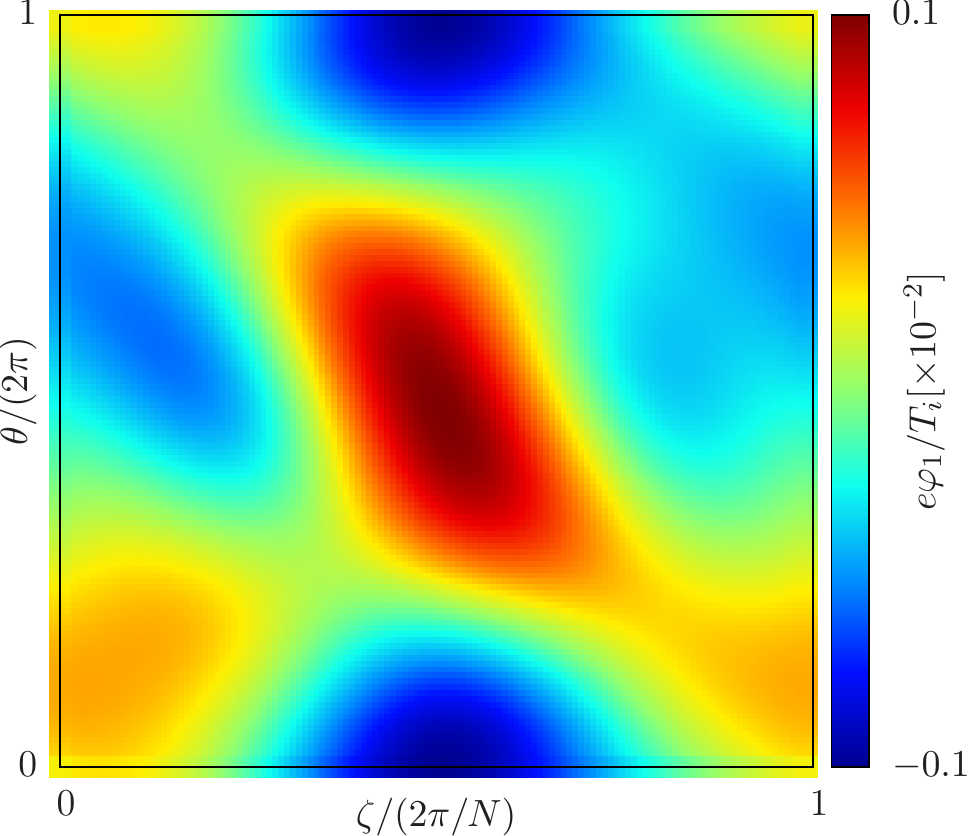}
\includegraphics[width=0.3\columnwidth,angle=0]{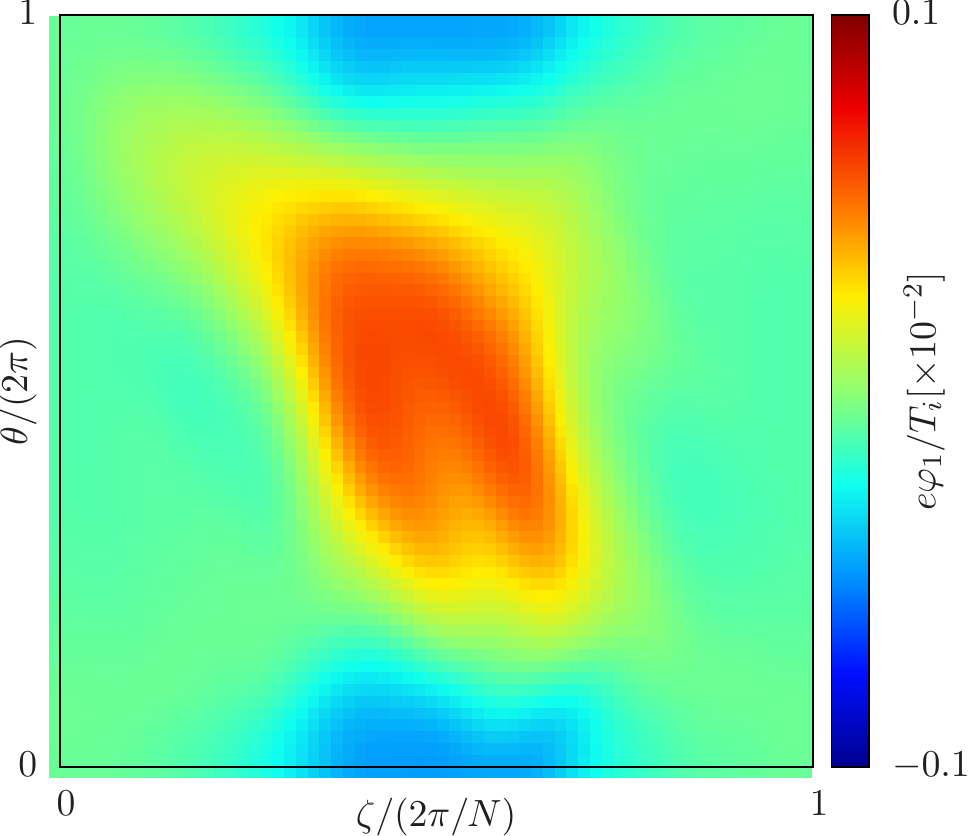}
\includegraphics[width=0.3\columnwidth,angle=0]{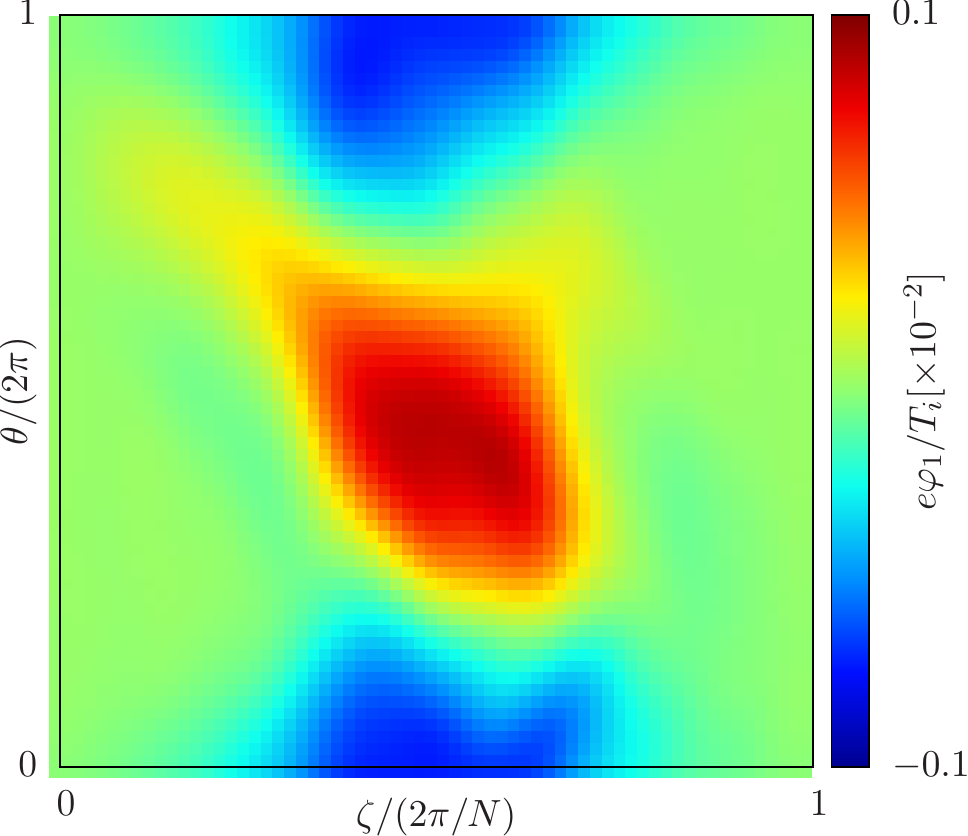}

\

\includegraphics[width=0.3\columnwidth,angle=0]{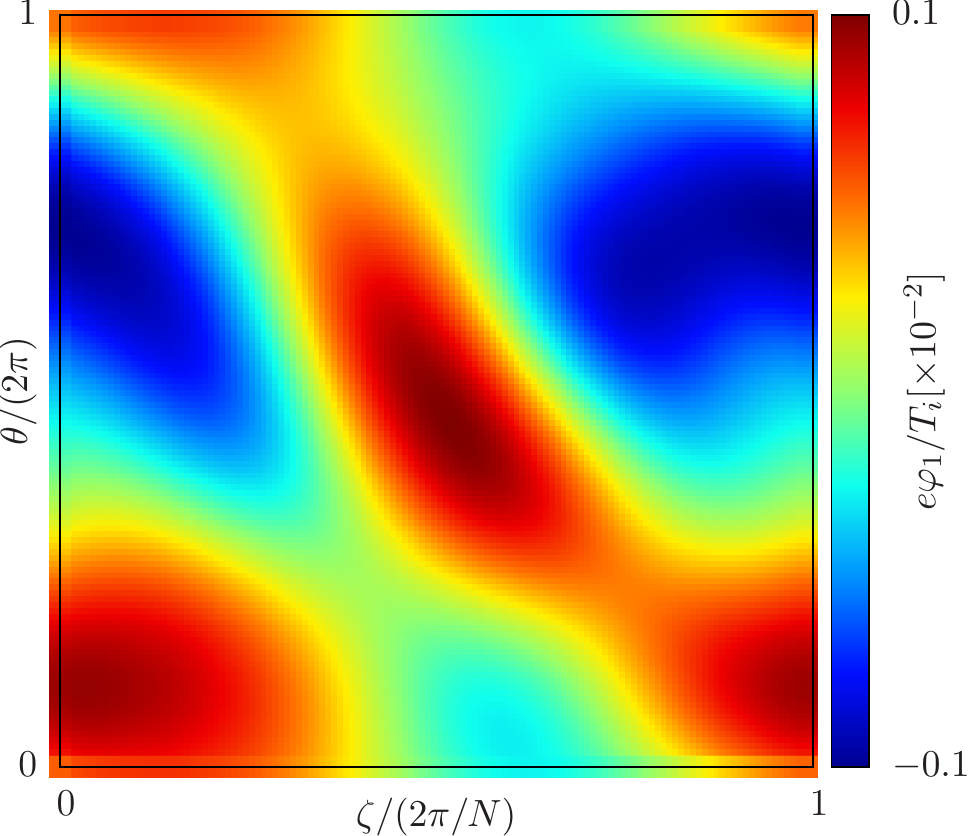}
\includegraphics[width=0.3\columnwidth,angle=0]{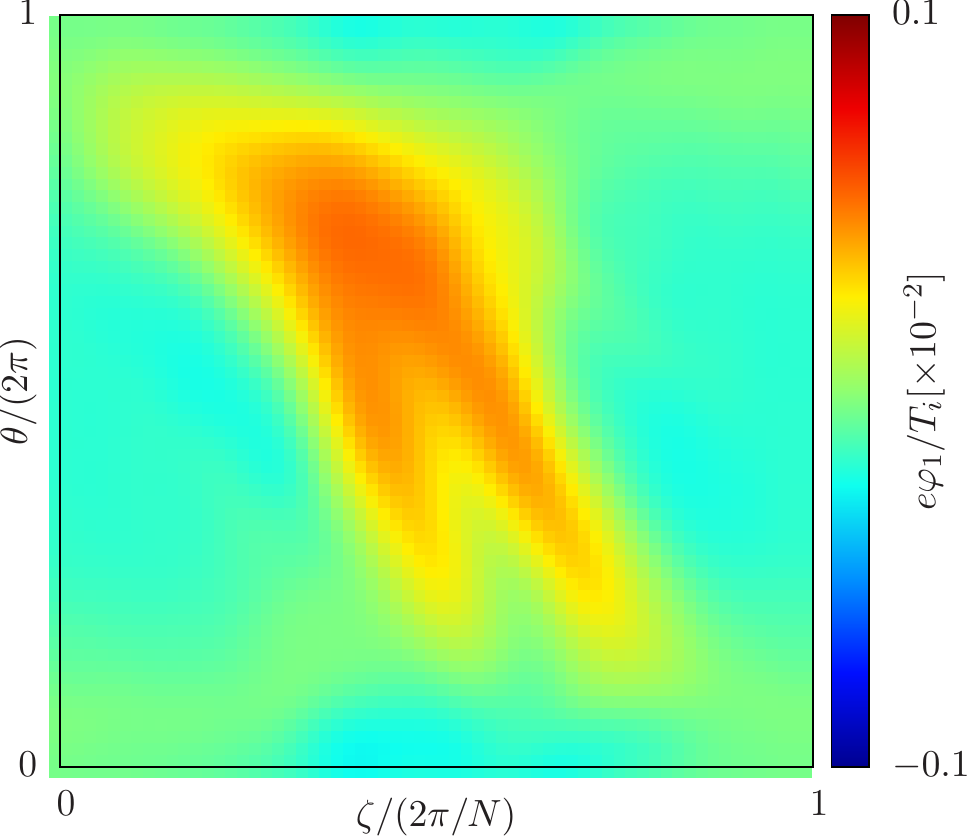}
\includegraphics[width=0.3\columnwidth,angle=0]{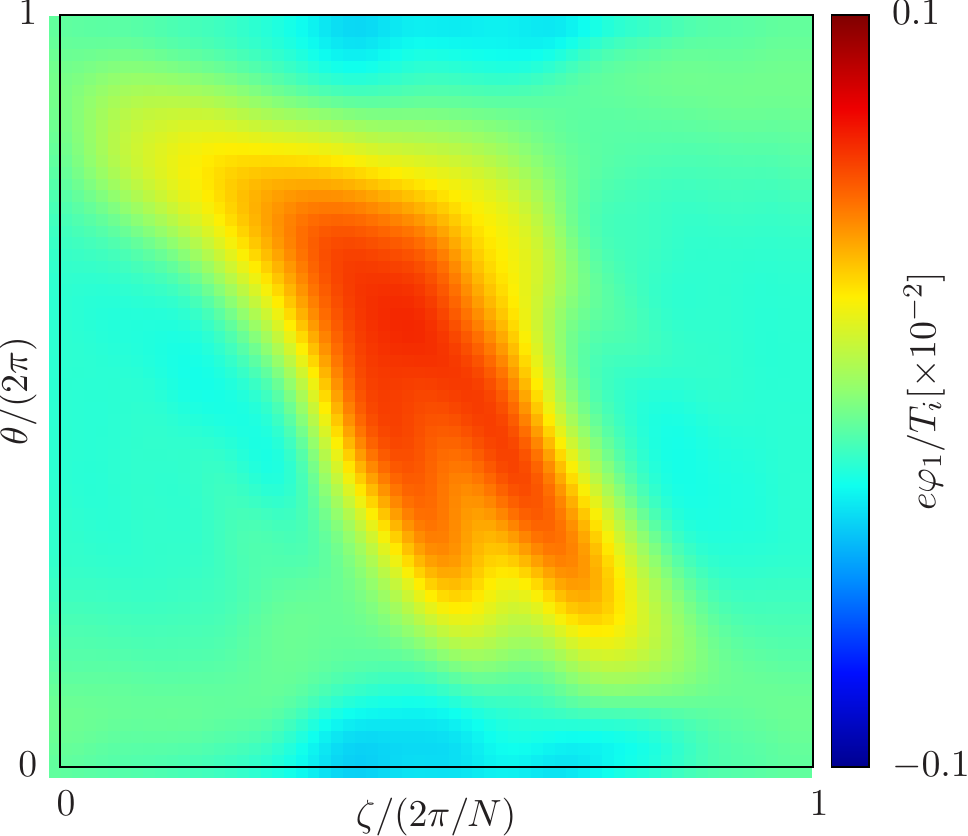}

\

\includegraphics[width=0.3\columnwidth,angle=0]{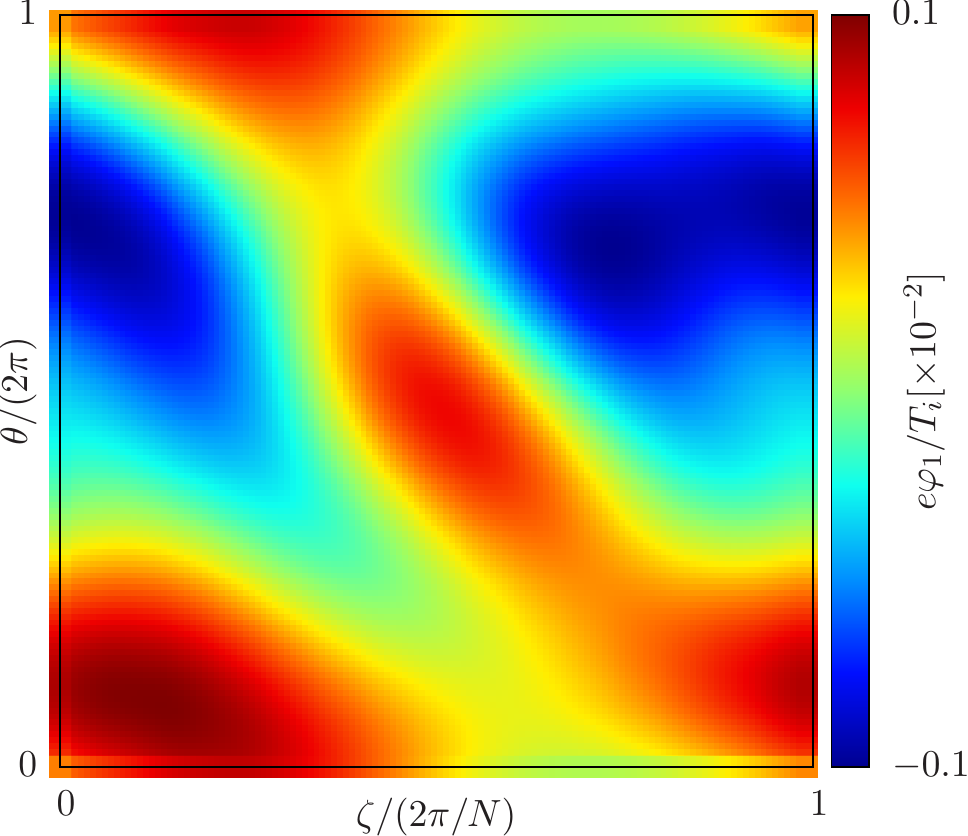}
\includegraphics[width=0.3\columnwidth,angle=0]{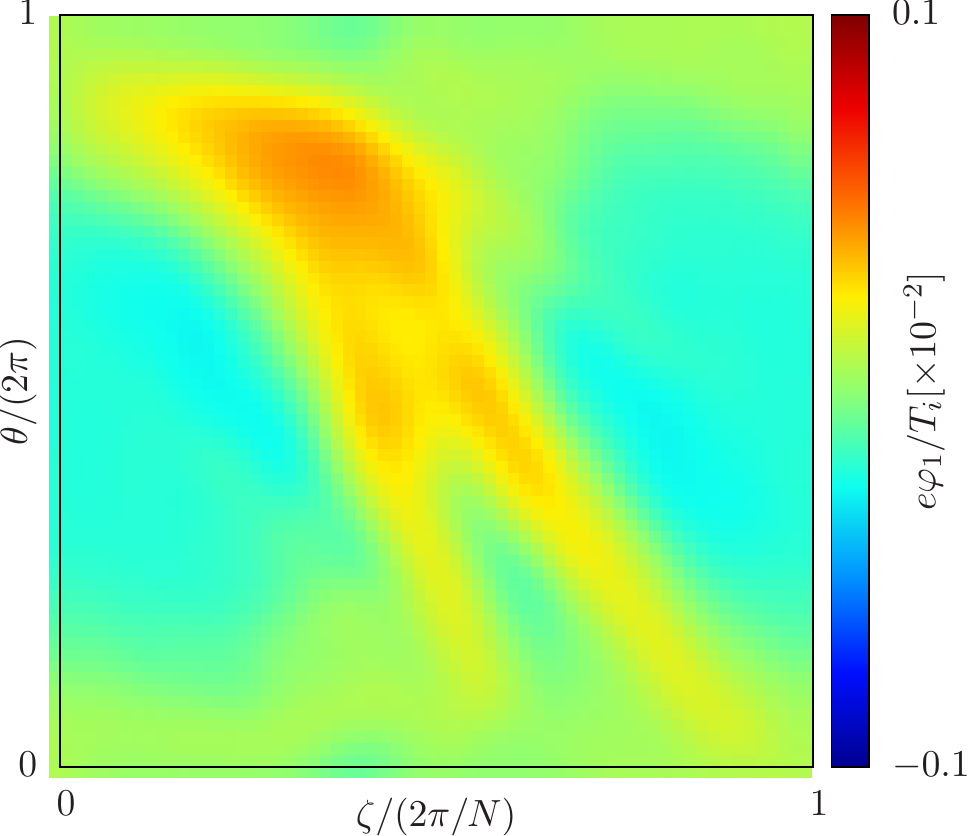}
\includegraphics[width=0.3\columnwidth,angle=0]{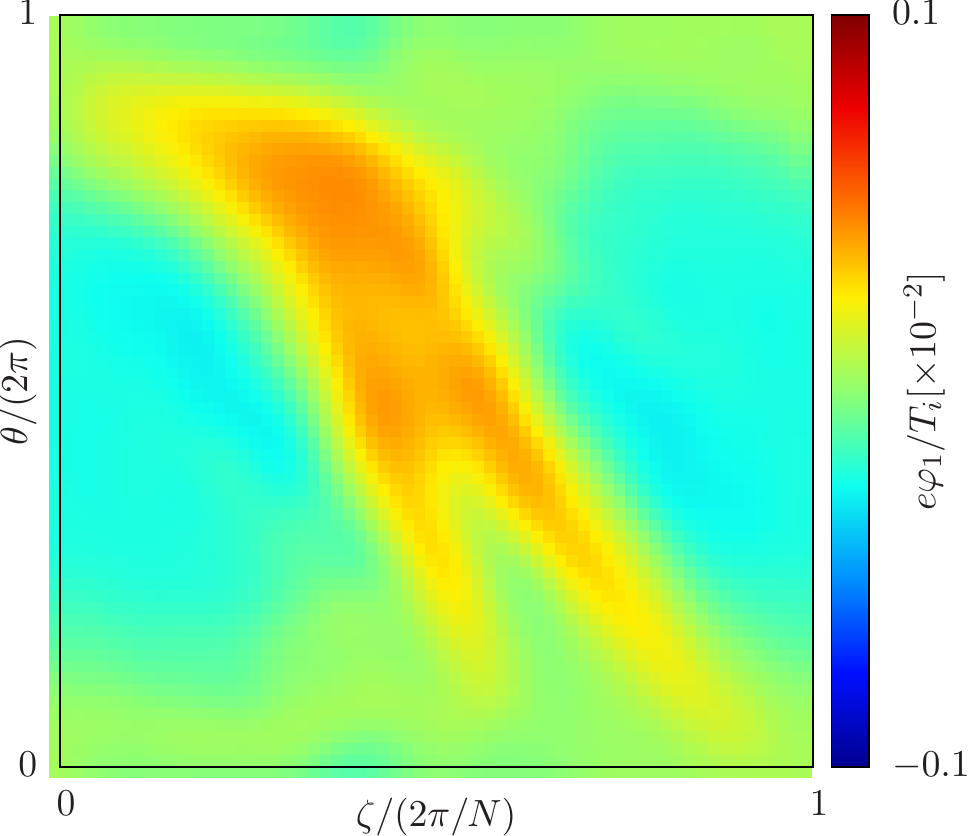}

\

\includegraphics[width=0.3\columnwidth,angle=0]{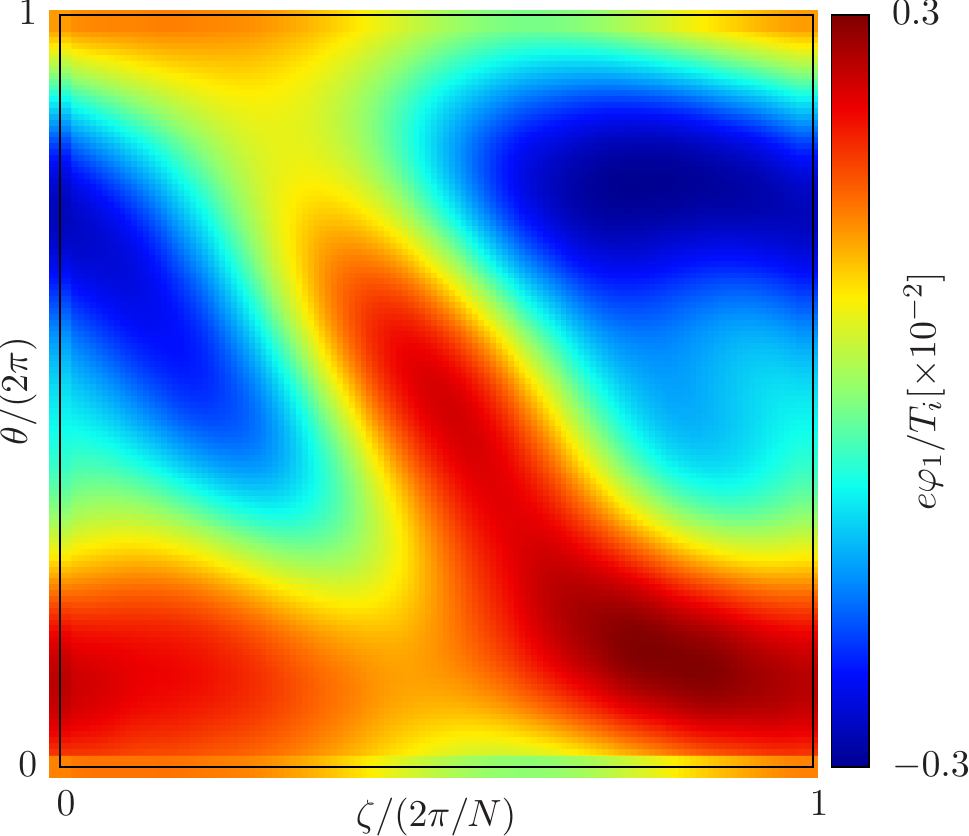}
\includegraphics[width=0.3\columnwidth,angle=0]{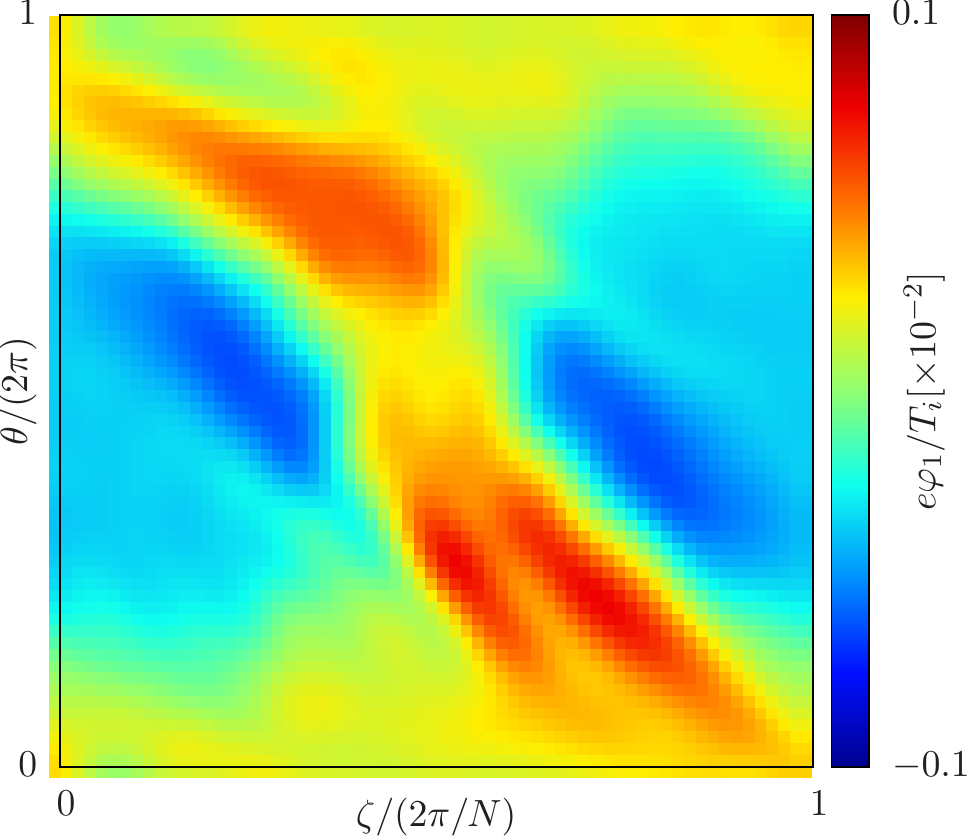}
\includegraphics[width=0.3\columnwidth,angle=0]{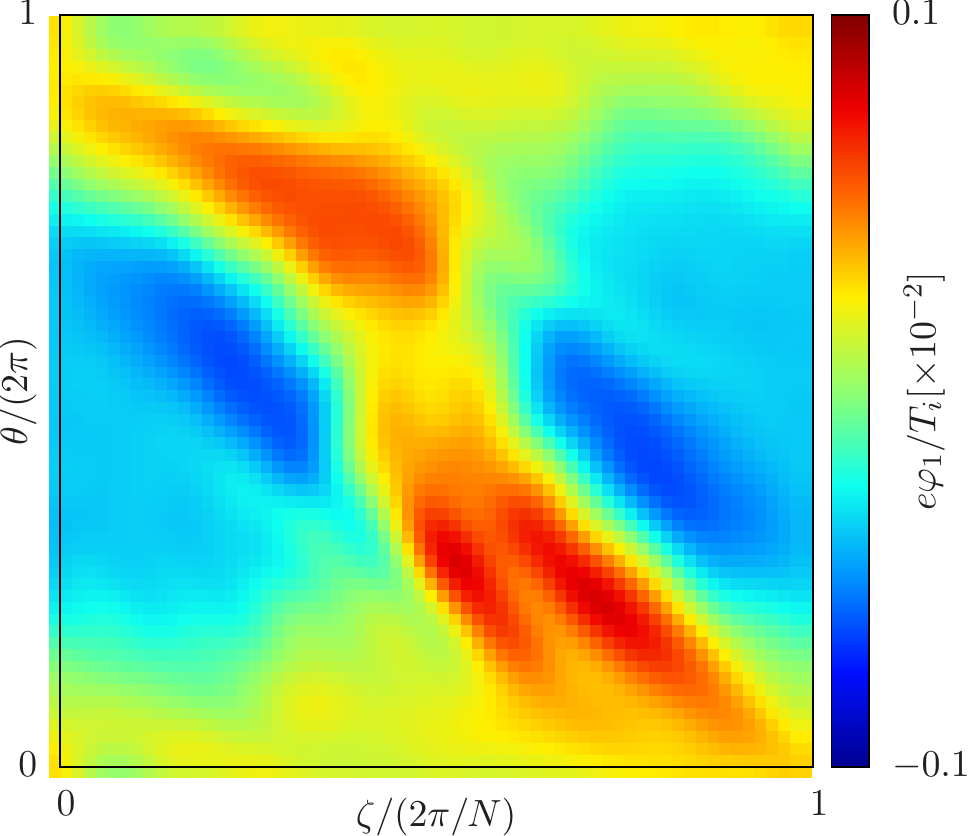}
\end{center}
\caption{Electrostatic potential variation on the flux surface calculated fort the W7X plasma with \texttt{EUTERPE} (left) and \texttt{KNOSOS} neglecting (center) and including (right) the tangential magnetic drift. The four rows correspond to radial positions $r/a\,=\,$0.2, 0.4, 0.6 and 0.8.}
\label{FIG_PHI1W7X}
\end{figure}

Neoclassical physics gives rise to $\varphi_1$, and the associated tangential electric field produces radial drifts in all species. This is the reason why we need to solve consistently the drift-kinetic equations of the bulk species and quasineutrality~\citep{calvo2017sqrtnu}, but the effect is more relevant for impurities, due to their larger charge number, changing even qualitatively transport~(e.g. making it depend on the radial electric field in the so-called mixed collisionality regime~\citep{calvo2018nf,buller2018jpp}). With impurity transport in mind, simulations of $\varphi_1$ for the stellarators W7-X, LHD and TJ-II have been performed in the last years with three codes,~\EUTERPE,~{\ttfamily SFINCS} and recently {\ttfamily FORTEC-3D}~\citep{regana2013euterpe,regana2017phi1,regana2018phi1,mollen2018phi1,fujita2019phi1}. Nevertheless, the number of simulations remains small because they are computationally very demanding, specially at low collisionalities. A more comprehensive study, including dependence on the configuration, collisionality, and bulk plasma profiles thus remains to be done. In this section, we will show that~\KNOSOS~can reproduce the results of~\EUTERPE~(with adiabatic electrons and no tangential magnetic drift) and, by accounting for the effect of the tangential magnetic drift, describe stellarator regimes only simulated before for simplified geometries~\citep{calvo2018jpp,velasco2018phi1}. Since it can do so while keeping the computing time low, this opens the door to a number of new impurity transport studies. 

We start by reproducing the results of~\citep{regana2017phi1}, specifically of two low-collisionality plasmas of LHD and W7-X. These are expected to be the plasma conditions of largest $e\varphi_1/T_i$ so that, even in optimized magnetic configurations, the effect on the radial transport of impurities may be large. It will be confirmed (as advanced in a previous work~\citep{velasco2018phi1} in a simplified calculation) that the inclusion of the tangential magnetic field leads to qualitative changes in $\varphi_1$, making it larger. Figure~\ref{FIG_PHI1LHD} shows the variation of the electrostatic potential on several flux surfaces of the inward-shifted configuration of LHD for a low-collisionality plasma (described in~\citep{regana2017phi1}), termed AIII, and characterized by a small negative $E_r$). Each row corresponds to a different flux surface, and each column to a different calculation method. Let us start by comparing the left column, calculated with~\EUTERPE, with the center column, calculated with~\KNOSOS~using equation (\ref{EQ_DKES}). The two methods should give the same results, and it can be observed that, although there are some differences \reft{(note the slightly different color scale)}, reasonable agreement between the two codes is obtained. It should be emphasized that differences in calculated values of $\varphi_1$ similar but smaller to those reported here, have been shown to produce negligible differences in impurity transport~\citep{velasco2018phi1}. If we now focus on the right column, we observe, as discussed in detail in~\citep{velasco2018phi1}, that the inclusion of the tangential magnetic drift produces relevant differences (in particular, more important than those between the left and center columns): the amplitude becomes larger, and the phase changes, with the angular dependence of $\varphi_1$ turning from being stellarator-symmetric (as expected for ions in the $\sqrt{\nu}$ regime), to not having definite symmetry (as corresponds to the superbanana-plateau regime~\citep{calvo2018jpp}).

Figure~\ref{FIG_PHI1W7X} contains a similar calculation performed for a low-collisionality plasma of W7-X (described in~\citep{regana2017phi1}, termed IV, and characterized by a larger negative $E_r$). Again, each row corresponds to a different flux surface, and each column to a different calculation method. The \reft{agreement} between~\EUTERPE~and~\KNOSOS~solving the same equation (left and center) is \reft{fair close to the core, since both show a similar angular dependence and a slightly different amplitude, but it becomes worse closer to the edge, where~\KNOSOS~clearly underestimates} the amplitude of $\varphi_1$. When the tangential magnetic drift is included (right), the results change very slightly in the core and do not change elsewhere. This feature is likely caused by the large radial electric field, which leaves the ions in the $\sqrt{\nu}$ regime (instead of the superbanana-plateau). \reft{As in subsection~\ref{SEC_TANGVM}, the large radial electric field, together with the lower level of $\varphi_1$ characteristic of optimized stellarators~\citep{calvo2018jpp,regana2017phi1}, may be behind the disagreement between~\EUTERPE~and~\KNOSOS, since the former includes the contribution of the plateau regime.}

The computing time for each of these \KNOSOS~simulations is of the order of a minute in a single processor. We note that including kinetic electrons (which may be necessary for high electron temperature) would roughly double the computing time. This is to be compared with the $(m_i/m_e)^{1/2} \approx 43$ factor in Monte Carlo codes such as~\EUTERPE~and {\ttfamily FORTEC-3D}.

Let us finally mention that the experimental validation of $\varphi_1$ predictions has drawn much attention in the last years: experimental measurements of $\varphi_1$ were first obtained at the edge of the TJ-II stellarator~\cite{pedrosa2015phi1}, and very recently in its core region~\citep{estrada2019phi1}. The validation of~\KNOSOS~predictions, including finer scans in the magnetic configuration, is left for a forthcoming work.


\section{Conclusions}\label{SEC_CONCLUSIONS}


{\ttfamily KNOSOS} is a freely-available open-source code that provides a fast computation of neoclassical transport at low collisionality in three-dimensional magnetic confinement devices, thanks to a rigorous application of the orbit-averaging technique to the \jlvg{drift-kinetic equation and an efficient solution of the quasineutrality equation}. We have shown that, when solving equivalent equations, {\ttfamily KNOSOS} reproduces the calculations of~\DKES~and {\ttfamily EUTERPE} in simulations that can be orders of magnitude faster. This makes it a tool that can be used for a variety of physics problems, that we summarize next.

As a first obvious application, it can provide a fast calculation of the level of transport of a magnetic configuration for low-collisionality transport regimes not usually considered in stellarator optimization, such as the $\sqrt{\nu}$ and superbanana-plateau regimes. Optimization \jlvg{programmes} are slowly starting to provide a more accurate characterization of transport by performing predictive simulations with prescribed sources and turbulent transport models. {\ttfamily KNOSOS} can also contribute to overcome two of the main limitations of this approach: the large computing time needed to create a database of monoenergetic neoclassical transport coefficients and/or the lack of accuracy involved in the monoenergetic approach itself.
 
But a fast neoclassical code can have uses beyond stellarator optimization. For instance, the transport of impurities caused by their interaction with the bulk ions (via $\varphi_1$ or through inter-species collisions) has drawn much attention in the last years; however, a systematic study of its dependence on the magnetic configuration, collisionality, and bulk plasma profiles remains to be done, due to the large computing resources needed for the combined solution of the quasineutrality and drift-kinetic equations of the bulk species. This will be addressed in forthcoming papers, in combination with analytical formulas for the radial flux of impurities in a variety of neoclassical regimes~\citep{calvo2018nf,calvo2019anis}.

Finally even in situations in which turbulence is dominant, a fast neoclassical code may be required. Its output (the radial electric field, the tangential electric field or the complete distribution function of the bulk species) can be read by gyrokinetic codes when studying the effect of neoclassical transport on turbulence. This effect is expected to be largest in those low-collisionality regimes in which the specificities of {\ttfamily KNOSOS} (very small computing time and inclusion of the tangential magnetic drift) are most relevant.


\section{Acknowledgments}



 
This work has been carried out within the framework of the EUROfusion Consortium and has received funding from the Euratom research and training programme 2014-2018 and 2019-2020 under grant agreement No. 633053. The views and opinions expressed herein do not necessarily reflect those of the European Commission. This research was supported in part by grant ENE2015-70142-P, Ministerio de Econom\'ia y Competitividad, Spain, by grant PGC2018-095307-B-I00, Ministerio de Ciencia, Innovaci\'on y Universidades, Spain, and by the Y2018/NMT [PROMETEO-CM] project of the Comunidad de Madrid, Spain.

\appendix


\section{Details of the collision operator}\label{AP0}


In \S\ref{SEC_EQUATIONS}, the pitch-angle-scattering collision operator has been employed, and its explicit expression has been provided in equation (\ref{EQ_COLOP}). As it has been discussed, this is a single-species collision operator, which is accurate for calculating ion transport, due to $\sqrt{m_e/m_i}\ll 1$. For electrons, however, electron-ion collisions need to be retained in the electron drift-kinetic equation. In order to overcome this limitation, an effective pitch-angle-scattering collision frequency $\nu_{\lambda,b}$ is employed in order to account for inter-species collisions. This is done for both species, although its effect will be negligible for the ions.

In this appendix, we provide the explicit expression of the pitch angle scattering frequency, given by the sum\footnote{We note that $\nu_{\lambda,b}=2\nu_b$, with the definition of $\nu_b$ of page 3 of~\cite{beidler2011icnts}.}
\begin{equation}
\nu_{\lambda,b}=\sum_{b'}\nu_0^{b/b'}\left[\mbox{erf} \left(\sqrt{m_{b'} v^2 /(2T_{b'})}\right) - \chi \left(\sqrt{m_{b'} v^2 /(2T_{b'})} \right)\right]\,,
\end{equation}
with
\begin{equation}
\nu_0^{b/b'}= \frac{8\pi n_{b'} {Z_b}^2{Z_{b'}}^2 e^4 \ln \Lambda^{b/b'}}{m_b^2 v^3} \,.
\end{equation}
Here, $\ln \Lambda^{b/b'}$ is the Coulomb logarithm, 
\begin{equation}
\chi (x) = \frac{\mbox{erf} (x) - (2 x/\sqrt{\pi})\exp(-x^2)}{2x^2}\,,
\end{equation}
and
\begin{equation}
\mbox{erf}(x) = \frac{2}{\sqrt{\pi}} \int_0^x \exp (- t^2)\, \mathrm{d} t
\end{equation}
is the error function.
 
 
\section{Analytical calculation of the divergences of equations~(\ref{EQ_BINT})}\label{AP1}


In this section, we discuss how integrals such as those in equations~(\ref{EQ_BINT}),
\begin{equation} 
I(\lambda)=\int_{l_{b_1}}^{l_{b_2}}\mathrm{d}l\,\frac{f(\lambda,l)}{\sqrt{1-\lambda B(l)}}\,,
\end{equation}
can be computed efficiently by removing the component that diverges close to bifurcations and solving it analytically. Since integration is done at fixed $\alpha$, we ease the notation by not making it explicit that $B$, $f$, and $I$ generally depend on the angular coordinate.

We first expand the magnetic field around the bounce point:
\begin{equation} 
B(l)=B(l_{b_1})+\partial_l B|_{l_{b_1}}(l-l_{b_1})+\frac{1}{2}\partial^2_l B|_{l_{b_1}}(l-l_{b_1})^2\,.
\end{equation}
Close to the bounce point, we have
\begin{align}
\frac{f(l)}{\sqrt{1-\lambda B(l)}} \approx \frac{f(l_{b_1})}{\sqrt{-\lambda (l-l_{b_1}) [\partial_l B|_{l_{b_1}}+\frac{1}{2}\partial^2_l B|_{l_{b_1}}(l-l_{b_1})]}}\,
\end{align}
since $\lambda B(l_{b_1})=1$. We can proceed exactly in the same way close to $l_{b_2}$, and similarly close to $\lambda_B$: there, $\lambda B(l_B)<1$ and the first derivative $\partial_l B|_{l_B}$ is zero, and we have
\begin{align}
\frac{f(l)}{\sqrt{1-\lambda B(l)}} \approx \frac{f(l_B)}{\sqrt{-(\lambda-\lambda_B) B(l_B)-\lambda_B\frac{1}{2}\partial^2_l B|_{l_B}(l-l_B)^2}}\,.
\end{align}
We can then split the integral in three contributions:
\begin{equation} 
I= I_0+I_1+I_2+I_B\,,
\end{equation}
with
\begin{align}
I_0 &= \int_{l_{b_1}}^{l_{b_2}}\mathrm{d}l\,\left(\frac{f(l)}{\sqrt{1-\lambda B(l)}}-\frac{f(l_{b_1})}{\sqrt{-\lambda (l-l_{b_1}) [\partial_l B|_{l_{b_1}}+\frac{1}{2}\partial^2_l B|_{l_{b_1}}(l-l_{b_1})]}}\right.\nonumber\\
 &-\left.\frac{f(l_{b_2})}{\sqrt{-\lambda (l-l_{b_2}) [\partial_l B|_{l_{b_2}}+\frac{1}{2}\partial^2_l B|_{l_{b_2}}(l-l_{b_2})]}}-\frac{f(l_B)}{\sqrt{-(\lambda-\lambda_B) B(l_B)-\lambda_B\frac{1}{2}\partial^2_l B|_{l_B}(l-l_B)^2}}\right)\,,
\end{align}
whose integrand does not diverge anywhere and
\begin{align} 
I_1 &= \int_{l_{b_1}}^{l_{b_2}}\mathrm{d}l\,\frac{\refm{f(l_{b_1})}}{\sqrt{-\lambda (l-l_{b_1}) [\partial_l B|_{l_{b_1}}+\frac{1}{2}\partial^2_l B|_{l_{b_1}}(l-l_{b_1})]}}\,,\nonumber\\
I_2 &= \int_{l_{b_1}}^{l_{b_2}}\mathrm{d}l\,\frac{\refm{f(l_{b_2})}}{\sqrt{-\lambda (l-l_{b_2}) [\partial_l B|_{l_{b_2}}+\frac{1}{2}\partial^2_l B|_{l_{b_2}}(l-l_{b_2})]}}\,,\nonumber\\
I_B &= \int_{l_{b_1}}^{l_{b_2}}\mathrm{d}l\,\frac{f(l_B)}{\sqrt{-(\lambda-\lambda_B) B(l_B)-\lambda_B\frac{1}{2}\partial^2_l B|_{l_B}(l-l_B)^2}}\,,
\end{align}
which can be solved analytically. The integral close to the bottom is
\begin{align} 
I_B = \refm{f(l_{B})} \sqrt{\frac{-2}{\lambda_B\partial^2_l B|_{l_B}}} \left[\mathrm{ln}\left(x+\sqrt{x^2+\frac{2(\lambda-\lambda_B) B(l_B)}{\lambda_B\partial^2_l B|_{l_B}}}\right)\right]^{l_B-l_{b_1}}_0+ \sqrt{\frac{-2}{\lambda_B\partial^2_l B|_{l_B}}} \left[\mathrm{ln}\left(x+\sqrt{x^2+\frac{2(\lambda-\lambda_B) B(l_B)}{\lambda_B\partial^2_l B|_{l_B}}}\right)\right]_0^{l_{b_2}-l_B}\,.
\end{align}

For the other two integrals, if $\partial^2_l B|_{l_{b_1}}<0$ and $\partial^2_l B|_{l_{b_2}}<0$, the solution is
\begin{align} 
I_1 &= \refm{f(l_{b_1})} \sqrt{\frac{-2}{\lambda\partial^2_l B|_{l_{b_1}}}} \left[\mathrm{ln}\left(2\lambda\sqrt{\frac{\partial^2_l B|_{l_{b_1}}}{-2}}\sqrt{ -\partial_l B|_{l_{b_1}}x -\frac{1}{2}\partial^2_l B|_{l_{b_1}}x^2} -\lambda \partial^2_l B|_{l_{b_1}} x -\lambda\partial_l B|_{l_{b_1}}\right)\right]_0^{l_{b_2}-l_{b_1}}\,,\nonumber\\
I_2 &= \refm{f(l_{b_2})} \sqrt{\frac{-2}{\lambda\partial^2_l B|_{l_{b_2}}}} \left[\mathrm{ln}\left(2\lambda\sqrt{\frac{\partial^2_l B|_{l_{b_2}}}{-2}}\sqrt{ -\partial_l B|_{l_{b_2}}x -\frac{1}{2}\partial^2_l B|_{l_{b_2}}x^2} -\lambda \partial^2_l B|_{l_{b_2}} x -\lambda\partial_l B|_{l_{b_2}}\right)\right]_{l_{b_1}-l_{b_2}}^0\hskip-0.5cm.
\end{align} 
These expressions are useful (in the sense of removing large analytical contributions to $I$) close enough to a bifurcation, where they can significantly accelerate the convergence of equations~(\ref{EQ_BINT}), but they are in principle valid for any $\lambda$ (far from bifurcations, when $\partial^2_l B|_{l_{b_1}}$ is positive, the expression within the square-root may become negative and cannot be used).



\section{Evaluation of the magnetic field strength along a field line}\label{AP2}


The fact that field lines are straight in magnetic coordinates can also be used to speed up the calculation of the coefficients of the drift-kinetic equation. We describe how in this appendix. 

The bounce-integrals are done, using the algorithm mentioned in \S\ref{SEC_COEFFICIENTS}, by following field lines using a fixed step in the Boozer angles given by $\Delta\zeta$ and $\Delta\theta=\iota\Delta\zeta$. After each step, the magnetic field can be calculated without loss of accuracy from its Fourier components
\begin{align}
B(\theta,\zeta)&=\sum_{m,n} B_{m,n}\cos[m\theta+nN\zeta]\,,\nonumber\\
B(\theta+\Delta\theta,\zeta+\Delta\zeta)&=\sum_{m,n} B_{m,n}\cos[m(\theta+\Delta\theta)+nN(\zeta+\Delta\zeta)]\,,\nonumber\\
B(\theta+2\Delta\theta,\zeta+2\Delta\zeta)&=\sum_{m,n} B_{m,n}\cos[m(\theta+2\Delta\theta)+nN(\zeta+2\Delta\zeta)]\,,\nonumber\\
...
\end{align}
Instead of calculating the cosines at every angular position, we can precalculate a few sines and cosines, $\cos(m\theta+nN\zeta)$, $\sin(m\theta+nN\zeta)$, $\cos(m\Delta\theta+nN\Delta\zeta)$ and $\sin(m\Delta\theta+nN\Delta\zeta)$, and use well-known trigonometric identities to iterate:
\begin{align}
\cos[m(\theta+\Delta\theta)+nN(\zeta+\Delta\zeta)] & = \cos(m\theta+nN\zeta)\cos(m\Delta\theta+nN\Delta\zeta)\nonumber\\
&- sin(m\theta+nN\zeta)\sin(m\Delta\theta+nN\Delta\zeta)\,,\nonumber\\
\sin[m(\theta+\Delta\theta)+nN(\zeta+\Delta\zeta)] & = \cos(m\theta+nN\zeta)\sin(m\Delta\theta+nN\Delta\zeta)\nonumber\\
&+ \sin(m\theta+nN\zeta)\cos(m\Delta\theta+nN\Delta\zeta)\,,
\end{align}
and
\begin{align}
\cos[m(\theta+2\Delta\theta)+nN(\zeta+2\Delta\zeta)] & = \cos[m(\theta+\Delta\theta)+nN(\zeta+\Delta\zeta)]\cos(m\Delta\theta+nN\Delta\zeta)\nonumber\\
&- sin[m(\theta+\Delta\theta)+nN(\zeta+\Delta\zeta)]\sin(m\Delta\theta+nN\Delta\zeta)\,,\nonumber\\
\sin[m(\theta+2\Delta\theta)+nN(\zeta+2\Delta\zeta)] & = \cos[m(\theta+\Delta\theta)+nN(\zeta+\Delta\zeta)]\sin(m\Delta\theta+nN\Delta\zeta)\nonumber\\
&+ \sin[m(\theta+\Delta\theta)+nN(\zeta+\Delta\zeta)]\cos(m\Delta\theta+nN\Delta\zeta)\,\nonumber\\
\end{align}
and so on.



\bibliographystyle{elsarticle-num}

\bibliography{/Users/velasco/Work/PAPERS/bibliography}

\begin{thebibliography}{10}
\expandafter\ifx\csname url\endcsname\relax
  \def\url#1{\texttt{#1}}\fi
\expandafter\ifx\csname urlprefix\endcsname\relax\def\urlprefix{URL }\fi
\expandafter\ifx\csname href\endcsname\relax
  \def\href#1#2{#2} \def\path#1{#1}\fi

\bibitem{dinklage2013ncval}
A.~Dinklage, M.~Yokoyama, K.~Tanaka, J.~L. Velasco, D.~L\'opez-Bruna, C.~D.
  Beidler, S.~Satake, E.~Ascas\'ibar, J.~Ar\'evalo, J.~Baldzuhn, Y.~Feng,
  D.~Gates, J.~Geiger, K.~Ida, M.~Jakubowski, A.~L\'opez-Fraguas, H.~Maassberg,
  J.~Miyazawa, T.~Morisaki, S.~Murakami, N.~Pablant, S.~Kobayashi, R.~Seki,
  C.~Suzuki, Y.~Suzuki, Y.~Turkin, A.~Wakasa, R.~Wolf, H.~Yamada, M.~Yoshinuma,
  {LHD Exp. Group}, {TJ-II Team}, {W7-AS Team},
  \href{http://stacks.iop.org/0029-5515/53/i=6/a=063022}{{Inter-machine
  validation study of neoclassical transport modelling in medium- to
  high-density stellarator-heliotron plasmas}}, Nuclear Fusion 53~(6) (2013)
  063022.
\newline\urlprefix\url{http://stacks.iop.org/0029-5515/53/i=6/a=063022}

\bibitem{dinklage2018np}
A.~Dinklage, {the W7-X Team},
  \href{https://doi.org/10.1038/s41567-018-0141-9}{{Magnetic configuration
  effects on the Wendelstein 7-X stellarator}}, Nature Physics 14~(8) (2018)
  855--860.
\newblock \href {http://dx.doi.org/10.1038/s41567-018-0141-9}
  {\path{doi:10.1038/s41567-018-0141-9}}.
\newline\urlprefix\url{https://doi.org/10.1038/s41567-018-0141-9}

\bibitem{hokulsrud1986neo}
D.~D. Ho, R.~M. Kulsrud,
  \href{https://aip.scitation.org/doi/abs/10.1063/1.866395}{Neoclassical
  transport in stellarators}, The Physics of Fluids 30~(2) (1987) 442--461.
\newblock \href {http://dx.doi.org/10.1063/1.866395}
  {\path{doi:10.1063/1.866395}}.
\newline\urlprefix\url{https://aip.scitation.org/doi/abs/10.1063/1.866395}

\bibitem{beidler2011icnts}
C.~D. Beidler, K.~Allmaier, M.~Y. Isaev, S.~V. Kasilov, W.~Kernbichler, G.~O.
  Leitold, H.~Maa{\ss}berg, D.~R. Mikkelsen, S.~Murakami, M.~Schmidt, D.~A.
  Spong, V.~Tribaldos, A.~Wakasa,
  \href{http://stacks.iop.org/0029-5515/51/i=7/a=076001}{{Benchmarking of the
  mono-energetic transport coefficients. Results from the International
  Collaboration on Neoclassical Transport in Stellarators (ICNTS)}}, Nuclear
  Fusion 51~(7) (2011) 076001.
\newline\urlprefix\url{http://stacks.iop.org/0029-5515/51/i=7/a=076001}

\bibitem{calvo2017sqrtnu}
I.~Calvo, F.~I. Parra, J.~L. Velasco, A.~Alonso,
  \href{http://stacks.iop.org/0741-3335/59/i=5/a=055014}{The effect of
  tangential drifts on neoclassical transport in stellarators close to
  omnigeneity}, Plasma Physics and Controlled Fusion 59~(5) (2017) 055014.
\newline\urlprefix\url{http://stacks.iop.org/0741-3335/59/i=5/a=055014}

\bibitem{cary1997omni}
J.~R. Cary, S.~G. Shasharina,
  \href{http://link.aps.org/doi/10.1103/PhysRevLett.78.674}{{Helical plasma
  confinement devices with good confinement properties}}, Physical Review
  Letters 78 (1997) 674--677.
\newline\urlprefix\url{http://link.aps.org/doi/10.1103/PhysRevLett.78.674}

\bibitem{parra2015omni}
F.~I. Parra, I.~Calvo, P.~Helander, M.~Landreman, {Less constrained omnigenous
  stellarators}, Nuclear Fusion 55 (2015) 033005.

\bibitem{boozer1983qs}
A.~H. Boozer,
  \href{https://aip.scitation.org/doi/abs/10.1063/1.864166}{Transport and
  isomorphic equilibria}, The Physics of Fluids 26~(2) (1983) 496--499.
\newblock \href
  {http://arxiv.org/abs/https://aip.scitation.org/doi/pdf/10.1063/1.864166}
  {\path{arXiv:https://aip.scitation.org/doi/pdf/10.1063/1.864166}}, \href
  {http://dx.doi.org/10.1063/1.864166} {\path{doi:10.1063/1.864166}}.
\newline\urlprefix\url{https://aip.scitation.org/doi/abs/10.1063/1.864166}

\bibitem{landreman2012omni}
M.~Landreman, P.~J. Catto, {Omnigenity as generalized quasisymmetry}, Physics
  of Plasmas 19 (2012) 056103.

\bibitem{klinger2017op11}
T.~Klinger, A.~Alonso, S.~Bozhenkov, R.~Burhenn, A.~Dinklage, G.~Fuchert,
  J.~Geiger, O.~Grulke, A.~Langenberg, M.~Hirsch, G.~Kocsis, J.~Knauer,
  A.~Kr{\"a}mer-Flecken, H.~Laqua, S.~Lazerson, M.~Landreman, H.~Maa{\ss}berg,
  S.~Marsen, M.~Otte, N.~Pablant, E.~Pasch, K.~Rahbarnia, T.~Stange,
  T.~Szepesi, H.~Thomsen, P.~Traverso, J.~L. Velasco, T.~Wauters, G.~Weir,
  T.~Windisch, {the Wendelstein 7-X Team},
  \href{http://stacks.iop.org/0741-3335/59/i=1/a=014018}{{Performance and
  properties of the first plasmas of Wendelstein 7-X}}, Plasma Physics and
  Controlled Fusion 59~(1) (2017) 014018.
\newline\urlprefix\url{http://stacks.iop.org/0741-3335/59/i=1/a=014018}

\bibitem{wolf2017op11}
R.~C. Wolf, {et al.},
  \href{http://stacks.iop.org/0029-5515/57/i=10/a=102020}{{Major results from
  the first plasma campaign of the Wendelstein 7-X stellarator}}, Nuclear
  Fusion 57~(10) (2017) 102020.
\newline\urlprefix\url{http://stacks.iop.org/0029-5515/57/i=10/a=102020}

\bibitem{takeiri2017iaea}
Y.~Takeiri, {the LHD team},
  \href{http://stacks.iop.org/0029-5515/57/i=10/a=102023}{{Extension of the
  operational regime of the LHD towards a deuterium experiment}}, Nuclear
  Fusion 57~(10) (2017) 102023.
\newline\urlprefix\url{http://stacks.iop.org/0029-5515/57/i=10/a=102023}

\bibitem{sunnpedersen2016nature}
T.~Sunn-Pedersen, M.~Otte, S.~Lazerson, P.~Helander, S.~Bozhenkov,
  C.~Biedermann, T.~Klinger, R.~Wolf, H.~S. Bosch, {the Wendelstein 7-X Team},
  {Confirmation of the topology of the Wendelstein 7-X magnetic field to better
  than 1:100,000}, Nature Communications 7~(13493).

\bibitem{mynick1982omni}
H.~E. Mynick, T.~K. Chu, A.~H. Boozer,
  \href{http://link.aps.org/doi/10.1103/PhysRevLett.48.322}{Class of model
  stellarator fields with enhanced confinement}, Phyics Review Letters 48
  (1982) 322--326.
\newline\urlprefix\url{http://link.aps.org/doi/10.1103/PhysRevLett.48.322}

\bibitem{yamada2005taue}
H.~Yamada, J.~Harris, A.~Dinklage, E.~Ascasibar, F.~Sano, S.~Okamura,
  J.~Talmadge, U.~Stroth, A.~Kus, S.~Murakami, M.~Yokoyama, C.~Beidler,
  V.~Tribaldos, K.~Watanabe, Y.~Suzuki,
  \href{http://stacks.iop.org/0029-5515/45/i=12/a=024}{{Characterization of
  energy confinement in net-current free plasmas using the extended
  International Stellarator Database}}, Nuclear Fusion 45~(12) (2005) 1684.
\newline\urlprefix\url{http://stacks.iop.org/0029-5515/45/i=12/a=024}

\bibitem{zarnstorff2001ncsx}
M.~C. Zarnstorff, L.~A. Berry, A.~Brooks, E.~Fredrickson, G.-Y. Fu,
  S.~Hirshman, S.~Hudson, L.-P. Ku, E.~Lazarus, D.~Mikkelsen, D.~Monticello,
  G.~H. Neilson, N.~Pomphrey, A.~Reiman, D.~Spong, D.~Strickler, A.~Boozer,
  W.~A. Cooper, R.~Goldston, R.~Hatcher, M.~Isaev, C.~Kessel, J.~Lewandowski,
  J.~F. Lyon, P.~Merkel, H.~Mynick, B.~E. Nelson, C.~Nuehrenberg, M.~Redi,
  W.~Reiersen, P.~Rutherford, R.~Sanchez, J.~Schmidt, R.~B. White,
  \href{https://doi.org/10.1088%2F0741-3335%2F43%2F12a%2F318}{Physics of the
  compact advanced stellarator {NCSX}}, Plasma Physics and Controlled Fusion
  43~(12A) (2001) A237--A249.
\newblock \href {http://dx.doi.org/10.1088/0741-3335/43/12a/318}
  {\path{doi:10.1088/0741-3335/43/12a/318}}.
\newline\urlprefix\url{https://doi.org/10.1088%2F0741-3335%2F43%2F12a%2F318}

\bibitem{sagara2010reactors}
A.~Sagara, Y.~Igitkhanov, F.~Najmabadi,
  \href{http://www.sciencedirect.com/science/article/pii/S0920379610001122}{Review
  of stellarator/heliotron design issues towards mfe demo}, Fusion Engineering
  and Design 85~(7) (2010) 1336 -- 1341, proceedings of the Ninth International
  Symposium on Fusion Nuclear Technology.
\newblock \href
  {http://dx.doi.org/https://doi.org/10.1016/j.fusengdes.2010.03.041}
  {\path{doi:https://doi.org/10.1016/j.fusengdes.2010.03.041}}.
\newline\urlprefix\url{http://www.sciencedirect.com/science/article/pii/S0920379610001122}

\bibitem{fuchert2018taue}
G.~Fuchert, S.~Bozhenkov, N.~Pablant, K.~Rahbarnia, Y.~Turkin, A.~Alonso,
  T.~Andreeva, C.~Beidler, M.~Beurskens, A.~Dinklage, J.~Geiger, M.~Hirsch,
  U.~H{\"o}fel, J.~Knauer, A.~Langenberg, H.~Laqua, H.~Niemann, E.~Pasch, T.~S.
  Pedersen, T.~Stange, J.~Svensson, H.~T. Mora, G.~Wurden, D.~Zhang, R.~W. and,
  \href{https://doi.org/10.1088%2F1741-4326%2Faad78b}{{Global energy
  confinement in the initial limiter configuration of Wendelstein 7-X}},
  Nuclear Fusion 58~(10) (2018) 106029.
\newblock \href {http://dx.doi.org/10.1088/1741-4326/aad78b}
  {\path{doi:10.1088/1741-4326/aad78b}}.
\newline\urlprefix\url{https://doi.org/10.1088%2F1741-4326%2Faad78b}

\bibitem{geiger2014w7x}
J.~Geiger, C.~D. Beidler, Y.~Feng, H.~Maa{\ss}berg, N.~B. Marushchenko,
  Y.~Turkin, \href{http://stacks.iop.org/0741-3335/57/i=1/a=014004}{{Physics in
  the magnetic configuration space of W7-X}}, Plasma Physics and Controlled
  Fusion 57~(1) (2015) 014004.
\newline\urlprefix\url{http://stacks.iop.org/0741-3335/57/i=1/a=014004}

\bibitem{alonso2017eps}
J.~A. Alonso, C.~D. Beidler, I.~Calvo, A.~Dinklage, Y.~Feng, G.~Fuchert,
  M.~Hirsch, M.~Landreman, A.~Langenberg, H.~Maa{\ss}berg, N.~Pablant,
  H.~Smith, J.~L. Velasco, G.~Weir, D.~Zhang, {the W7-X Team}, {Ion heat
  transport in low-density Wendelstein 7-X plasmas}, in: 44th European Physical
  Society Conference on Plasma Physics, Belfast, United Kingdom, 2017.

\bibitem{nemov1999neo}
V.~V. Nemov, S.~V. Kasilov, W.~Kernbichler, M.~F. Heyn, {Evaluation of 1/$\nu$
  neoclassical transport in stellarators}, Physics of Plasmas 6 (1999) 4622.

\bibitem{hirshman1986dkes}
S.~P. Hirshman, K.~C. Shaing, W.~I. van Rij, C.~O. Beasley, E.~C. Crume,
  \href{http://link.aip.org/link/?PFL/29/2951/1}{{Plasma transport coefficients
  for nonsymmetric toroidal confinement systems}}, Physics of Fluids 29~(9)
  (1986) 2951--2959.
\newline\urlprefix\url{http://link.aip.org/link/?PFL/29/2951/1}

\bibitem{turkin2011predictive}
Y.~Turkin, C.~D. Beidler, H.~Maa{\ss}berg, S.~Murakami, V.~Tribaldos,
  A.~Wakasa, \href{http://link.aip.org/link/?PHP/18/022505/1}{{Neoclassical
  transport simulations for stellarators}}, Physics of Plasmas 18~(2) (2011)
  022505.
\newline\urlprefix\url{http://link.aip.org/link/?PHP/18/022505/1}

\bibitem{beidler2007icnts}
C.~D. Beidler, M.~Y. Isaev, S.~V. Kasilov, W.~Kernbichler, H.~M.~S. Murakami,
  V.~V. Nemov, D.~Spong, V.~Tribaldos,
  \href{http://www.nifs.ac.jp/itc/itc17/file/PDF_proceedings/poster2/P2-031.pdf}{{ICNTS-Impact
  of Incompressible E$\times$ B Flow in Estimating Mono-Energetic Transport
  Coefficients}}, in: Proceedings of the 16th Int. Stellarator/Heliotron
  Workshop, Toki, Vol. NIFS-PROC-69, 2007, p. P2.O31.
\newline\urlprefix\url{http://www.nifs.ac.jp/itc/itc17/file/PDF_proceedings/poster2/P2-031.pdf}

\bibitem{calvo2018jpp}
I.~Calvo, J.~L. Velasco, F.~I. Parra, J.~A. Alonso, J.~M. Garc\'ia-Regana,
  Electrostatic potential variations on stellarator magnetic surfaces in low
  collisionality regimes, Journal of Plasma Physics 84~(4) (2018) 905840407.

\bibitem{landreman2014sfincs}
M.~Landreman, H.~Smith, A.~Moll\'en, P.~Helander,
  \href{http://scitation.aip.org/content/aip/journal/pop/21/4/10.1063/1.4870077}{{Comparison
  of particle trajectories and collision operators for collisional transport in
  nonaxisymmetric plasmas}}, Physics of Plasmas 21~(4) (2014) 042503.
\newline\urlprefix\url{http://scitation.aip.org/content/aip/journal/pop/21/4/10.1063/1.4870077}

\bibitem{regana2013euterpe}
J.~M. Garc\'ia-Rega{\~n}a, R.~Kleiber, C.~D. Beidler, Y.~Turkin, H.~Maassberg,
  P.~Helander, \href{http://stacks.iop.org/0741-3335/55/i=7/a=074008}{{On
  neoclassical impurity transport in stellarator geometry}}, Plasma Physics and
  Controlled Fusion 55~(7) (2013) 074008.
\newline\urlprefix\url{http://stacks.iop.org/0741-3335/55/i=7/a=074008}

\bibitem{regana2017phi1}
J.~Garc{\'\i}a-Rega{\~n}a, C.~Beidler, R.~Kleiber, P.~Helander, A.~Moll{\'e}n,
  J.~Alonso, M.~Landreman, H.~Maa{\ss}berg, H.~Smith, Y.~Turkin, J.~Velasco,
  \href{http://stacks.iop.org/0029-5515/57/i=5/a=056004}{Electrostatic
  potential variation on the flux surface and its impact on impurity
  transport}, Nuclear Fusion 57~(5) (2017) 056004.
\newline\urlprefix\url{http://stacks.iop.org/0029-5515/57/i=5/a=056004}

\bibitem{satake2006fortec3d}
S.~Satake, M.~O. an~N~Nakajima, H.~Sugama, M.~Yokoyama,
  \href{https://www.jstage.jst.go.jp/article/pfr/1/0/1_0_002/_article}{{Non-local
  simulation of the formation of neoclassical ambipolar electric field in
  non-axisymmetric configurations}}, Plasma and Fusion Research 1 (2006) 002.
\newline\urlprefix\url{https://www.jstage.jst.go.jp/article/pfr/1/0/1_0_002/_article}

\bibitem{calvo2013er}
I.~Calvo, F.~I. Parra, {J L Velasco}, J.~A. Alonso,
  \href{http://stacks.iop.org/0741-3335/55/i=12/a=125014}{{Stellarators close
  to quasisymmetry}}, Plasma Physics and Controlled Fusion 55~(12) (2013)
  125014.
\newline\urlprefix\url{http://stacks.iop.org/0741-3335/55/i=12/a=125014}

\bibitem{calvo2014er}
I.~Calvo, F.~I. Parra, J.~A. Alonso, {J L Velasco},
  \href{http://stacks.iop.org/0741-3335/56/i=9/a=094003}{{Optimizing
  stellarators for large flows}}, Plasma Physics and Controlled Fusion 56~(9)
  (2014) 094003.
\newline\urlprefix\url{http://stacks.iop.org/0741-3335/56/i=9/a=094003}

\bibitem{calvo2015flowdamping}
I.~Calvo, F.~I. Parra, {J L Velasco}, J.~A. Alonso,
  \href{http://stacks.iop.org/0741-3335/57/i=1/a=014014}{{Flow damping in
  stellarators close to quasisymmetry}}, Plasma Physics and Controlled Fusion
  57~(1) (2015) 014014.
\newline\urlprefix\url{http://stacks.iop.org/0741-3335/57/i=1/a=014014}

\bibitem{velasco2018phi1}
J.~L. Velasco, I.~Calvo, J.~M. Garc{\'\i}a-Rega{\~n}a, F.~I. Parra, S.~Satake,
  J.~A. Alonso, the LHD~team,
  \href{http://stacks.iop.org/0741-3335/60/i=7/a=074004}{Large tangential
  electric fields in plasmas close to temperature screening}, Plasma Physics
  and Controlled Fusion 60~(7) (2018) 074004.
\newline\urlprefix\url{http://stacks.iop.org/0741-3335/60/i=7/a=074004}

\bibitem{numericalrecipes}
W.~H. Press, S.~A. Teukolsky, W.~T. Vetterling, B.~P. Flannery, {Numerical
  Recipes in Fortran 77: the Art of Scientific Computing. Second Edition},
  Vol.~1, Cambridge University Press, 1996.

\bibitem{kernbichler2016neo2}
W.~Kernbichler, S.~Kasilov, G.~Kapper, A.~Martitsch, V.~Nemov, C.~Albert,
  M.~Heyn, {Solution of drift kinetic equation in stellarators and tokamaks
  with broken symmetry using the code NEO-2}, Plasma Physics and Controlled
  Fusion 58 (2016) 104001.

\bibitem{velasco2011bootstrap}
J.~L. Velasco, K.~Allmaier, A.~L. Fraguas, C.~D. Beidler, H.~Maa{\ss}berg,
  W.~Kernbichler, F.~Castej\'on, J.~A. Jim\'enez,
  \href{http://stacks.iop.org/0741-3335/53/i=11/a=115014}{{Calculation of the
  bootstrap current profile for the TJ-II stellarator}}, Plasma Physics and
  Controlled Fusion 53~(11) (2011) 115014.
\newline\urlprefix\url{http://stacks.iop.org/0741-3335/53/i=11/a=115014}

\bibitem{barnes2019stella}
M.~Barnes, F.~Parra, M.~Landreman,
  \href{http://www.sciencedirect.com/science/article/pii/S002199911930066X}{stella:
  An operator-split, implicit--explicit delta-f-gyrokinetic code for general
  magnetic field configurations}, Journal of Computational Physics 391 (2019)
  365 -- 380.
\newblock \href {http://dx.doi.org/https://doi.org/10.1016/j.jcp.2019.01.025}
  {\path{doi:https://doi.org/10.1016/j.jcp.2019.01.025}}.
\newline\urlprefix\url{http://www.sciencedirect.com/science/article/pii/S002199911930066X}

\bibitem{landreman2013intv}
M.~Landreman, D.~R. Ernst,
  \href{http://www.sciencedirect.com/science/article/pii/S0021999113001605}{{New
  velocity-space discretization for continuum kinetic calculations and
  Fokker--Planck collisions}}, Journal of Computational Physics 243 (2013) 130
  -- 150.
\newblock \href {http://dx.doi.org/https://doi.org/10.1016/j.jcp.2013.02.041}
  {\path{doi:https://doi.org/10.1016/j.jcp.2013.02.041}}.
\newline\urlprefix\url{http://www.sciencedirect.com/science/article/pii/S0021999113001605}

\bibitem{mollen2018phi1}
A.~Moll{\'e}n, M.~Landreman, H.~M. Smith, J.~M. Garc{\'\i}a-Rega{\~n}a,
  M.~Nunami,
  \href{http://stacks.iop.org/0741-3335/60/i=8/a=084001}{Flux-surface
  variations of the electrostatic potential in stellarators: impact on the
  radial electric field and neoclassical impurity transport}, Plasma Physics
  and Controlled Fusion 60~(8) (2018) 084001.
\newline\urlprefix\url{http://stacks.iop.org/0741-3335/60/i=8/a=084001}

\bibitem{paul2017iter}
E.~J. Paul, M.~Landreman, F.~M. Poli, D.~A. Spong, H.~M. Smith, W.~Dorland,
  \href{https://doi.org/10.1088\%2F1741-4326\%2Faa7fa4}{Rotation and
  neoclassical ripple transport in {ITER}}, Nuclear Fusion 57~(11) (2017)
  116044.
\newblock \href {http://dx.doi.org/10.1088/1741-4326/aa7fa4}
  {\path{doi:10.1088/1741-4326/aa7fa4}}.
\newline\urlprefix\url{https://doi.org/10.1088\%2F1741-4326\%2Faa7fa4}

\bibitem{matsuoka2015tangential}
S.~Matsuoka, S.~Satake, R.~Kanno, H.~Sugama,
  \href{http://dx.doi.org/10.1063/1.4923434}{Effects of magnetic drift
  tangential to magnetic surfaces on neoclassical transport in non-axisymmetric
  plasmas}, Physics of Plasmas 22~(7) (2015) 072511.
\newblock \href {http://dx.doi.org/10.1063/1.4923434}
  {\path{doi:10.1063/1.4923434}}.
\newline\urlprefix\url{http://dx.doi.org/10.1063/1.4923434}

\bibitem{dherbemont2020fow}
V.~d\'\ Herbemont, F.~I. Parra, I.~Calvo, J.~L. Velasco, {Finite orbit width
  effects in large aspect ratio stellarators,} in preparation.

\bibitem{petsc-efficient}
S.~Balay, W.~D. Gropp, L.~C. McInnes, B.~F. Smith, Efficient management of
  parallelism in object oriented numerical software libraries, in: E.~Arge,
  A.~M. Bruaset, H.~P. Langtangen (Eds.), Modern Software Tools in Scientific
  Computing, Birkh{\"{a}}user Press, 1997, pp. 163--202.

\bibitem{petsc-web-page}
S.~Balay, S.~Abhyankar, M.~F. Adams, J.~Brown, P.~Brune, K.~Buschelman,
  L.~Dalcin, A.~Dener, V.~Eijkhout, W.~D. Gropp, D.~Karpeyev, D.~Kaushik, M.~G.
  Knepley, D.~A. May, L.~C. McInnes, R.~T. Mills, T.~Munson, K.~Rupp, P.~Sanan,
  B.~F. Smith, S.~Zampini, H.~Zhang, H.~Zhang,
  \href{https://www.mcs.anl.gov/petsc}{{PETS}c {W}eb page},
  \url{https://www.mcs.anl.gov/petsc} (2019).
\newline\urlprefix\url{https://www.mcs.anl.gov/petsc}

\bibitem{petsc-user-ref}
S.~Balay, S.~Abhyankar, M.~F. Adams, J.~Brown, P.~Brune, K.~Buschelman,
  L.~Dalcin, A.~Dener, V.~Eijkhout, W.~D. Gropp, D.~Karpeyev, D.~Kaushik, M.~G.
  Knepley, D.~A. May, L.~C. McInnes, R.~T. Mills, T.~Munson, K.~Rupp, P.~Sanan,
  B.~F. Smith, S.~Zampini, H.~Zhang, H.~Zhang,
  \href{https://www.mcs.anl.gov/petsc}{{PETS}c users manual}, Tech. Rep.
  ANL-95/11 - Revision 3.11, Argonne National Laboratory (2019).
\newline\urlprefix\url{https://www.mcs.anl.gov/petsc}

\bibitem{kotschenreuther1995response}
M.~Kotschenreuther, G.~Rewoldt, W.~Tang,
  \href{http://www.sciencedirect.com/science/article/pii/001046559500035E}{Comparison
  of initial value and eigenvalue codes for kinetic toroidal plasma
  instabilities}, Computer Physics Communications 88~(2) (1995) 128 -- 140.
\newblock \href
  {http://dx.doi.org/https://doi.org/10.1016/0010-4655(95)00035-E}
  {\path{doi:https://doi.org/10.1016/0010-4655(95)00035-E}}.
\newline\urlprefix\url{http://www.sciencedirect.com/science/article/pii/001046559500035E}

\bibitem{sunnpedersen2015op11}
T.~S. Pedersen, T.~Andreeva, H.-S. Bosch, S.~Bozhenkov, F.~Effenberg,
  M.~Endler, Y.~Feng, D.~A. Gates, J.~Geiger, D.~Hartmann, H.~H\"olbe,
  M.~Jakubowski, R.~K\"onig, H.~Laqua, S.~Lazerson, M.~Otte, M.~Preynas,
  O.~Schmitz, T.~Stange, Y.~Turkin, {the W7-X Team},
  \href{http://stacks.iop.org/0029-5515/55/i=12/a=126001}{{Plans for the first
  plasma operation of Wendelstein 7-X}}, Nuclear Fusion 55~(12) (2015) 126001.
\newline\urlprefix\url{http://stacks.iop.org/0029-5515/55/i=12/a=126001}

\bibitem{ascasibar2019iaea}
E.~Ascas{\'{\i}}bar, D.~Alba, D.~Alegre, A.~Alonso, J.~Alonso,
  F.~de~Arag{\'{o}}n, A.~Baciero, J.~Barcala, E.~Blanco, J.~Botija, L.~Bueno,
  S.~Cabrera, E.~de~la Cal, I.~Calvo, A.~Cappa, D.~Carralero, R.~Carrasco,
  B.~Carreras, F.~Castej{\'{o}}n, R.~Castro, A.~de~Castro, G.~Catal{\'{a}}n,
  A.~Chmyga, M.~Chamorro, A.~Cooper, A.~Dinklage, L.~Eliseev, T.~Estrada,
  M.~Ezzat, F.~Fern{\'{a}}ndez-Marina, J.~Fontdecaba, L.~Garc{\'{\i}}a,
  I.~Garc{\'{\i}}a-Cort{\'{e}}s, R.~Garc{\'{\i}}a-G{\'{o}}mez,
  J.~Garc{\'{\i}}a-Rega{\~{n}}a, A.~Gonz{\'{a}}lez-Jerez, G.~Grenfell,
  J.~Guasp, J.~Hern{\'{a}}ndez-S{\'{a}}nchez, J.~Hernanz, C.~Hidalgo,
  E.~Hollmann, A.~Jim{\'{e}}nez-Denche, P.~Khabanov, N.~Kharchev,
  I.~Kirpitchev, R.~Kleiber, A.~Kozachek, L.~Krupnik, F.~Lapayese, M.~Liniers,
  B.~Liu, D.~L{\'{o}}pez-Bruna, A.~L{\'{o}}pez-Fraguas, B.~L{\'{o}}pez-Miranda,
  J.~L{\'{o}}pez-R{\'{a}}zola, U.~Losada, E.~de~la Luna, A.~M. de~Aguilera,
  F.~Mart{\'{\i}}n-D{\'{\i}}az, M.~Mart{\'{\i}}nez-Fuentes,
  G.~Mart{\'{\i}}n-G{\'{o}}mez, A.~Mart{\'{\i}}n-Rojo,
  J.~Mart{\'{\i}}nez-Fern{\'{a}}ndez, K.~McCarthy, F.~Medina, M.~Medrano,
  L.~Mel{\'{o}}n, A.~Melnikov, P.~M{\'{e}}ndez, R.~Merino, F.~Miguel, B.~van
  Milligen, A.~Molinero, B.~Momo, P.~Monreal, S.~Mulas, Y.~Narushima,
  M.~Navarro, M.~Ochando, S.~Ohshima, J.~Olivares, E.~Oyarz{\'{a}}bal,
  J.~de~Pablos, L.~Pacios, N.~Panadero, F.~Parra, I.~Pastor, A.~de~la
  Pe{\~{n}}a, A.~Pereira, J.~Pinz{\'{o}}n, A.~Portas, E.~Poveda, J.~Quintana,
  F.~Ramos, G.~Ratt{\'{a}}, M.~Redondo, E.~Rinc{\'{o}}n, L.~R{\'{\i}}os,
  C.~Rodr{\'{\i}}guez-Fern{\'{a}}ndez, L.~Rodr{\'{\i}}guez-Rodrigo, B.~Rojo,
  A.~Ros, E.~Rosa, E.~S{\'{a}}nchez, J.~S{\'{a}}nchez, M.~S{\'{a}}nchez,
  E.~S{\'{a}}nchez-Sarabia, S.~Satake, J.~Sebasti{\'{a}}n, R.~Sharma, C.~Silva,
  E.~Solano, A.~Soleto, B.~Sun, F.~Tabar{\'{e}}s, D.~Tafalla, H.~Takahashi,
  N.~Tamura, A.~Tolkachev, J.~Vega, G.~Velasco, {{J. L. Velasco}}, S.~Yamamoto,
  B.~Z. and, \href{https://doi.org/10.1088%2F1741-4326%2Fab205e}{Overview of
  recent {TJ}-{II} stellarator results}, Nuclear Fusion 59~(11) (2019) 112019.
\newblock \href {http://dx.doi.org/10.1088/1741-4326/ab205e}
  {\path{doi:10.1088/1741-4326/ab205e}}.
\newline\urlprefix\url{https://doi.org/10.1088%2F1741-4326%2Fab205e}

\bibitem{klinger2019op12}
T.~Klinger, {the W7-X Team},
  \href{https://doi.org/10.1088%2F1741-4326%2Fab03a7}{{Overview of first
  Wendelstein 7-X high-performance operation}}, Nuclear Fusion 59~(11) (2019)
  112004.
\newblock \href {http://dx.doi.org/10.1088/1741-4326/ab03a7}
  {\path{doi:10.1088/1741-4326/ab03a7}}.
\newline\urlprefix\url{https://doi.org/10.1088%2F1741-4326%2Fab03a7}

\bibitem{pablant2019ionroot}
N.~A. Pablant, {et al}, {Investigations of the role of neoclassical transport
  in ion-root plasmas on W7-X}, Nuclear Fusion submitted.

\bibitem{carralero2019irw}
D.~Carralero, T.~Estrada, T.~Windisch, J.~L. Velasco, J.~A. Alonso,
  M.~Beurskens, S.~Bozhenkov, H.~Damm, G.~Fuchert, E.~Pasch, G.~Weir, {the W7-X
  Team}, {First V-band Doppler reflectometer results from the OP1.2b campaign
  in Wendelstein 7-X}, in: 14th International Reflectometry Workshop, Lausanne,
  Switzerland, 2019.

\bibitem{calvo2018nf}
I.~Calvo, F.~I. Parra, J.~L. Velasco, J.~A. Alonso, J.~M.
  Garc{\'\i}a-Rega{\~n}a,
  \href{http://stacks.iop.org/0029-5515/58/i=12/a=124005}{Stellarator impurity
  flux driven by electric fields tangent to magnetic surfaces}, Nuclear Fusion
  58~(12) (2018) 124005.
\newline\urlprefix\url{http://stacks.iop.org/0029-5515/58/i=12/a=124005}

\bibitem{buller2018jpp}
S.~Buller, H.~M. Smith, P.~Helander, A.~Moll{\'e}n, S.~L. Newton, I.~Pusztai,
  Effects of flux-surface impurity density variation on collisional transport
  in stellarators, Journal of Plasma Physics 84~(4) (2018) 905840409.

\bibitem{regana2018phi1}
J.~M. Garc{\'\i}a-Rega{\~n}a, T.~Estrada, I.~Calvo, J.~L. Velasco, J.~A.
  Alonso, D.~Carralero, R.~Kleiber, M.~Landreman, A.~Moll{\'e}n,
  E.~S{\'a}nchez, C.~Slaby, T.-I. Team, W.-X. Team,
  \href{http://stacks.iop.org/0741-3335/60/i=10/a=104002}{On-surface potential
  and radial electric field variations in electron root stellarator plasmas},
  Plasma Physics and Controlled Fusion 60~(10) (2018) 104002.
\newline\urlprefix\url{http://stacks.iop.org/0741-3335/60/i=10/a=104002}

\bibitem{fujita2019phi1}
K.~Fujita, S.~Satake, R.~Kanno, M.~Nunami, M.~Nakata, J.~M. G.-R. na, Global
  effects on the variation of ion density and electrostatic potential on the
  flux surface in helical plasmas, Plasma and Fusion Research 14 (2019)
  3403102.

\bibitem{pedrosa2015phi1}
M.~A. Pedrosa, J.~A. Alonso, J.~M. Garc\'ia-Rega{\~n}a, C.~Hidalgo, J.~L.
  Velasco, I.~Calvo, C.~Silva, P.~Helander,
  \href{http://stacks.iop.org/0029-5515/55/i=5/a=052001}{{Electrostatic
  potential variations along flux surfaces in stellarators}}, Nuclear Fusion
  55~(5) (2015) 052001.
\newline\urlprefix\url{http://stacks.iop.org/0029-5515/55/i=5/a=052001}

\bibitem{estrada2019phi1}
T.~Estrada, E.~S\'anchez, J.~M. G.-R. na, J.~A. Alonso, E.~Ascas\'ibar,
  I.~Calvo, A.~Cappa, D.~Carralero, C.~Hidalgo, M.~Liniers, I.~Pastor, J.~L.
  Velasco, \href{https://doi.org/10.1088%2F1741-4326%2Fab1940}{Turbulence and
  perpendicular plasma flow asymmetries measured at {TJ}-{II} plasmas}, Nuclear
  Fusion 59 (2019) 076021.
\newblock \href {http://dx.doi.org/10.1088/1741-4326/ab1940}
  {\path{doi:10.1088/1741-4326/ab1940}}.
\newline\urlprefix\url{https://doi.org/10.1088%2F1741-4326%2Fab1940}

\bibitem{calvo2019anis}
I.~Calvo, F.~I. Parra, J.~L. Velasco, J.~M. Garc{\'\i}a-Rega{\~n}a,
  \href{https://arxiv.org/abs/1907.08482}{Impact of main ion pressure
  anisotropy on stellarator impurity transport}, Nuclear Fusion
  https://arxiv.org/abs/1907.08482.
\newline\urlprefix\url{https://arxiv.org/abs/1907.08482}

\end{thebibliography}

\end{document}